# Graph Iterators:
# Decoupling Graph Structures from Algorithms

## Master Thesis

Marco Nissen

# CONTENTS

# 1. INTRODUCTION

I will present a way to implement graph algorithms which is different from traditional methods. It will be illustrated by implementing five graph algorithms and one rather complex one: a maximum cardinality matching algorithm for non-bipartite graphs. This work was motivated by the belief that some ideas from software engineering should be applied to graph algorithms. I will explain how these ideas can be established in graph algorithm implementation. Part of this work was done in Konstanz[1] (see [NW96]).

*Re-usability* of software is an important and difficult problem in general, and this is particularly true for graph[2] algorithms. The scientific literature demonstrates plenty of applications of graph algorithms as subroutines for other algorithms. Moreover, many practical problems from various domains may be modeled as graph problems and hence solved by means of graph algorithms.

Here, some techniques will be discussed, which can be used to solve some of the problems that arise in the re-use of graph algorithms. It is not easy to understand, but after using iterators in algorithms several times, it becomes clearer.

Chapter 2 introduces some data structures that will be used in 5 basic graph algorithms in chapter 3. Chapter 4 discusses an implementation of a maximum cardinality matching algorithm for general graphs. Chapter 5 explains some techniques in `C++`, which are useful to implement the data structures and algorithms in an efficient way. Finally chapter 6 contains some concluding remarks.

The standard template library[3] demonstrates a mechanism for decoupling

---

[1] While visiting the algorithms and data structures group of the university of Konstanz (Germany) in the project "efficient, reusable implementations of graph algorithms" in summer 1996. This project is supported by DFG special program on "Efficient Algorithms for Discrete Problems and Their Applications" within the project, "Robust algorithms for path and flow problems" (responsible: Dorothea Wagner).

[2] A graph is informally a network structure; for a complete definition see section 2.

[3] The Standard Template Library is a library consisting of data structures and algorithms, implemented in `C++`. It tries to bring together flexibility with efficiency using advanced techniques in `C++`-programming. For further reference, see [MS96]

data representations from actual algorithms. For example, a quick sort algorithm does not work on a concrete list, but it uses objects, which are called **iterators** and can traverse the list. This makes it easy to customize the algorithm to work with different data structures, e.g. an array can be sorted by using the same algorithm, but with iterators traversing the array.

Unfortunately, in the STL there is no graph representation class like the `graph` class of LEDA[4]. It contains a large collection of graph algorithms based on this class. An attempt to exchange the underlying data structure or to change the functionality would cause additional overhead either in time spent for programming or in performance of program execution. For example, if you would like to use LEDA's shortest path algorithm, which is based on the LEDA `graph` class, with another graph representation you might have to convert the new graph object to a LEDA `graph` object. This is not always straightforward, and it may heavily increase the algorithm execution time. A second solution is to write a **wrapper class** that adapts the syntax and semantics of the interface (all methods and their signatures) from the new graph class to the LEDA graph class, but this can be very complex depending on which conversion we have to do.

As a concrete example I can mention a GIS-product (GIS: geographical information system) of a company that is located in the Saarland. It was presented 1996 at the EITC in Brussels (European Information Technology Congress) by Christian Uhrig and myself as an example for adding new functionality to existing software. This company implemented topological structures and databases but they did not know how to compute the fastest bus route for two locations.

In theory, this is no problem: Dijkstra's shortest path algorithm can do it. Unfortunately they had their own data representation, which was completely different from a LEDA graph. The use of LEDA's shortest path algorithm led to conversion of the data, because the algorithm implementation was too inflexible.

In terms of software engineering it is desirable to avoid the need for such conversions or adaptation of algorithms. The approach in the following chapters can be summarized as adding flexibility to algorithms, especially in the realm of graph algorithms.

In the following I will try to develop a complete picture of the approach:

---

[4] LEDA means library of efficient data types and algorithms. It consists of both simple and higher level data structures together with combinatorial and geometrical algorithms. see [MN95] and [LEDA]

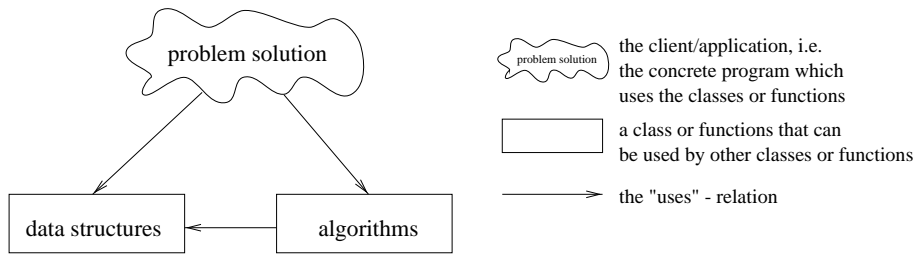

*Fig. 1.1:* traditional way to develop problem solutions: write algorithms that work for specific data structures.

A quick sort might be implemented as an algorithm that operates directly on a C-array. Suppose that the user wants to reuse the code to sort a self-defined structure, which actually represents an array of integers, but is completely different from a C-array. Then, the algorithm must be adaptable to the different data structure.

**Observation 1** *The algorithm has to be adapted because the algorithm must work on different data structures.*

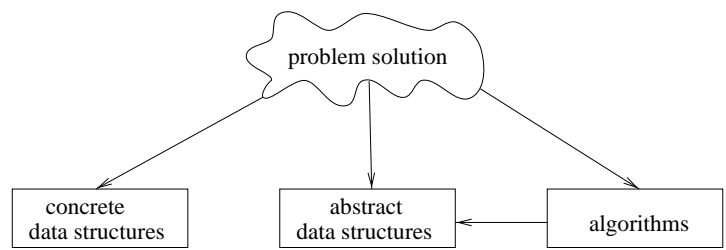

*Fig. 1.2:* first abstraction on data structures: algorithms work on several data structures (e.g. lists, vectors, graphs)

At a first glance, a solution to the problem in the first observation is to standardize the interface of the algorithm (i.e., name and signature of the function) and to let it operate on abstract data structures such as lists, vectors, matrices or graphs.

LEDA is an example of a library that permits reuse of complex algorithms that work with LEDA data structures; with that, a project that has been developed from the beginning with LEDA, can take advantage of its re-usability, because user specific structures can be easily mapped to, or designed with LEDA structures.

In practice, it cannot be assumed that a project was designed with LEDA from scratch. Converting a user defined graph structure to the LEDA graph structure might cause significant runtime overhead, as mentioned above.
Suppose, we want to run a shortest path algorithm such as Dijkstra on it: we
may try to adapt the algorithm, i.e. copy the LEDA source code and adapt
the copy — this is not reuse in its true sense and is highly error-prone. This
is clearly seen if we want to reuse an algorithm of the LEDA library on the
self–defined graph structure "fast graph" (section 5.5) as an example: if we
only have an implementation of Dijkstra's algorithm like LEDA's, we had to
adapt the internals of the algorithm.

**Observation 2** *A library may not be generic enough to handle data repre-
sentations that are completely different in the interface (i.e. different names
or signatures) and functionality.*

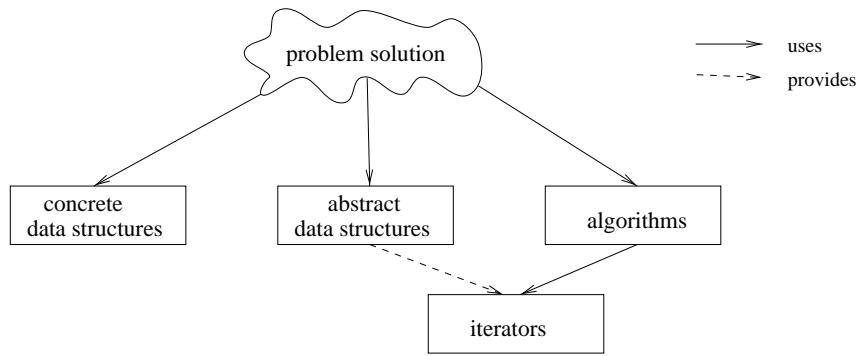

*Fig. 1.3:* more abstraction: algorithms may work on different data structures that
         provide uniform access through iterators

An advanced solution appears in the design of the STL, where algorithms
and data structures are decoupled.
Containers, such as lists, vectors or queues give **uniform access** to the
items in the sequences by providing **iterators**. For example, lists provide
bi-directional iterators, which can move forward or backward through the
structure.
Algorithms are not based on having certain structures such as lists or vectors,
but they need iterators with certain requirements. For example, quick sort
needs iterators that can access the items of the sequence to be sorted in a
random manner.
The result is that the algorithms are independent of the concrete realization
of the underlying data structure – they only need iterators that are able to
traverse a structure, retrieve elements from it or change the value of them.
It is now easy to exchange the underlying data structures as long as they
provide the right iterators.

Unfortunately, though algorithms are independent of data structures, they are not very flexible. If an algorithm is realized as a function, it may return the complete result in form of a single data structure. For example, LEDA provides an algorithm that computes a depth first search in a graph and records each visited node in a list that will be returned after the search has been completed. Suppose we want to give each node a certain color. We can do that by first running the dfs-algorithm and then traversing the resulting list. The additional storage of nodes as a result of a depth first search will be redundant if the algorithm is designed in a flexible way where the coloring is done while traversing the graph that means while the algorithm is running. This is even more important if are interested in intermediate results of an algorithm, which are not accessible if we used only single functions.

Another example: suppose we have a function that computes shortest paths in graphs, and we want to know the smallest distance between two locations. A common implementation to do that is the computation of the predecessor tree. But note that we could suspend the algorithm once the second location is reached.

**Observation 3** *Algorithms must be re-implemented or adapted if the functionality has to be changed.*

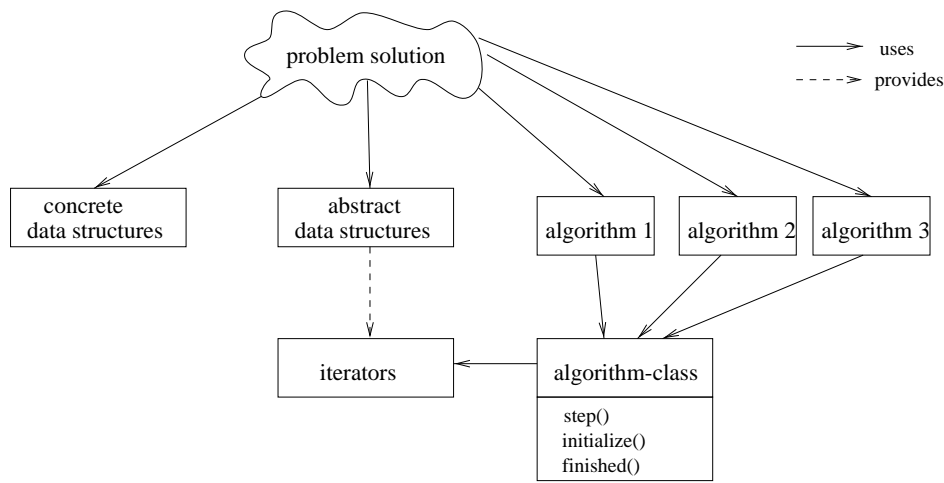

*Fig. 1.4:* abstraction of the algorithm: an algorithm class is more flexible than a function — functions may use this class and provide the same interface (i.e. equal signature) to the problem solution as functions that do not use this class

More flexibility in the functionality of algorithms can be achieved by introducing **algorithm classes** that can be instantiated. An object of such a class has parameter values and an internal state. In chapter 3 the advantages of this design will be discussed in detail.

If algorithms are defined as abstract algorithms, they can be specialized to solve more concrete problems. For example, let some train stations and routes be represented by a graph. Then, an abstract algorithm would for instance be the computation of shortest paths in a more general sense, i.e. a very generic implementation of shortest paths. A specialization would be the computation of shortest paths between two train stations.

Although this design adds flexibility and re-usability to algorithms it is not clear if they still remain efficient, but this can achieved by using the capability of `C++` to write generic classes with the template mechanism (see chapter 5).

Imagine now that there is a sequence of items, each consisting of several attributes, with one iterator that represents an item in a sequence. In STL, every kind of iterator always refer to only one distinct attribute, but not several ones. Therefore we need a more powerful design. Graph algorithms need auxiliary structures like LEDA's `node_array<T>` (or `GRAPH<T1,T2>`) that yields for a given node a value of an arbitrary but fixed type `T`.

For example, the Dijkstra shortest path algorithm operates with distances and lengths. Different algorithms could be implemented to work on distances, stored in an array, stored directly in a node or somewhere else. The same cases apply to the length values and edges, or they might be computed on–line. We can dramatically simplify the effort needed for realizing the different cases and the usage by having an abstract view on data access. We will be rewarded by increased simplicity and flexibility in the design of the algorithm.

**Observation 4** *Items of sequences may be associated with several attributes.*

The solution given here was first mentioned in [KW97b] and introduces an indirection into the access to the data, called **data accessor**.

Algorithms will not access directly the underlying data, but an intermediate structure, which has a standardized interface (i.e. standardized function names with uniform signatures).

For example in a depth first search the Boolean value which marks visited nodes can be stored with a data accessor *mark*: given an iterator *it* that represents a node *v*, a value can be set or read with *mark(it)*. This remains the same even if we store the values in an array, in a list, a database, if we compute them online or if we have them in a self-defined structure.

To summarize, this thesis presents an applicable implementation of the last picture that is — as we will see — flexible, reusable and efficient.

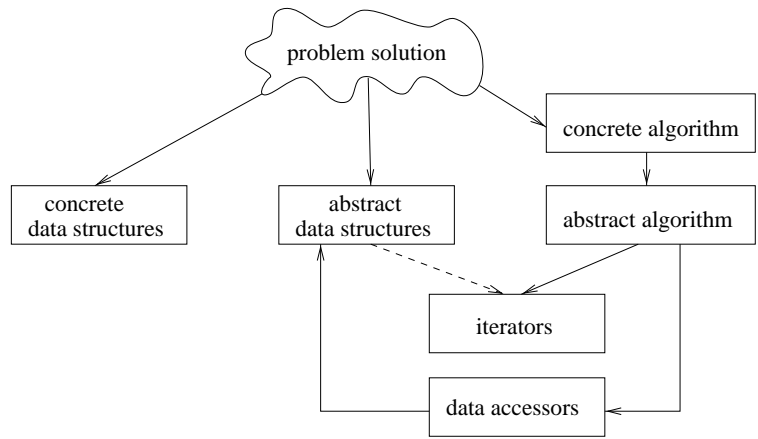

*Fig. 1.5:* abstraction of data access: algorithm classes use the iterators and data accessors to the access the graph structure and associated values

Some parts of the results were integrated into LEDA: release 3.6. contains iterators and data accessors already specialized for LEDA types, which helps new users to learn the "iterator-driven" style of writing graph algorithms. In addition to the specialized data structures, the graph-iterator LEP [5] has templatized versions, which can be adapted to other graph representations.

---

[5] Leda Extension Package: a package that contains classes that are designed in the same way as LEDA but is not intended to be used by all LEDA users. Additionally, the graph-iterator LEP contains all examples of graph algorithms presented here, both as algorithm-classes or algorithm-iterators (for example, `DFS_It` traverses the nodes of a graph in depth first order). See [GitLep] for further reference.

# 2. DATA STRUCTURES

In this chapter we define the terminology used in this thesis: A **(directed) graph** $G = (V_G, E_G)$ consists of a finite node set $V_G$ and an edge set $E_G \subseteq V_G \times V_G$. For an edge $(v, w) \in E_G$, $v$ is the **source node** and $w$ the **target node**.

$v, w \in V_G$ are **adjacent** if and only if there is an edge $(v, w)$ or $(w, v) \in E_G$. The edge $(v, w) \in E_G$ is called **incident** to $v$ and $w$.

We refer to $Adj(v) = \{w | \exists (v, w) \in E_G\}$ as the **adjacency list** of $v$.

The edge $(v, w) \in E_G$ is an **outgoing edge** of $v \in V_G$ and an **incoming edge** of $w \in V_G$.

Accordingly, $out(v) = \{(v, w) | (v, w) \in E_G\}$ is the list of outgoing edges of $v \in V_G$ and $in(w) = \{(v, w) | (v, w) \in E_G\}$ is the list of incoming edges of $w \in V_G$.

$P = (e_1, \ldots e_n)$ is a **path** in $G = (V_G, E_G)$ if and only if $e_j \in V_G$ (for $0 \leq j \leq n$) and $e_j = (s_j, t_j) \Rightarrow t_j = s_{j+1}$ (for $0 \leq j < n$). $distance(v, w) = c$ is called the distance of $v, w \in V_G$ if and only if $c = \min\{k | P = (e_1, \ldots e_k) \text{ is a path in } G \wedge e_1 = v \wedge e_k = w\}$.

A path $P = (e_1, \ldots e_n)$ **connects** two vertices $v, w$ if and only if $v = source(e_1) \wedge w = target(e_n)$ or $w = source(e_1) \wedge v = target(e_n)$. $C \subseteq V_G$ is a **connected component** of $G = (V_G, E_G)$ if and only if $\forall v, w \in C \Rightarrow \exists$ path $p$ in $G$ that connects $v$ and $w$.

## 2.1 Overview

The data structures presented here serve as an indirection to the underlying data representation[6].

The advantage of this design is that algorithms can be written, which can easily be adapted to other data representations without causing too much performance overhead.

---

[6] Only for motivation: by using the template mechanism of `C++`, it was possible to implement this indirection without adding large constant factors to the runtime (see chapter 5).

The first part will be **iterators**[7] that replace the concrete graph primitives node and edge. Nodes and edges will still be used, but only internally, and may be exchanged by the user to objects in completely different graph representations. Iterators can **traverse the different domains** of a graph, i.e. nodes, edges or adjacent nodes of a fixed node.

An iterator *it* provides the following functionality:

- *it* refers to an item in a sequence of graph primitives (node or edge)

- *it* can be updated to an arbitrary location

- *it* can be sequentially moved in the sequence

- test for validity of *it*, i.e. does *it* refer to a valid item in the sequence

- test for equality, i.e. do $it_1$ and $it_2$ refer to the same items in the sequence

The second part consists of a collection of data accessors that decouple the parameters associated with nodes and edges. We will use the term of *attributes* to denote a set of values, and an attribute value is one element of this set. A data accessor *da* is able to do the following:

- *da* represents an attribute of nodes or edges

- uniform read and / or write access is done by iterators

Figure 2.1 shows the domain of a node iterator. For example, it can point to the node "c" and after application of the increment-method[8] it will point to the next item of the sequence, which is node "d". In the abstract view, we maintain two sets of nodes: the first one contains all nodes which have been seen and the second one contains all nodes which have not been seen yet. Every time we move to a new node, we delete the new node from the second set and insert it to the first one. In practice, it is more efficient to think of lists of nodes and traversing this list linearly from the beginning to the end. Edge iterators are defined analogously, i.e. the predecessor for edge 5 is edge 6.

**Adjacency iterators** such as in figure 2.2 are different. They can iterate through the set of outgoing or incoming edges of a node, which is seen as "fixed". Additionally, they can follow the current path by taking the target node of the current incident edge as the new fixed node. Every time an adjacency iterator is instantiated or assigned a new node, the edge is set to the first outgoing (or incoming) edge of this node.

---

[7] Iterators were introduced in [GHJV95] as a design pattern, see there for a more abstract view on this topic.

[8] This is for example in STL the pre-increment operator.

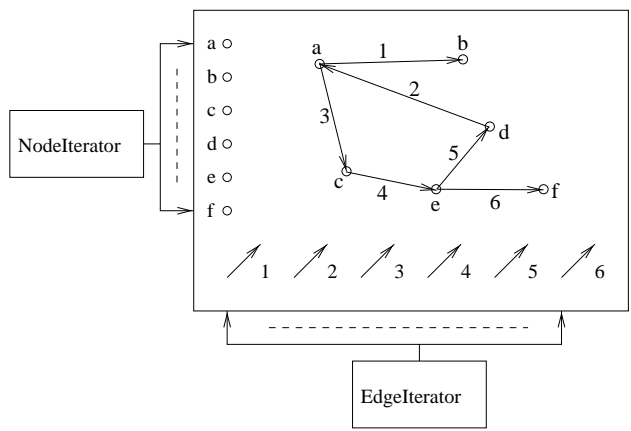

*Fig. 2.1:* Node iterators (edge iterators) can iterate linearly over the node set (edge set) of a graph. Here, the node iterator can iterator through nodes 'a' to 'f' and the edge iterator from edge '1' to edge '6'.

Moreover, additional iterator classes are provided, which can be "wrapped around" existing iterator classes and change the iterator's functionality.

## 2.2 Preliminaries

In the following, the concept of *language independent class definitions* will be introduced, which will help us define the syntax and semantics of e.g. iterators. First we define the data structures and functions used in preceding chapters. Note that $A^*$ denotes all ordered subsets $a = (a_1, a_2, \dots a_k)$ for all $k \geq 0$ where $a_i \in A$.

$$
\begin{aligned}
Id &= \text{\textit{set of identifiers}} \\
char &= \text{\textit{set of characters}} \\
name &: Id \rightarrow char^* \text{\textit{(name for an identifier)}} \\
state &: Id \rightarrow char^* \text{\textit{(state name for an identifier)}} \\
arglist &: Id \rightarrow Id^* \text{\textit{(argument list)}} \\
nexpr &: Id \rightarrow \mathcal{N} \text{\textit{(length of ordered list)}} \\
expr &: Id \rightarrow Expr \text{\textit{(single expression)}}
\end{aligned}
$$

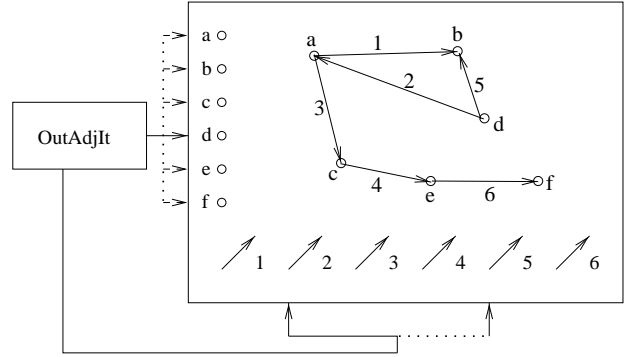

*Fig. 2.2:* Adjacency iterators iterate over the incident edges of a fixed node: For example, the fixed node of the iterator is currently *d* and it can iterate over the incident edges ('2' and '5').

$$lcondition, rcondition \quad : \quad Id \rightarrow BoolExpr \, (conditions \ for \ invariant)$$

$BoolExpr$ and $Expr$ are string expressions (according to the syntax definition in 7.1): $BoolExpr, Expr \subset char^*$.

**Definition 2.1** *A language independent class definition* (LICD) *is a collection of state variables, methods that can modify them, or functions that return values together with a name for that class. Creation methods are mandatory. State modifying methods are defined for each state that is going to be modified. Invariants may be checked as well. It consists of the following sections:*

1. **states:** $S = \{s_1, s_2, \ldots s_{k_1}\}$: $\forall s \in S : s \in Id \wedge name(s) \neq \varepsilon$
   [*a class contains $k_1$ states that have names*]

2. **methods:** $M = \{m_1, m_2, \ldots m_{k_2}\}$ :
   $\forall m \in M : name(m) \neq \varepsilon \wedge \forall a \in arglist(m) : name(a) \neq \varepsilon \wedge$
   $\exists s \in S : name(s) = state(m) \wedge$
   *after application of $m \in M$: $state(m) \equiv expr(m)$*
   [*a method has $k_2$ arguments and changes one state of the class*]

3. **functions:** $F = \{f_1, f_2, \ldots f_{k_3}\}$ :
   $\forall f \in F : name(f) \neq \varepsilon \wedge \forall a \in arglist(f) : name(a) \neq \varepsilon \wedge$
   *after application of $f \in F$: $f \equiv expr(m)$*
   [*a function has $k_3$ arguments and returns an expression*]

4. **invariants:** $I = \{i_1, i_2, \ldots i_{k_4}\}$ :
   $\forall i \in I : lcondition(i) \Rightarrow rcondition(i)$

5. $\exists m \in M : name(m) =$`Creation`
   [*creation methods are mandatory*]

6. $\forall s \in S : \exists m \in M : name(m) =$`Creation` $\wedge state(m) = s$
   [*for each state, there must be a creation method that initializes it*]

A LICD is written as follows:

- If the class name is introduced with "*interface*", the described class is only a description of the interface, i.e. the listed methods are required for a class of that type. It can be compared to the interface concept of Java, where no variables or method definitions appear, but only method declarations. If the keyword "*interface*" does not appear, each method and function, as well as the states, must be defined.

- To add convenience to the description of a LICD, we omit the definition of the underlying graph as state $G$, since it appears in every class description. It will be initialized automatically with the underlying graph.

- The keyword "*invariant*" introduces an invariant that may be checked for runtime correctness.

- The creation methods initialize all states.

- A method looks like this: `Methodname`.$state(PL) \equiv expression$

  That means, after applying method `Methodname` with the list of arguments according to $PL$, *state* will be equivalent with *expression* (here, equivalence means they are semantically identical)

- A Function looks like this: `Functionname`$(PL) \equiv expression$

  That means, function `Functionname` with the list of arguments according to $PL$ will compute a result that is equivalent to *expression*

- An expression (like above "*expression*") is any expression (arithmetic, set, boolean) or application of a mathematical function. It may contain applications of methods or functions of:

  - The use of a method $O.$`method`$(p_1, \ldots p_k)$ yields $O$ after applying the method description.

- The use of a method $O.\texttt{method}(p_1, \ldots p_k).s$ yields the state $s$ of $O$ after applying the method description.

- The use of a function $O.\texttt{function}(p_1, \ldots p_k)$ yields the result of the evaluation of the function description.

- $it.a()^7$ means application of method $a()$ of class $it$ 7 times

- $\texttt{this}$ refers to the object of a class itself.

**Example:**

---

**natural number** for $N \subset \mathcal{N}$:

1. $number \in N$ (internal state)

2. $valid \in boolean$ (internal state)

3. $\texttt{Creation}(n).number \equiv n$

4. $\texttt{Creation}(n).valid$ $\equiv$
$$\begin{cases} \texttt{true} & \text{if } number \in N \\ \texttt{false} & \text{otherwise} \end{cases}$$

5. $\texttt{Succ}().number \equiv number + 1$

6. $\texttt{Succ}().valid \equiv \begin{cases} \texttt{true} & \text{if } number + 1 \in N \\ \texttt{false} & \text{otherwise} \end{cases}$

7. $\texttt{Valid}() \equiv valid$

---

This describes a class, where an instance of it may be created using the creation method, which initializes the internal state $\texttt{number}$. The internal number can be modified using method $\texttt{Succ()}$ and can be tested for membership in $N$ using method $\texttt{Valid()}$.

Other descriptions may be more powerful than LICD, but as we will see, this will be enough for defining iterators.

## 2.3   Iterators

We will discuss a more *formal approach* according to the informal description of iterators (section 2.1).

The semantics of iterators will be described by several definitions, where a special notation for classes of objects was used. We look at iterators as objects with certain internal states and methods that modify these states. We will see a new simple model that can easily be transferred to any object oriented language such as `C++` or Java.

Now for each type of iterator we need its own LICD, first first must define the interface of iterators:

**L.I.C.Definition 2.1** *interface* **iterator**:

1. `Succ()`

2. `Valid()`

Informally, this means: if an iterator refers to an object of a sequence, `Succ()` "moves" the iterator to the next object in the sequence. If there is no successor object, it will be seen as "invalid". `Valid()` returns `true` if the iterator refers to an object of the sequence and false otherwise. Of course, this is not described by the interface above, but this is what we informally associate with the notion of an iterator. We will be more precise when we introduce node, edge and adjacency iterators.

Again some definitions: $G = (V, E)$ formulates the graph, while $V$ is the node set and $E$ its edge set. Assume that $V \cap E = \emptyset$. Let $V^\varepsilon = V \cup \{\varepsilon\}$ and $E^\varepsilon = E \cup \{\varepsilon\}$. We have two bijective ordering functions $\sigma_V : \mathcal{N} \to V$ and $\sigma_E : \mathcal{N} \to E$, where $\mathcal{N}$ is the set of natural numbers $\{1, 2, \ldots\}$. Consequently, we have two additional ordering functions $\sigma_{in} : V \times \mathcal{N} \to E^\varepsilon$ for incoming edges and $\sigma_{out} : V \times \mathcal{N} \to E^\varepsilon$ for outgoing edges. *source, target* $: E \to V^\varepsilon$ yield for a given edge the source or target node. Let $m$ be the number of edges and $n$ the number of nodes in $G$. With these mappings we can formulate additional functions to compute the successor object of a given node or edge:

$$advance_v(x) = \begin{cases} \sigma_V(i+1) & \text{if } \exists i : 1 \le i < n \wedge \sigma_V(i) = x \\ \varepsilon & \text{otherwise} \end{cases} \qquad (2.1)$$

$$advance_e(x) = \begin{cases} \sigma_E(i+1) & \text{if } \exists i : 1 \le i < m \wedge \sigma_E(i) = x \\ \varepsilon & \text{otherwise} \end{cases} \qquad (2.2)$$

| Function | Purpose |
|---|---|
| $advance_v : V^\varepsilon \mapsto V^\varepsilon$ | for each node: computes successor node |
| | $\varepsilon$ if last node |
| $advance_e : E^\varepsilon \mapsto E^\varepsilon$ | for each edge: computes successor edge |
| | $\varepsilon$ if last edge |
| $advance_{out} : E^\varepsilon \mapsto E^\varepsilon$ | for each edge: computes successor edge $e$ in the list of incident outgoing edges of source of $e$ |
| | $\varepsilon$ if last edge |
| $advance_{in} : E^\varepsilon \mapsto E^\varepsilon$ | for each edge: computes successor edge $e$ in the list of incident incoming edges of target of $e$ |
| | $\varepsilon$ if last edge |

*Tab. 2.1:* An overview of the different functions that define the abstract interface to the underlying graph representation.

Note that in the sections, which introduce safe iterators (section **??**) and contraction iterators (section **??**), we omit the definition of method `Valid()` because they remain completely identical. For example, the valid-method of a node contraction iterator is the same as the valid-method of (simple) node iterators.

**L.I.C.Definition 2.2 node iterator** *for $G = (V, E)$:*

1. $v \in V \cup \{\varepsilon\}$

2. `Creation`$(v').v \equiv v'$

3. `Succ`$().v \equiv advance_v(v)$

4. `Valid`$() \equiv \begin{cases} \texttt{true} & \textit{if } v \in V \\ \texttt{false} & \textit{otherwise} \end{cases}$

**L.I.C.Definition 2.3 edge iterator** *for $G = (V, E)$:*

1. $e \in E \cup \{\varepsilon\}$

2. `Creation`$(e').e \equiv e'$

3. `Succ`$().e \equiv advance_e(e)$

4. `Valid`$() \equiv \begin{cases} \texttt{true} & \textit{if } e \in E \\ \texttt{false} & \textit{otherwise} \end{cases}$

Informally, this means: the iterator refers either to an element of the node set $V$ or is invalid ($v = \varepsilon$). It is created by initializing the object $v$ with a given node. $\texttt{Succ}()$ moves the object to the next node in the node set or makes it invalid if there is no such successor node. For edge iterators, it follows the same completely analogous.

Since we have now defined the iterators, which iterate through the sequences of nodes and edges linearly, we want to define the adjacency iterators. These need other advance-functions:

$$advance_{out}(e) = \begin{cases} \sigma_{out}(source(e), i+1) & \text{if } \exists i : 1 \leq i \wedge \sigma_{out}(source(e), i) = e \\ \varepsilon & \text{otherwise} \end{cases}$$
$$(2.3)$$

$$advance_{in}(e) = \begin{cases} \sigma_{in}(target(e), i+1) & \text{if } \exists i : 1 \leq i \wedge \sigma_{in}(target(e), i) = x \\ \varepsilon & \text{otherwise} \end{cases}$$
$$(2.4)$$

Now we are able to define adjacency iterators, which are a bit more complicated:

**L.I.C.Definition 2.4 adjacency iterator for outgoing edges** *for $G = (V, E)$:*

1. *$v \in V \cup \{\varepsilon\}$*

2. *$e \in E \cup \{\varepsilon\}$*

3. *invariant $e \neq \varepsilon \Rightarrow source(e) = v$*

4. *$\texttt{Creation}(v').v \equiv v'$ and $\texttt{Creation}(v').e \equiv \sigma_{out}(v', 1)$*

5. *$\texttt{Creation}(v', e').v \equiv v'$ and $\texttt{Creation}(v', e').e \equiv$*
   *$\begin{cases} e' & \text{if } source(e') = v' \\ \varepsilon & \text{otherwise} \end{cases}$*

6. *$\texttt{Succ}().e \equiv advance_{out}(e)$ and $\texttt{Succ}().v \equiv v$*

7. *$\texttt{Valid}().e \equiv \begin{cases} \texttt{true} & \text{if } e \in E \\ \texttt{false} & \text{otherwise} \end{cases}$*

8. *$\texttt{CurrAdj}().v \equiv target(e)$ and $\texttt{CurrAdj}().e \equiv \sigma_{out}(target(e), 1)$*

Informally, this means: an adjacency iterator of this kind refers to a node and an edge of the node set and edge set, respectively. If the iterator refers to a valid edge, the node must be the source node of the edge. An iterator may be initialized with a node, and the edge will be the first outgoing edge. If the iterator is initialized with a node and an edge, it will be checked if the invariant is true. Here, we have an additional function `CurrAdj()` that initializes the node with the target node of the edge. The edge will then be the first outgoing edge of that node. This will be useful for traversing graphs in depth.

**L.I.C.Definition 2.5 adjacency iterator for incoming edges** *for $G = (V, E)$:*

1. *$v \in V \cup \{\varepsilon\}$*

2. *$e \in E \cup \{\varepsilon\}$*

3. *invariant $e \neq \varepsilon \Rightarrow target(e) = v$*

4. *`Creation(v').v` $\equiv v'$ and `Creation(v').e` $\equiv \sigma_{in}(v', 1)$*

5. *`Creation(v', e').v` $\equiv$ $v'$ and `Creation(v', e').e` $\equiv$*
   $\begin{cases} e' & \text{if } target(e') = v' \\ \varepsilon & \text{otherwise} \end{cases}$

6. *`Succ().e` $\equiv advance_{in}(e)$ and `Succ().v` $\equiv v$*

7. *`Valid()` $\equiv$* $\begin{cases} \texttt{true} & \text{if } e \in E \\ \texttt{false} & \text{otherwise} \end{cases}$

8. *`CurrAdj().v` $\equiv source(e)$ and `CurrAdj().e` $\equiv \sigma_{in}(source(e), 1)$*

This kind of iterator is again defined completely analogous to the previous adjacency iterator, except that we consider incoming edges instead of outgoing edges.

There are four iterator classes, which are already specialized for the LEDA environment (see figure 2.3).

This scheme is orthogonal to the STL iterator taxonomy, because every graph iterator can be used as an input or output iterator (with the appropriate wrapper class[9]), forward or backward iterator, or random access iterator.

---

[9] Wrapper classes can be "wrapped around" existing classes and attach additional responsibilities and functionalities. It is a design pattern. They document specific reoccurring problems and solutions and are an abstraction of common design occurrences (see [GHJV95]). The iterators discussed here are realizations of the pattern iterator, applied to graph structures (see [GHJV95]).

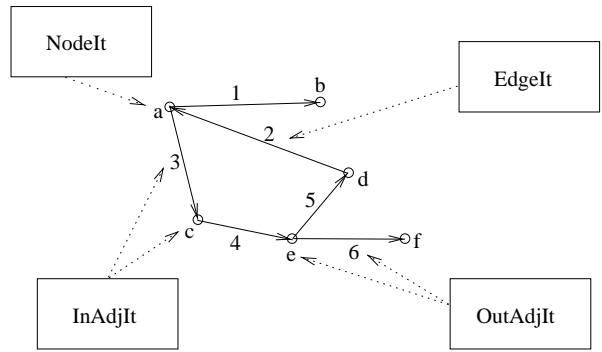

*Fig. 2.3:* These are the four basic iterator classes node iterator, edge iterator, adjacency iterator for outgoing edges and adjacency iterator for incoming edges.

For example, random access iterators provide random access to the items of a sequence. If we provide for node iterators a method that sets the internal node to a given node, then we have random access for nodes.

Unfortunately, if we provide such a function, we cannot guarantee anymore that the sequence is traversed consecutively. Suppose, if a graph is too large for the main memory, this could end up in random access to the hard disc data. This is true in general, because we do not know how the graph is stored.

### *2.3.1 Safe Iterators*

Still, it is not clear what happens if nodes are deleted, because iterators could refer to nodes that do not exist anymore. To prevent this, we introduce a second kind of iterators, iterators that are "safe" in the following sense: if any node or edge is deleted, every **safe iterator** that uses the object will be informed to search for a new object ("`Refresh()`").

Again there will be some terminology to define what we mean by deleting and inserting nodes or edges:

An insertion of a node $v$ into the given graph $G$ yields a graph $G'$ where $G = (V_G, E_G), G' = (V_G \cup \{v\}, E_G)$. Analogously, an insertion of an edge $e$ into the given graph $G$ yields a graph $G'$ where $G = (V_G, E_G), G' = (V_G, E_G \cup \{e\})$.

A deletion of a node $v \in V_G$ from the given graph $G$ yields a graph $G'$ where $G = (V_G, E_G), G' = (V_G - \{v\}, E_G)$. Analogously, a deletion of an edge $e \in E_G$ from the given graph $G$ yields a graph $G'$ where $G = (V_G, E_G), G' = (V_G, E_G - \{e\})$.

**L.I.C.Definition 2.6** *interface* **safe iterator***:*

1. `Succ()`

2. `Valid()`

3. `Delete`($x$)

4. `Insert`($x$)

5. `Refresh`($x$)

These iterators operate on a class, which is "wrapped around" the underlying graph representation. This wrapper class implements an administrative structure *nmap* (*emap*) that knows at any time for any node (edge) every iterator, which refers to the node (the edge).

**L.I.C.Definition 2.7 safe graph** *for* $G = (V, E)$*:*

1. *nmap* $: V \rightarrow \{$*safe iterator*$\}^*$

2. *emap* $: E \rightarrow \{$*safe iterator*$\}^*$

3. *in the following: if a given node $v$ or edge $e$ equals $\varepsilon$, nmap and emap remains the same (only for convenience)*

4. `Creation`($G$)*.nmap* $\equiv \forall v \in V : nmap(v) = \emptyset$

| Function | Purpose |
|----------|---------|
| $refresh(e, v)$ | computes for a given edge $e$ the next successor edge in the sequence of edges, where the source and target of the new edge are different from $v$. |

*Tab. 2.2:* The refresh function refreshes iterators.

5. $\texttt{Creation}(G).emap \equiv \forall e \in E : emap(e) = \emptyset$

6. $\texttt{InsertNode}(v, it).nmap \equiv nmap \cup \{(v, nmap(v) \cup \{it\})\}$

7. $\texttt{InsertEdge}(e, it).emap \equiv emap \cup \{(e, emap(e) \cup \{it\})\}$

8. $\texttt{RemoveNode}(v, it).nmap \equiv nmap \cup \{(v, nmap(v) - \{it\})\}$

9. $\texttt{RemoveEdge}(e, it).emap \equiv emap \cup \{(e, emap(e) - \{it\})\}$

10. $\texttt{RemoveAllNodes}(v).nmap \equiv nmap \cup \{(v, \emptyset)\}$

11. $\texttt{RemoveAllEdges}(v).nmap \equiv emap \cup \{(e, \emptyset)\}$

12. $\texttt{GraphInsertNode}(v).nmap \equiv nmap \cup \{(v, \emptyset)\}$

13. $\texttt{GraphNewNode}() \equiv v \notin V$

14. $\texttt{GraphInsertEdge}(e).emap \equiv emap \cup \{(e, \emptyset)\}$

15. $\texttt{GraphNewEdge}(v_1, v_2) \equiv e \notin E$ , where $source(e) = v_1$ and $target(e) = v_2$

16. $\texttt{RefreshDeleteNode}[1](v, it').nmap \equiv$
    $nmap \cup \{(v, \{it.\texttt{Refresh}(v) | it \in nmap(v) \wedge it \neq it'\}$

17. $\texttt{RefreshDeleteNode}[2](v, it').nmap \equiv \texttt{RemoveAllNodes}(v).nmap$

18. $\texttt{RefreshDeleteEdge}[1](e, it').emap \equiv$
    $emap \cup \{(e, \{it.\texttt{Refresh}(\varepsilon) | it \in emap(e) \wedge it \neq it'\}$

19. $\texttt{RefreshDeleteEdge}[2](e, it').emap \equiv \texttt{RemoveAllEdges}(e).emap$

Sometimes, edges need to be recomputed if they are not valid anymore. This will be done with the following function (see table 2.2):

$$refresh(e, v) = \begin{cases} advance_e^j(e) & \text{if } \exists j : 1 \leq j \leq n \\ & \text{with } source(advance_e^j(e)) \neq v \wedge \\ & target(advance_e^j(e)) \neq v \\ \varepsilon & \text{otherwise} \end{cases}$$

(2.5)

| Event | Reaction |
|-------|----------|
| new node iterator | save pointer to this iterator for current node |
| new edge iterator | save pointer to this iterator for current edge |
| delete node $v$ | for every save iterator $it$: $it.\texttt{Refresh}(v)$ |
| delete edge | for every save iterator $it$: $it.\texttt{Refresh}(\varepsilon)$ |
| refresh node iterator | search other node |
| refresh edge iterator | **if** deleted object was source or target node |
| |     **then** search other edge with different nodes |
| | **otherwise** search any other edge |
| refresh adjacency iterator | **if** deleted object was source node |
| |     **then** search new source node and ready; |
| | **otherwise** advance edge and |
| |     **if** there is no incident edge anymore, |
| |         **then** search new source node |

*Tab. 2.3:* Overview of all possible reactions on events. This is only an informal discussion of the reactions, a more formal version follows in the class definitions.

In table 2.3 all possible reactions on events (e.g. deletion of nodes) are listed - it is obvious that the definition of safe iterators must be more complicated than simple iterators. Here, we need an extended version of LICD, because there are some instructions, which have to be executed in a defined order. This is covered by the following definition:

**Definition 2.2** *An* **ordered language independent class definition** *(*OLICD*) is a* **language independent class definition** *that has an order criteria for the execution of methods, i.e. for each name of a method there is a distinct order for the modification of the states. It is the following:*

1. *it is a LICD*

2. $\forall x \in M : order(x) \geq 0$

3. $x_1, x_2 \in M \wedge name(x_1) = name(x_2)$ *and*
   $x_1 \neq x_2 \wedge order(x_1) + 1 = order(x_2) \Rightarrow state_a(x_1) \equiv state_b(x_1)$,
   *where* $state_a$ *means state after execution of method* $x_1$ *and* $state_b$ *means state before execution of method* $x_2$

*3. means the following: if there are two methods with equal name* $m_1$ *and* $m_2$, *and the first appears in the ordering before the second then the modified state of method* $m_1$ *after execution of it equals the state of method* $m_2$ *referenced by the same name.*

**Example:**

**natural number** for $N \subset \mathcal{N}$:

   1. $number \in N$ (internal state)

   2. $valid \in boolean$ (internal state)

   3. $\texttt{Creation}[1](n).number \equiv n$

   4. $\texttt{Creation}[2](n).valid$              $\equiv$
$$\begin{cases} \texttt{true} & \text{if } number \in N \\ \texttt{false} & \text{otherwise} \end{cases}$$

   5. $\texttt{Succ}[1]().number \equiv number + 1$

   6. $\texttt{Succ}[2]().valid$              $\equiv$
$$\begin{cases} \texttt{true} & \text{if } number + 1 \in N \\ \texttt{false} & \text{otherwise} \end{cases}$$

   7. $\texttt{Valid}() \equiv valid$

This describes the same class as in the former example for definition 2.1, except that we fix the execution order of the methods.

In the next four iterator definitions, we have two "escape-modes", i.e. if a node was deleted, a node iterator with escape-mode *forward* tries to search a new node in the list of nodes in that direction. If the escape-mode of that node iterator was *none*, it simply becomes invalid. This is only exemplary and can be extended in concrete implementations [10].

**ordered L.I.C. Definition 2.8 safe node iterator** *for safe graph* $G = (V, E)$:

   *1.* $v \in V \cup \{\varepsilon\}$

   *2.* $escape\_mode \in \{forward, none\}$

   *3.* $\texttt{Creation}(v').v \equiv v'$

   *4.* $\texttt{Creation}(v').G.nmap \equiv G.\texttt{InsertNode}(v', this).nmap$

   *5.* $\texttt{Succ}[1]().G.nmap \equiv G.\texttt{RemoveNode}(v, this).nmap$

   *6.* $\texttt{Succ}[2]().v \equiv advance_v(v)$

---

[10] In the current implementation which is included into LEDA, I used two directions, *forward* and *backward* and one additional mode *none*.

7. $\texttt{Succ}[3]().G.nmap \equiv G.\texttt{InsertNode}(v, this).nmap$

8. $\texttt{Delete}().G \equiv G.\,\texttt{RefreshDeleteNode}(v, this)$

9. $\texttt{Insert}[1]().v \equiv G.\texttt{GraphNewNode}()$

10. $\texttt{Insert}[2]().G.nmap \equiv G.\texttt{InsertNode}(v, this).nmap$

11. $\texttt{Refresh}(v').v \equiv \begin{cases} advance_v(v') & if\ escape\_mode = forward \\ \varepsilon & otherwise \end{cases}$

Informally, this means: a safe node iterator is a node iterator that is informed to refresh itself if the node, which it refers to, was deleted. Additionally, it has methods for deletion and insertion of nodes.

**ordered L.I.C.Definition 2.9 safe edge iterator** *for safe graph* $G = (V, E)$:

1. $e \in E \cup \{\varepsilon\}$

2. $escape\_mode \in \{forward, none\}$

3. $\texttt{Creation}(e').e \equiv e' \wedge\ G.\texttt{InsertEdge}(e', this)$

4. $\texttt{Succ}[1]().G.emap \equiv G.\texttt{RemoveEdge}(e, this).emap$

5. $\texttt{Succ}[2]().e \equiv advance_e(e))$

6. $\texttt{Succ}[3]().G.emap \equiv G.\texttt{InsertEdge}(e, this).emap$

7. $\texttt{Delete}().G \equiv G.\,\texttt{RefreshDeleteEdge}(e, this)$

8. $\texttt{Insert}[1](v_1, v_2).e \equiv G.\texttt{GraphNewEdge}(v_1, v_2)$

9. $\texttt{Insert}[2](v_1, v_2).G.emap \equiv G.\texttt{InsertEdge}(e, this).emap$

10. $\texttt{Refresh}(v').e \equiv \begin{cases} refresh(e, v') & if\ escape\_mode = forward \wedge v' \neq \varepsilon \\ advance_e(e) & if\ escape\_mode = forward \wedge v' = \varepsilon \\ \varepsilon & otherwise \end{cases}$

Informally, this means: a safe edge iterator is an edge iterator that is informed to refresh itself if the edge, which the iterator refers to, was deleted. The same applies if the start or end node of the edge was deleted. Additionally, it has method for deletion and insertion of edges.

**ordered L.I.C.Definition 2.10 safe adjacency iterator** *for outgoing edges for safe graph* $G = (V, E)$:

*1.* $v \in V \cup \{\varepsilon\}$

*2.* $e \in E \cup \{\varepsilon\}$

*3.* $escape\_mode \in \{forward, none\}$

*4.* *invariant* $e \neq \varepsilon \Rightarrow source(e) = v$

*5.* $\texttt{Creation}[1](v').v \equiv v'$

*6.* $\texttt{Creation}[2](v').e \equiv \sigma_{out}(v', 1)$

*7.* $\texttt{Creation}[3](v').emap \equiv G.\texttt{InsertEdge}(e, this).emap$

*8.* $\texttt{Creation}[1](v', e').v \equiv v'$

*9.* $\texttt{Creation}[2](v', e').e \equiv \left\{ \begin{array}{ll} e' & \textit{if } source(e') = v' \\ \varepsilon & \textit{otherwise} \end{array} \right.$

*10.* $\texttt{Creation}[3](v', e').G.emap \equiv G.\texttt{InsertEdge}(e', this).emap$

*11.* $\texttt{Succ}[1]().G.emap \equiv G.\texttt{RemoveEdge}(e, this).emap$

*12.* $\texttt{Succ}[2]().e \equiv advance_{out}(e)$

*13.* $\texttt{Succ}[3]().G.emap \equiv G.\texttt{InsertEdge}(e, this).emap$

*14.* $\texttt{Succ}[4]().v \equiv v$

*15.* $\texttt{Delete}().e \equiv G.\,\texttt{RefreshDeleteEdge}(e, this)$

*16.* $\texttt{CurrAdj}[1]().v \equiv target(e)$

*17.* $\texttt{CurrAdj}[2]().G.emap \equiv G.\texttt{RemoveEdge}(e, this).emap$

*18.* $\texttt{CurrAdj}[3]().e \equiv \sigma_{out}(v, 1)$

*19.* $\texttt{CurrAdj}[4]().G.emap \equiv G.\texttt{InsertEdge}(e, this).emap$

*20.* $\texttt{Refresh}(v').v \equiv \left\{ \begin{array}{ll} v & \textit{if } escape\_mode = forward \wedge v \neq v' \\ source(refresh(e, v')) & \textit{if } escape\_mode = forward \\ \varepsilon & \textit{otherwise} \end{array} \right.$

*21.* $\texttt{Refresh}(v').e \equiv \left\{ \begin{array}{ll} advance_{out}(e) & \textit{if } escape\_mode = forward \wedge v \neq v' \\ refresh(e, v') & \textit{if } escape\_mode = forward \wedge v = v' \\ \varepsilon & \textit{otherwise} \end{array} \right.$

Informally, this means: a safe adjacency iterator for outgoing edges is an
adjacency iterator for outgoing edges that will be informed to refresh itself
in the same cases like above. The refreshing procedure is different: if the
start node of the edge was deleted, a new start node will be searched; if the
end node of the edge was deleted, a new edge in the adjacency list will be
searched.

**ordered L.I.C.Definition 2.11 safe adjacency iterator** *for incoming
edges for safe graph $G = (V, E)$:*

1. *$v \in V \cup \{\varepsilon\}$*

2. *$e \in E \cup \{\varepsilon\}$*

3. *$escape\_mode \in \{forward, none\}$*

4. *invariant $e \neq \varepsilon \Rightarrow target(e) = v$*

5. $\mathtt{Creation}[1](v').v \equiv v'$

6. $\mathtt{Creation}[2](v').e \equiv \sigma_{in}(v', 1)$

7. $\mathtt{Creation}[3](v').emap \equiv G.\mathtt{InsertEdge}(e, this).emap$

8. $\mathtt{Creation}[1](v', e').v \equiv v'$

9. $\mathtt{Creation}[2](v', e').e \equiv \begin{cases} e' & \text{if } target(e') = v' \\ \varepsilon & \text{otherwise} \end{cases}$

10. $\mathtt{Creation}[3](v', e').G.emap \equiv G.\mathtt{InsertEdge}(e', this).emap$

11. $\mathtt{Succ}[1]().G.emap \equiv G.\mathtt{RemoveEdge}(e, this).emap$

12. $\mathtt{Succ}[2]().e \equiv advance_{in}(e)$

13. $\mathtt{Succ}[3]().G.emap \equiv G.\mathtt{InsertEdge}(e, this).emap$

14. $\mathtt{Succ}[4]().v \equiv v$

15. $\mathtt{Delete}().e \equiv G.\mathtt{RefreshDeleteEdge}(e, this)$

16. $\mathtt{CurrAdj}[1]().v \equiv source(e)$

17. $\mathtt{CurrAdj}[2]().G.emap \equiv G.\mathtt{RemoveEdge}(e, this).emap$

18. $\mathtt{CurrAdj}[3]().e \equiv \sigma_{in}(v, 1)$

*19.* `CurrAdj[4]()`.$G.emap \equiv G.$`InsertEdge`$(e, this).emap$

*20.* `Refresh`$(v').v \equiv \begin{cases} v & \textit{if escape\_mode} = forward \wedge v \neq v' \\ target(refresh(e, v')) & \textit{if escape\_mode} = forward \\ \varepsilon & \textit{otherwise} \end{cases}$

*21.* `Refresh`$(v').e \equiv \begin{cases} advance_{in}(e) & \textit{if escape\_mode} = forward \wedge v \neq v' \\ refresh(e, v') & \textit{if escape\_mode} = forward \wedge v = v' \\ \varepsilon & \textit{otherwise} \end{cases}$

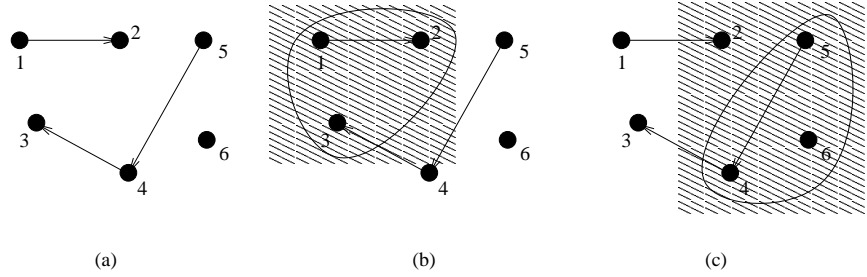

**Fig. 2.4:**  *Contraction:* Examples of three views on a graph: (a) no nodes are
treated as a single node, i.e. no node is contracted (b) nodes 1,2,3 are
treated as a single node (c) nodes 4,5,6 are treated as a single node. In
the first case, every iterator sees the original graph. In the second case, a
node iterator sees only the nodes $c$,4,5,6 where $c$ represents the contracted
nodes. An adjacency iterator for outgoing edges sees for the fixed node 4
an incident edge that terminates in $c$, i.e. the adjacent node is $c$. In the
third case, an adjacency for outgoing edges sees for $c$ as fixed node only
the edge (4,3), i.e. the adjacent node is 3.

### 2.3.2   Contraction Iterators

Some algorithms require that two nodes or a list of nodes can be treated
as a single one. This is called *contraction of nodes*: A **contraction** of two
nodes $v, w \in V_G$ yields a graph $G'$ where $G = (V_G, E_G)$, $G' = (V_{G'}, E_{G'})$
and $V_{G'} = V_G - \{w\}$, $E_{G'} = E_G - in(w) - out(w) \cup \{(v, x) | (w, x) \in E_G\} \cup
\{(x, v) | (x, w) \in E_G\})$. An **expansion** is the reverse process of contraction.
In figure 2.4 there are three different views on a sample graph.

In this chapter, we discuss how iterators can be extended to be able to do con-
traction and expansion. For simplicity, we speak only about contraction, but
expansion must be done analogous to contraction. We will assume that the
underlying graph structure does not provide any mechanism for contraction.
A straightforward approach is to modify the graph $G = (V, E)$, i.e. if we
want to contract a subset of nodes $V_S \subseteq V$ we could delete the nodes $\in V_S$
and introduce a new "pseudo-node" $x \notin V$. After contraction the original
graph is modified and we get a new "contracted graph" $G_C = (V_C, E_C)$:

$$
\begin{aligned}
G' &= (V - V_S \cup \{x\}, \{(v, w) | (v, w) \in E \wedge v, w \notin V_S\} \cup E_S) \\
E_S &= E_{in} \cup E_{out} \\
E_{in} &= \{(v, x) | (v, w) \in E \wedge v \notin V_S \wedge w \in V_S\} \\
E_{out} &= \{(x, w) | (v, w) \in E \wedge v \in V_S \wedge w \notin V_S\}
\end{aligned}
$$

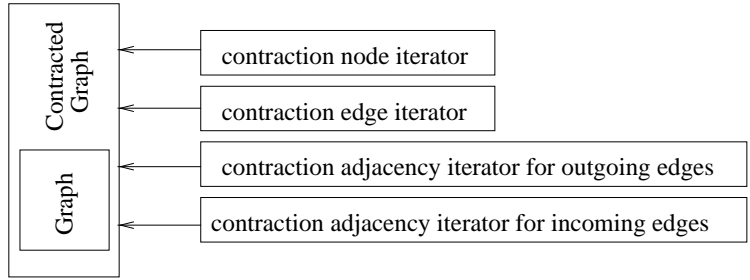

*Fig. 2.5:* **Contraction Iterators:** for the underlying graph, we introduce a wrapper structure, the "contracted graph" which provide all administrative functions that we need for handling contraction and expansion. It cannot be in one iterator, because all iterator must know the same about the contraction process.

Informally, we modify the node set and the adjacency lists of the nodes of the new node set such that each node of $V - V_S$, which was connected with some node of $V_S$ will now be connected with $x$.

In the following, we develop "contraction iterators" that satify these aims:

1. A contraction iterator provides this "contracted view" without modifying the underlying graph structure.

2. Automatic detection of substructures in the graph should be possible.

3. Different contraction representations should be possible.

### Graph Preserving Contraction

Again, as for safe iterators we need a wrapper structure for the underlying graph structure since we need a central place to decide e.g. which node is the successor node of a given node.

**L.I.C.Definition 2.12** *interface* **contracted graph**:

1. $\texttt{Contract}(v_1, v_2)$

2. $\texttt{Expand}(v)$

3. $\texttt{NextNode}(v)$

4. $\texttt{NextEdge}(e)$

5. `FirstOutgoingEdge`$(v)$ *and* `NextOutgoingEdge`$(e)$

6. `FirstIncomingEdge`$(v)$ *and* `NextIncomingEdge`$(e)$

7. `Nodes`$(v)$

8. `Equal`$(v_1, v_2)$

9. `Contracted`$(v)$

Informally, this means: a contracted graph is able to contract two nodes, expand contracted nodes, it knows successor nodes and edges and adjacency lists of nodes. It knows if two nodes are equal or if a node is a contracted node.

Since we have this intermediate structure, every contraction iterators questions it to behave correctly. For the straightforward approach it would be sufficient to provide a single wrapper class which adds the desired functionality to the iterators, like it was done for filter predicates.

With this approach, we are able implement contraction iterators which preserve the underlying graph structure.

### Automatic Substructure Detection

Suppose that an algorithm wants to contract the node set $V_S \subseteq V$ of a graph $G = (V, E)$. Then, there are for example these possibilites:

1. $V_S$ is an explicit list structure that stores all nodes

2. $V_S$ is a tree that is rooted in a single node of $V$

For (1.) we can simply develop a class that traverses this list by using an iterator that traverses this list. For (2.) we can develop a class that traverses the tree from its leaves to its root node.

Here, we look at the second case. We will assume that the algorithm provides a predecessor array, i.e. an array that stores for each node of $V_S$ its predecessor (or parent) node in the tree. All nodes of $V - V_S$ will be seen as undefined.

If we have implemented a class "contractor" that traverses the tree from its leaves to its root node and contracts the nodes it encounters, there remains an open possibility:

1. contractor contracts all nodes at the end of the traversal in one step

*Fig. 2.6:* Here, the task is to contract node 1, 2, 3 (marked in (a)). Suppose we
have discovered these three nodes in an arbitrary algorithm that provides
a predecessor structure that describes these nodes (which is here equiv-
alent to a doubly-connected list). This structure is symbolized by the
dotted arrows in (b). A graph may provide a mechanism to contract two
nodes (nodes 1,2 or 2,3 or 1,3) or a sequence of nodes (nodes 1,2,3). The
result will be a graph like (c). Expansion of contracted nodes means here,
restauration of the graph such that nodes 1,2 and 3 are visible through
the iterators as before.

2. contractor contracts two nodes a a time until the traversal is done

This can be done by designing "contractor" in a way that it needs a class
which coordinates the contraction in the desired fashion.

### Different Contraction Representations

Contraction can be represented in many ways. In the mathematical descrip-
tion from above there is nothing said about how contracted structures can
be restored in the expansion process, but this is very important.

Suppose the algorithm is not interested in restoring the graph, we can use
the straightforward approach from above. If the algorithm needs the original
graph to be restored, we can use the graph wrapper class or we make a copy
of the graph. In the difficult case, if one single "contracted node" must be
expanded, we have to store additional administrative information about the
contraction history in the graph.

Since administrative structures are located in the graph wrapper class, it is
possible to provide different wrapper classes for different realizations.

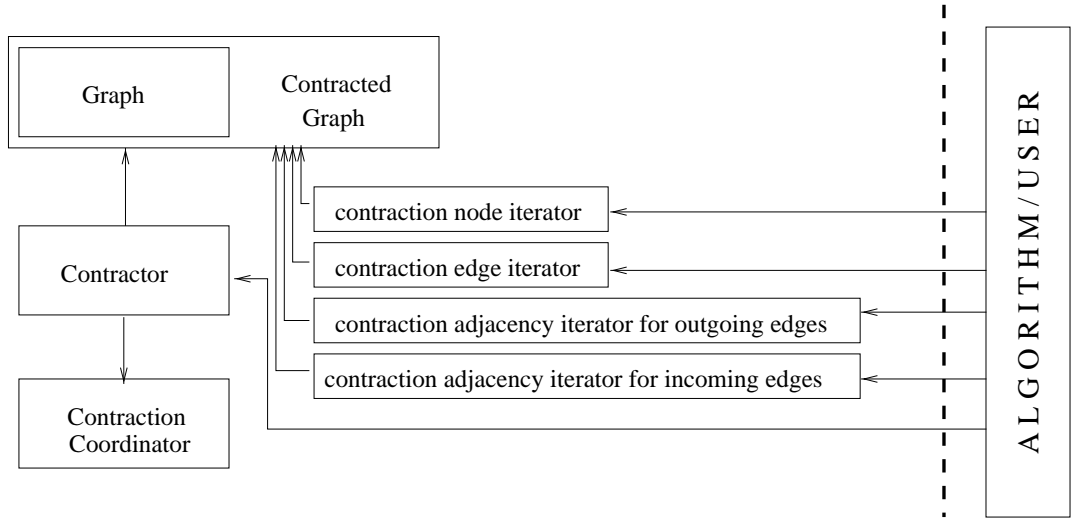

*Fig. 2.7:* Big Picture

### Big Picture

We collect the ideas from above and integrate them into an overview of the design: the underlying graph structure ("Graph") is adapted by a graph wrapper class "Contracted Graph". The iterators are specially adapted for the use with the graph wrapper class and provide the same functionality as the (simple) iterators from section 2.3.

An automatic detection an contraction of substructures of a graph happens in class "Contractor", since it may be more convenient to use this class than invoking the contraction method of the iterators.

An algorithm may use this "Contractor" for contracting substructures and access the contracted graph with the iterators. Coordination will be done with "Contraction Coordinators".

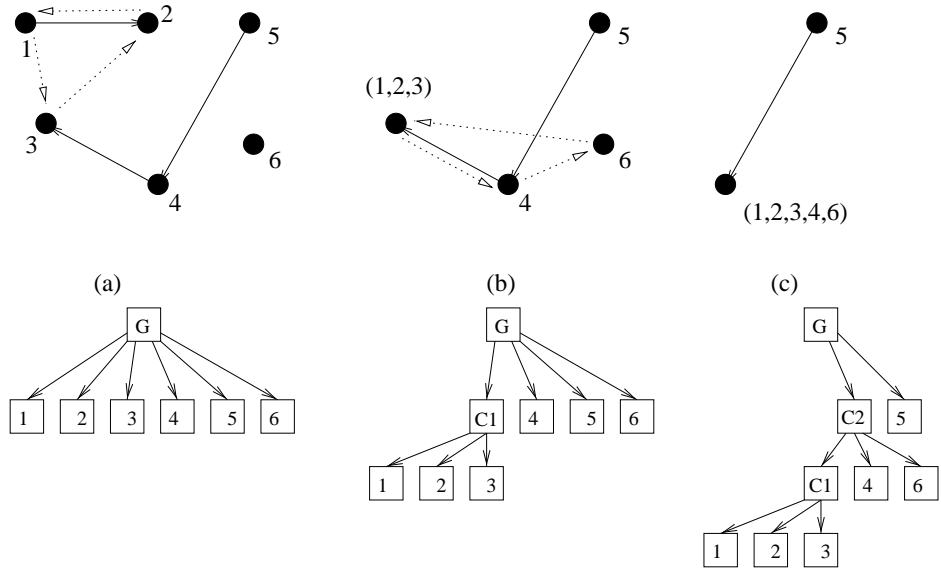

**Fig. 2.8:** Contraction with *contraction trees*: (a) nodes 1,2,3 have to be contracted - all nodes are at the moment visible (b) nodes 1,2,3 are contracted - only nodes 4,5,6 and (1,2,3) are visible (c) nodes (1,2,3), 4 and 6 are contracted - only (1,2,3,4,6) and 5 are visible (all tree-nodes that are children of the root node are visible)

### Concrete Design of Contraction Iterators

**Representation:** Suppose, we would like to perform a sequence of contraction operations on a graph. It may happen that we encounter a node, which represents contracted nodes and must think about different possibilities. See figure 2.8 for an example of nested contraction.

We may represent contracted nodes in the graph wrapper class in the following ways:

- each graph is represented by a *contraction tree* (see figure 2.8) A contraction tree is a tree, rooted in a distinct node, which represents the visible graph, because all children of this node are visible through the contraction iterators. Tree-nodes are either nodes of the underlying graph or contraction nodes. Nodes of the first type have no children. The children of contraction nodes were all contracted at the same time.

  Each time a list of nodes will be contracted, we create a new contraction node $C_i$ and add the nodes to $C_i$. The benefit is: we are able to handle nested contraction and can expand these contraction nodes one by one (in this example: e.g. expand $C_2$ and then $C_1$ or in figure 2.8 expand (1,2,3) ). Observe that we have to expand top–down, because all nodes that are in $C_1$ are in $C_2$, as well.

- each graph is represented by a *reduced contraction tree*. A reduced contraction tree is a tree, rooted in a distinct node, which represents the visible graph, because all children of this node are visible through the contraction iterators. Tree-nodes are either nodes of the underlying graph or contraction nodes. Nodes of the first type have no children. *Contraction nodes do not have contraction nodes as ancestors.*

  The difference is that we reduce the complexity of the model. The disadvantage is that we have no contraction hierarchy: if perform nested contraction we are only able to expand all nested contraction in one step.

In the following, we will only describe the second (flat) approach, because the extent of description is much less[11].

**Graph Wrapper Class: "Contracted Graph"**

**L.I.C.Definition 2.13 simple contracted graph** *for $G = (V, E)$:*

1. *$nodes : V \rightarrow V^{\varepsilon}$*

2. *$rep : V \rightarrow V$*

3. *$\texttt{Creation}(G).nodes \equiv nodes \cup \{(v, \{v\})\}$*

4. *$\texttt{Creation}(G).rep \equiv rep \cup \{(v, v)\}$*

5. *$\texttt{Contract}(v_1, v_2).nodes \equiv$*
   *$nodes \cup \{(v_1, nodes(v_1) \cup nodes(v_2))\} - \{(v_2, nodes(v_2))\}$*

6. *$\texttt{Contract}(v_1, v_2).rep \equiv rep \cup \{(v', v_1) | v' \in nodes(rep(v_2))\}$*

7. *$\texttt{Expand}(v).nodes \equiv$*
   *$nodes \cup \{(v', \{v'\}) | v' \in nodes(rep(v))\}$*

8. *$\texttt{Expand}(v).rep \equiv$*
   *$rep \cup \{(v', v') | v' \in nodes(rep(v))\}$*

---

[11] The non-bipartite matching algorithm of chapter 4 uses contraction iterators that are based on a graph wrapper class, which stores contracted nodes in contraction trees.

9. $\texttt{NextNode}(v) \equiv \begin{cases} advance_v^k(v) & \textit{if } \exists k : k \textit{ minimal} \land \neg\texttt{Equal}(advance_v^k(v), v) \\ \varepsilon & \textit{otherwise} \end{cases}$

10. $\texttt{NextEdge}(e)$ *defined analogously to* $\texttt{NextNode}(v)$

11. $\texttt{FirstOutgoingEdge}(v) \equiv$

$$\begin{cases} \sigma_{out}(v,1) & \textit{if } target(e) \neq v \\ advance_{out}^k(\sigma_{out}(v,1)) & \textit{if } target(e) = v \land \exists k : k \textit{ minimal} \land \\ & \neg\texttt{Equal}(target(advance_{out}^k(\sigma_{out}(v,1))), v) \\ \varepsilon & \textit{otherwise} \end{cases}$$

12. $\texttt{NextOutgoingEdge}(e) \equiv$

$$\begin{cases} advance_{out}^k(e) & \textit{if } \exists k : k \textit{ minimal} \land \\ & \neg\texttt{Equal}(source(advance_{out}^k(e)), source(e)) \\ \varepsilon & \textit{otherwise} \end{cases}$$

13. $\texttt{FirstIncomingEdge}(v)$ *defined analogously to* $\texttt{FirstOutgoingEdge}(v)$

14. $\texttt{NextIncomingEdge}(e)$ *defined analogously to* $\texttt{NextOutgoingEdge}(e)$

15. $\texttt{Nodes}(v) \equiv nodes(v)$

16. $\texttt{Equal}(v_1, v_2) \equiv \begin{cases} \texttt{true} & \textit{if } rep(v_1) = rep(v_2) \\ \texttt{false} & \textit{otherwise} \end{cases}$

17. $\texttt{Contracted}(v) \equiv \begin{cases} \texttt{true} & \textit{if } |nodes(v)| > 1 \\ \texttt{false} & \textit{otherwise} \end{cases}$

Now follows the definitions for iterators that traverse the different domains of a contracted graph. These definitions are very similar to the (simple) iterator definitions, because the advance-functions were replaced by methods of the underlying graph.

**L.I.C.Definition 2.14 node contraction iterator** *for contracted graph* $G = (V, E)$:

1. $v \in V \cup \{\varepsilon\}$

2. $\texttt{Creation}(v').v \equiv v'$

3. $\texttt{Succ}().v \equiv G.\texttt{NextNode}(v)$

**L.I.C.Definition 2.15 edge contraction iterator** *for contracted graph* $G = (V, E)$:

1. $e \in E \cup \{\varepsilon\}$

2. $\texttt{Creation}(e').e \equiv e'$

3. $\texttt{Succ}().e \equiv G.\texttt{NextEdge}(e)$

**L.I.C.Definition 2.16 adjacency contraction iterator for outgoing edges** *for contracted graph $G = (V, E)$:*

1. $v \in V \cup \{\varepsilon\}$

2. $e \in E \cup \{\varepsilon\}$

3. *invariant $e \neq \varepsilon \Rightarrow source(e) = v$*

4. $\texttt{Creation}(v').v \equiv v'$ *and* $\texttt{Creation}(v').e \equiv G.\texttt{FirstOutgoingEdge}(v')$

5. $\texttt{Creation}(v', e').v \equiv v'$ *and* $\texttt{Creation}(v', e').e \equiv$
   $\begin{cases} e' & \textit{if } source(e') = v' \\ \varepsilon & \textit{otherwise} \end{cases}$

6. $\texttt{Succ}().e \equiv G.\texttt{NextOutgoingEdge}(e)$ *and* $\texttt{Succ}().v \equiv v$

7. $\texttt{CurrAdj}().v \equiv target(e)$

8. $\texttt{CurrAdj}().e \equiv G.\texttt{FirstOutgoingEdge}(target(e))$

**L.I.C.Definition 2.17 adjacency contraction iterator for incoming edges** *for contracted graph $G = (V, E)$:*

1. $v \in V \cup \{\varepsilon\}$

2. $e \in E \cup \{\varepsilon\}$

3. *invariant $e \neq \varepsilon \Rightarrow target(e) = v$*

4. $\texttt{Creation}(v').v \equiv v'$ *and* $\texttt{Creation}(v').e \equiv G.\texttt{FirstIncomingEdge}(v')$

5. $\texttt{Creation}(v', e').v \equiv v'$ *and* $\texttt{Creation}(v', e').e \equiv$
   $\begin{cases} e' & \textit{if } target(e') = v' \\ \varepsilon & \textit{otherwise} \end{cases}$

6. $\texttt{Succ}().e \equiv G.\texttt{NextIncomingEdge}(e)$ *and* $\texttt{Succ}().v \equiv v$

7. $\texttt{CurrAdj}().v \equiv source(e)$

8. $\texttt{CurrAdj}().e \equiv G.\texttt{FirstIncomingEdge}(source(e))$

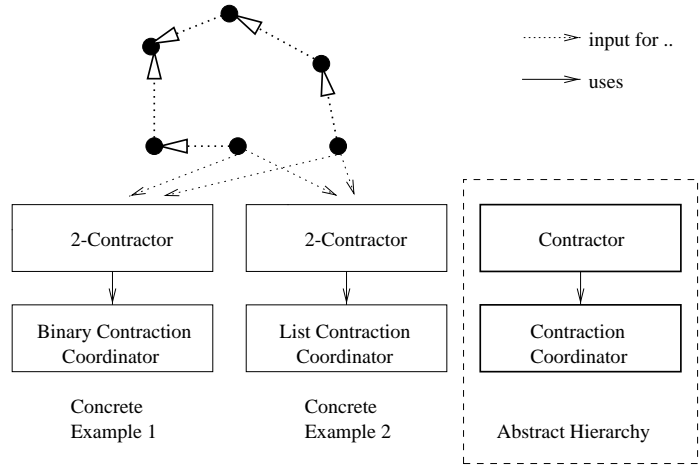

**Fig. 2.9:** A *contractor* decides which nodes has to be contracted and a *contraction coordinator* coordinates the process, for example by invoking the contraction method of the underlying graph for each two nodes or for a list of nodes. On the right, there is the abstract hierarchy, taken from figure 2.7.

### Performing Contractions: "Contractor"

There are different ways **how contraction can be done** in algorithms. If the list of nodes that has to be contracted are described by a predecessor array we can contract the component by traversing it with iterators (see figure 2.10):

1. predecessor array describes **a cycle**

2. predecessor array describes **two paths**

3. predecessor array describes $n$ **paths**

Now we will discuss each case:

*Ad 1.:* If the predecessor array describes **a cycle** and we have an iterator *it* that refers to a node in that cycle (see figure 2.10(a)), we can contract the whole cycle by calling the method `Contract` with *it* as a parameter.

**L.I.C.Definition 2.18** *interface* **cycle contractor***:*

1. `Contract`(*it*)

The implementation of this method could follow this piece of pseudo-code:

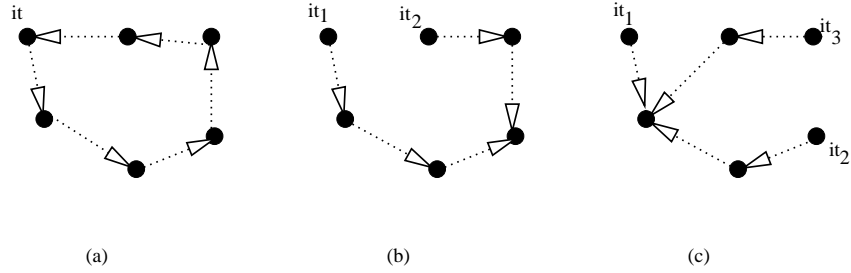

Fig. 2.10:  *Examples for performing contraction using a predecessor array:* (a) *pred* is a cycle, *it* traverses the cycle (b) *pred* consists of two paths (c) *pred* consists of three paths

> *it* node iterator that refers to a node of the cycle
> *pred* predecessor array
> $it_{old} = it$
> **do**
>     $G$.Contract($it.v, pred(it).v$)
>     $it = pred(it)$
> **while** $it \neq it_{old}$

*Ad 2.:* If the predecessor array describes **two paths** that only intersect in a node $r$ and each path begins with a different node (see figure 2.10(b)), then we can use two iterators $it_1$, $it_2$ which refer to these nodes and can contract the whole tree. Contract($it_1, it_2$) will contract these two paths.

**L.I.C.Definition 2.19** *interface* **two path contractor**:

1. Contract($it_1, it_2$)

The implementation of this method could follow this piece of pseudo-code:

> $it_1$, $it_2$ node iterators referring to $i_1$, $i_2$
> *pred* predecessor array
> **while** $it_1$.Valid() and $pred(it_1)$.Valid()
>     $G$.Contract($it_1.v, pred(it_1).v$)
> **while** $it_2$.Valid() and $pred(it_2)$.Valid()
>     $G$.Contract($it_2.v, pred(it_2).v$)

Moreover, if both paths are of equal length, the loops can be rewritten as:

$it_1, it_2$ node iterators referring to $i_1, i_2$

*pred* predecessor array

**while** $it_1$.`Valid()` and $pred(it_1)$.`Valid()`

$\quad\quad$ $G$.`Contract`$(it_1.v, pred(it_1).v)$

$\quad\quad$ $G$.`Contract`$(it_2.v, pred(it_2).v)$

*Ad 3.:* If the predecessor array describes $n$ **paths** that only intersect in a node $r$ and each path begins with a different node, then we can use $n$ iterators $it_1, it_2, \ldots it_n$ which refer to these nodes and can contract the whole tree analogously to 2.

To decouple the contraction process from the algorithm we could use classes that behave differently, depending on the number of iterators they get.

### *Coordinating Contraction: "Contraction Coordinator"*

At this stage, we are interested in how the **contraction process** should be coordinated:

- contract 2 nodes at a time or

- contract a list of nodes

This should be independent of the graph wrapper class or the kind of contractor, i.e. a contractor class should be generic enough that coordination can be changed later without changing the code of the contractor.

For this purpose we introduce a contraction hierarchy — the idea is that the contractor uses contraction coordinators.

**L.I.C.Definition 2.20** *interface* **contraction coordinator***:*

*1.* `Init`$(iterator, predecessor, coord\_base)$

*2.* `StartContraction()`

*3.* `Succ()`

The different contractors have to be adapted to use coordinators, which will be done now:

1. **cycle contractor:** the coordinator will be initialized with *it* and
   *pred*.   Every time `Succ()` is called, *coord.it* will be moved to
   *coord.pred(coord.it)*. This will be done until the original iterator is
   reached. Then we have discovered the complete cycle. Then, method
   `StartContraction()`[12] is called. *coord_base*[13] is a data structure that
   stores information during a complete contraction process.

   > *it* node iterator that refers to a node of the cycle
   > *pred* predecessor array
   > *coord* contraction coordinator
   > *coord_base* coordination base
   > *coord*.`Init`($it, pred, coord\_base$)
   > **do**
   >     *coord*.`Succ()`
   > **while** $it \neq coord.it$
   > *coord*.`StartContraction()`

2. **two-path contractor:**   coordinator $coord_1$ (or $coord_2$) will be
   initialized with $it_1$ ($it_2$)and *pred*.   Every time `Succ()` is called,
   $coord_1.it$ ($coord_2.it$ )will be moved to $coord_1.pred(coord_1.it)$ (
   $coord_2.pred(coord_2.it)$ ). This will be done for each path until the itera-
   tor has become invalid. Then, a method is called, `StartContraction()`
   that terminates the contraction. Here we can imagine why *coord_base*
   was introduced: as a common place for complex coordinator to store
   data, but this will be explained in the following.

   > $it_1, it_2$ node iterators referring to $i_1$, $i_2$
   > *pred* predecessor array
   > $coord_1, coord_2$ contraction coordinator
   > *coord_base* coordination base
   > $coord_1$.`Init`($it_1, pred, coord\_base$)
   > $coord_2$.`Init`($it_2, pred, coord\_base$)
   > **while** $coord_1.it$.`Valid()` and $coord_1.pred(coord_1.it_1)$.`Valid()`
   >     $coord_1$.`Succ()`
   > **while** $coord_2.it$.`Valid()` and $coord_2.pred(coord_2.it_2)$.`Valid()`
   >     $coord_2$.`Succ()`

---

[12] If nodes are contracted during the while-loop, it does nothing. If nodes are stored in
a list and contracted at the end in one step, it starts the actual contraction-process.

[13] For example, it can be a list of nodes and it contains all nodes of a contracted cycle
such that we can contract the cycle in one step by processing the list.

$coord_1.\texttt{StartContraction}()$

Moreover, if both paths are of equal length, the loops can be rewritten as:

$it_1, it_2$ node iterators referring to $i_1$, $i_2$
*pred* predecessor array
$coord_1, coord_2$ contraction coordinator
*coord\_base* coordination base
$coord_1.\texttt{Init}(it_1, pred, coord\_base)$
$coord_2.\texttt{Init}(it_2, pred, coord\_base)$
**while** $coord_1.it\,.\texttt{Valid}()$ and $coord_1.pred(coord_1.it_1)\,.\texttt{Valid}()$
$\quad$ $coord_1.\texttt{Succ}()$
$\quad$ $coord_2.\texttt{Succ}()$
$coord_1.\texttt{StartContraction}()$

3. $n$-**path contractor:** analogous to **two-path contractor**

Since we have adapted the contractors, we are now able to discuss coordinators that fit into the interface *contraction coordinator*, given in LICD 2.20. At first, a language independent definition of contraction coordinators for the case in which every two nodes are going to be contracted in the graph. The main idea is to contract in each successor step. Here, *coord\_base* has void functionality.

**L.I.C.Definition 2.21 contraction coordinator** *for pair-contraction in a contracted graph* $G = (V, E)$:

1. *it is an iterator*

2. *pred is a predecessor array*

3. *coord\_base is a coordination base*

4. $\texttt{Creation}(it, pred, cb).it \equiv it$

5. $\texttt{Creation}(it, pred, cb).pred \equiv it$

6. $\texttt{Creation}(it, pred, coord\_base).coord\_base \equiv coord\_base$

7. $\texttt{Init}(it, pred, cb).it \equiv it$

8. $\mathtt{Init}(it, pred, cb).pred \equiv it$

9. $\mathtt{Init}(it, pred, coord\_base).coord\_base \equiv coord\_base$

10. $\mathtt{Succ}().G.nodes \equiv G.\mathtt{Contract}(it.v, pred(it.v))$

11. $\mathtt{StartContraction}().it \equiv it$

Now we need a language independent class definition of contraction coordinators for the case in which a complete list of nodes will be contracted in the graph in one single step. The main idea is to notify all nodes in the successor steps in a list and to contract the whole list in method $\mathtt{StartContraction}$. Here, $coord\_base$ is an external list and stores all nodes during the traversal, i.e. coordination base is a list of nodes (observe that the contracted graph structure must provide a method that contracts a list of nodes at once).

**L.I.C.Definition 2.22 contraction coordinator** *for list-contraction in a contracted graph* $G = (V, E)$:

1. *it is an iterator*

2. *pred is a predecessor array*

3. *coord_base is a coordination base*

4. $\mathtt{Creation}(it, pred, cb).it \equiv it$

5. $\mathtt{Creation}(it, pred, cb).pred \equiv it$

6. $\mathtt{Creation}(it, pred, coord\_base).coord\_base \equiv coord\_base$

7. $\mathtt{Init}(it, pred, cb).it \equiv it$

8. $\mathtt{Init}(it, pred, cb).pred \equiv it$

9. $\mathtt{Init}(it, pred, coord\_base).coord\_base \equiv coord\_base$

10. $\mathtt{Succ}[1]().coord\_base = coord\_base \cup \{it.v\}$

11. $\mathtt{Succ}[2]().it \equiv pred(it)$

12. $\mathtt{StartContraction}().G.nodes \equiv G.\mathtt{Contract}(coord\_base)$

### 2.3.3 Wrapper classes for Iterators

#### Filter iterators

Filter iterators are wrapper classes for any other iterator. They take an additional predicate.

The internal predicate controls the access of the iterators: Nodes that do not fulfill the Boolean condition represented by the predicate are skipped by the successor method `Succ()`.

**L.I.C.Definition 2.23** *interface* **predicate**:

1. `Ask`(*iterator*)

**L.I.C.Definition 2.24 filter iterator** *for* $G = (V, E)$:

1. *it is an iterator*

2. *pred is a predicate*

3. `Creation`$(it', pred').it \equiv it'$

4. `Creation`$(it', pred').pred \equiv pred'$

5. $\texttt{Succ}().it \equiv \begin{cases} it.\texttt{Succ}^j() & \textit{if } \exists j : 1 \leq j \leq n \wedge \\ & pred.Ask(it.\texttt{Succ}^j()) \equiv \texttt{true} \wedge \\ & j \textit{ minimal} \\ \varepsilon & \textit{otherwise} \end{cases}$

Suppose that each node is associated with a color, and we only want to see those with a specific color, for example red. At first, the solution in pseudo-code with a function *color* associates colors with the nodes of iterators.

> $it :=$ node iterator for $G = (V, E)$
> $it.\texttt{Creation}(\text{first node of } G)$
> **while** $it.v \neq \varepsilon$
>       **if** $color(it) \equiv red$
>         do something with $it$
>       $it.\texttt{Succ}()$

With the filter iterator class, we can wrap the test, if the node is red, in the iterator itself. Now we have a version with silent, transparent incorporation of predicates:

$pred$ := predicate that is true for red nodes
$it$ := filter iterator for $G = (V, E)$
$n$ := node iterator for $G = (V, E)$
$n$.**Creation**(first node of $G$)
$it$.**Creation**($n$, $pred$)
**while** $it.v \neq \varepsilon$
      do something with $it$
      $it$.**Succ**()

By using this design, filter predicates can be decoupled from the actual algorithm implementation.

For example, consider a graph traversal algorithm. If we are only interested in a restricted area in the graph, we can use filter iterators and do not need to change the code of the algorithm.

### Observer iterators

Observer iterators are wrapper classes for any iterator that take an additional class where the observation is coordinated. The observer class contains methods that will be invoked if the iterator moves forward, or at creation time of the iterator: `Notify_Succ(it)` and `Notify_Creation(it)`.

**L.I.C.Definition 2.25** *interface* **observer***:*

 1. `Notify_Creation`(*iterator*)

 2. `Notify_Succ`(*iterator*)

These observer iterators can be used, e.g. , for operation–counting, on–line checking, or animation. The creation methods get as parameters the current iterator and an observer class. For example, the graph is embedded in a two–dimensional plane, an algorithm always highlights the current investigated node. The observer object may "ask" the current point's coordinate from the iterator to highlight the corresponding point on the screen.

**L.I.C.Definition 2.26 observer iterator** *for $G = (V, E)$:*

 1. *it is an iterator for $G = (V, E)$*

 2. *obs is an observer*

 3. `Creation`($it'$, $obs'$)$.it \equiv it'$

*4.* $\mathtt{Creation}(it', obs').obs \equiv obs'.\mathtt{Notify\_Creation}(it)$

*5.* $\mathtt{Succ}().it = it.\mathtt{Succ}()$

*6.* $\mathtt{Succ}().obs = obs.\mathtt{Notify\_Succ}(it)$

The next example shows how to use observer iterators. Suppose that *obs* is an object of an observer class that increases a counter each time *obs*.$\mathtt{Notify\_Succ(it)}$ is invoked and does nothing when *obs*.$\mathtt{Notify\_Creation(it)}$ is invoked. Then the following pseudo-code example computes the number of nodes in $V$ minus 1.

**L.I.C.Definition 2.27 observer counter class**

*1.* $counter \in \mathcal{N}$

*2.* $\mathtt{Notify\_Creation}(iterator).counter \equiv 0$

*3.* $\mathtt{Notify\_Succ}(iterator).counter \equiv counter + 1$

*4.* $\mathtt{GetCounter}() \equiv counter$

> $obs :=$ observer counter class
> $it :=$ observer iterator for $G = (V, E)$
> $n :=$ node iterator for $G = (V, E)$
> $n.\mathtt{Creation}(\text{first node of } G)$
> $it.\mathtt{Creation}(n, obs)$
> **while** $it.v \neq \varepsilon$
>       do something with $it$
>       $it.\mathtt{Succ}()$
> $Print(obs.\mathtt{GetCounter}())$

## 2.4 Data Accessors

In general, algorithms *cannot make any assumption* about how the data is organized:

- *materialization*: data can be represented by stored values in the memory or can be computed from other values, i.e. is not actually materialized in the memory.

- *data structure*: different data structures can be used to store values, e.g. hash tables, fixed arrays in the memory, database tables . . .

- *interface*: the access to the data may not be uniform, e.g. one data structure provides direct access, the other needs to have index variables

- *localization*: data can be directly associated to nodes or edges. For example, structures are sometimes extended with a small information object that stores additional values.

- *variability*: data can be constant

In common libraries, it seems that the problem of organising item attributes is often underestimated or even overlooked:

- *STL:* Item attributes can be accessed through a member function of an iterator, i.e. each item is associated with exactly one attribute. This is not practical in e.g. graph algorithms. Suppose we want to compute a Dijkstra shortest path, then we can have a distance value and a predecessor node for each node.

- *LEDA:* Nodes and edges can be associated with attributes by using e.g. node arrays or edge arrays, respectively. Unfortunately, the signature of the algorithm determines which data structure is suitable. Moreover, if we have self–defined data structures, we have to adapt the algorithm (e.g., if the edge length is computed online as the euclidean length of two points in an embedding)

The concept of data accessors (introduced first in [KW97b]) is designed to overcome these problems. A data accessor class is responsible for the access to a single item attribute. For example, if we have an iterator as defined in section 2.3, we can use the data accessor to access the attribute value associated with the item which the iterator refers.

A **data accessor** $da$ is responsible for the access to the attribute value of an attribute $att$ associated to an item $c$ of a collection $C$. Let $x$ be an object that refers to an item $c$ of a collection $C$. Then $da(x)$ is the attribute value denoted by $att(c)$, where $att(c)$ is the attribute value associated to $c$.

We will separate the access to data in two methods, $\texttt{Get}(x)$ and $\texttt{Set}(x, value)$, where the former one retrieves values and the second stores values.

**L.I.C.Definition 2.28** *interface* **data accessor***:

1. $\texttt{Get}(x)$

2. $\texttt{Set}(x, val)$

There may be different types of data accessors, but each of them have the same interface which is very useful for the design of algorithms, because we do not have to think about the realization.

Now the different ideas:

- *arrays*: the attributes are stored explicitly in form of arrays (or something similar)

- *member variables*: the attributes are stored in the items themselves (e.g. if nodes in a graph are represented by a structure where the location in an embedding is stored, as well)

- *method invocation*: the attributes will be computed by calling a method from a class

- *constant value*: the attributes are equal and constant

Since attributes are not associated to iterators, but to items the iterators refer to, it might happen that there are multiple possibilities for selecting the right item. For example, colors can be associated with nodes and edges. An adjacency iterator has both, a fixed node and an incident edge. Here, there are two possibilities and it is not clear which one of the items is the right one. This has to be determined in the creation phase of a data accessor, and can be done by using object accessors:

**L.I.C.Definition 2.29** *interface* **object accessor***:*

1. $\texttt{GetObject}(x)$

Informally, this means: an object accessor has a method that $\texttt{GetObject(x)}$ returns a specified item of e.g. an iterator $x$ (for iterators, this would be the node or the edge).

**L.I.C.Definition 2.30 node object accessor***:*

1. $\texttt{GetObject}(it) \equiv it.v$

This version of an object accessor returns the node that the iterator $it$ refers to.

**L.I.C.Definition 2.31 edge object accessor***:*

1. $\texttt{GetObject}(it) \equiv it.e$

This version of an object accessor returns the edge that the iterator *it* refers to.

Obviously, other items could be supported (not only nodes and edges) without any problem, but for graph algorithms, this should be sufficient.

In the discussion above we discussed different ideas of data realization. Each case can be handled by one of the following data accessor versions. If the data realization changes, only the data accessor has to be exchanged. The uniform access to the data by two simple methods, `Get` and `Set`, allows us to keep the part of the code that uses data accessors without modification.

The first data accessor is designed for an array-like realization of the data, which is called for simplicity *handler*. For any item object, *handler* returns direct access to the associated data.

**L.I.C.Definition 2.32 handler accessor** *for item class C and attribute class A:*

1. *handler* : $C \to A$

2. *oa is an object accessor*

3. `Creation`$(handler, oa).handler \equiv handler$

4. `Creation`$(handler, oa).oa \equiv oa$

5. `Get`$(x) \equiv handler(oa.\texttt{GetObject(x)})$

6. `Set`$(x, value).handler \equiv handler \cup \{(oa.\texttt{GetObject(x)}, value)\}$

For example, a data accessor *color* is responsible for the access to the color *color'*. For an iterator *it*, *color*(*it*) will be exactly the value referenced by *color'*(*it.v*). Observe that we associate colors to nodes, and *color'* yields for each node a color.

The second data accessor is designed for data that is directly associated to the items. Here we know that the item object has a state, which is the desired data.

**L.I.C.Definition 2.33 member accessor** *for item class C and attribute class A:*

1. *oa is an object accessor*

2. *varname is a member variable name of A*

*3.* $\mathtt{Creation}(oa, varname).oa \equiv oa$

*4.* $\mathtt{Creation}(oa, varname).varname \equiv varname$

*5.* $\mathtt{Get}(x) \equiv oa.\mathtt{GetObject(x)}.varname$

*6.* $\mathtt{Set}(x, value).G.x.varname \equiv value$

This gives access to a member variable in the class that is associated to an item.

The third data accessor is designed for the situation that the item class has a method that will be called. For any item, we invoke the method *met* and get the data.

**L.I.C.Definition 2.34 method accessor** *for item class C and attribute class A:*

*1. oa is an object accessor*

*2. met is a method of A*

*3.* $\mathtt{Creation}(oa, met).oa \equiv oa$

*4.* $\mathtt{Creation}(oa, met).met \equiv met$

*5.* $\mathtt{Get}(x) \equiv oa.\mathtt{GetObject(x)}.\mathtt{met()}$

In this data acessor, a method can be called to compute a value. For the `Set`- method there are two possibilities: provide a method with same name and additional parameters to set a value or do not provide `Set`.

The last data accessor is designed for the situation in which data is constant.

**L.I.C.Definition 2.35 constant accessor** *for attribute class A:*

*1. cval is a variable of A*

*2.* $\mathtt{Creation}(cval).cval \equiv cval$

*3.* $\mathtt{Get}(x) \equiv cval$

Here, no `Set`-method is provided, because the value must be constant.

This is only a short list of data accessors, but it is possible to write for each situation special variants. Algorithms are possible, which do not have to assume anything about the organization of the data, since the process — how data is retrieved or stored — is always equal. Only some differences
appear if the possibility to store values is required, because not all of the
listed data accessors are able to store values.

In section 5 there are examples of the usage of data accessors in `C++`, but the
usage is completely analogous for other languages. Also, all listed variants
of data accessors are revisited to see how they can be realized with `C++`.

# 3. BASIC GRAPH ALGORITHMS

Here, two ideas will be applied to each algorithm realization: algorithms as objects[14] with the ability to "advance" an algorithm and access to the internal state of an algorithm.

## 3.1 Preliminaries

An algorithm that contains $n$ loops (**while**, **repeat**, ...) may be structured as a hierarchy of loops. Let *level* be the level of a loop in this hierarchy: if a loop $l_1$ contains another loop $l_2$, then $level(l_2) = level(l_1) + 1$. We call the loops $L_M = \{l | level(l) = 0\}$ **core loops**. Two loops $l_1$ and $l_2$ are nested, if one contains the other.

Now we define *loop kernels*[15]:

**Definition 3.1** *A **loop kernel** is a class that implements an algorithm with $|L_M| \geq 0$, i.e. at least one core loop. Objects of this class can iterate through all loops $l \in L_M$.*

*Comments on loop kernels:* here, a loop kernel implements no pre or post-processing operations. Probably most of the (non-trivial) algorithms can be seen as a hierarchy of several loops. For example, the Dijkstra algorithm contains only one core loop (for other scenarios see figure 3.1).

Another problem in algorithm class design is that we iterate only over one single loop. Suppose we have a nested loop but would to iterate over each item of the nested loop. To do this, we might use flags that control the iteration process.

For example, if we have one core loop $A$ and one nested loop $B$ we can use a flag $c$ that indicates if we have completed a whole nested loop. Each time we move in $A$, $c$ is set to `false` and we know that we have to move in $B$ in the next iteration. If $B$ is completed, we set $c$ to `true`. The next time we iterate, we know that $B$ is completed and we move in $A$ and reset $c$ to `false`.

---

[14] This seems to be similar to "algorithmic generators" as described in chapter 3 of [F97], but is different (see also discussion in [W97]).

[15] First introduced in [KW96b].

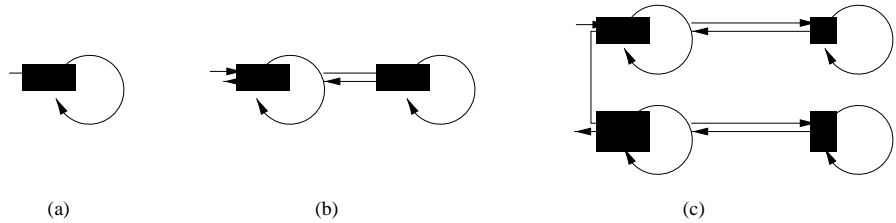

**Fig. 3.1:** *Restricted view on the loops in algorithms:* (a) shows an algorithms
that contains only one core loop (b) is an algorithm that has two nested
loops, for example a square search (for each node $v, w$: if $v \neq w$ ...)
(c) is an algorithm that has two core loops while each one contains one
loop (can be seen as algorithm that works in two phases, in each phase a
sub-algorithm is invoked)

By this technique, we are able to iterate in a simple way over nested loops.
It is clearly possible to expand this technique to arbitrary nested loops.

**Property 3.2** *A loop kernel class LK has the following properties:*

- *core loops are basic units of reuse*

- *core loops are implemented as classes*

- *nested loops are sub-algorithms[16]*

- *LK has at least one method which executes one iteration for each core
  loop*

- *pre- and post-processing operations are left to the user*

The following definition is based on the loop kernel principle:

**L.I.C.Definition 3.1** *interface* **algorithm class***:*

1. `Creation()`

2. `Next()`

3. `FinishAlgo()`

---

[16] And implemented as polymorphic members of *LK* (pattern *strategy*[GHJV95]); this
can be done in `C++` using the template mechanism, i.e. the algorithm strongly connected
components can get as template parameter an algorithm that computes a depth first
search.

*4.* `Finished()`

*5.* `Current()`

If a loop kernel provides methods to read all details of the current logical state of an object, we speak of **full logical inspect-ability**. For example, a depth first search algorithm provides access to the stack that is used in the algorithm. When the algorithm starts, the stack will be initialized to be empty. During the execution of the algorithm, the element on top of the stack refers to the current investigated node in the search. Full logical inspect-ability can be used for pre-processing operations such as initializing data structures or extension of the functionality of loop kernel classes:

- *initialization:* distance values for Dijkstra may be initialized only for a fraction of the whole node set, since we are only interested in a restricted area for a shortest path computation. Here, we can leave initialization to the user so that this will be done outside the algorithm. The user is free to modify initialization.

- *extending functionality:* algorithms may be animated in a graphical output window such that the current activity of the algorithm can be visualized without changing the code of the algorithm.

It might not always be straightforward to convert recursive functions into iterative ones, since we are sometimes not able to "iterate" simply through a recursive function. For example, the effort for developing an iterative version of the depth first numbering which is normally implemented with the recursive approach was quite high[17]. This additional effort will be rewarded by increased flexibility because new functionality can be wrapped around the algorithm class.

### 3.1.1 *Algorithmic Notation*

For better understanding the algorithms will be presented by control flow graphs.
Figure 3.2 is an example for a (very) simple algorithm.

---

[17] It fortunately turned out that this version can be used not only for numbering but also in different areas, for instance evaluating attribute grammars. Suppose we associated with each node of a derivation tree the correct production and there are no dependency cycles. Then we are able to compute the correct attributes, since each time we visit a new node, its inherited attributes are filled with values and if we leave an already visited node, we apply the associated semantic rules to the stored attributes and get the synthesized attributes as a result. This may be used in semantic analysis in compilers, for example type-checking (see [WM97] for a definition of attribute grammar and semantic analysis).

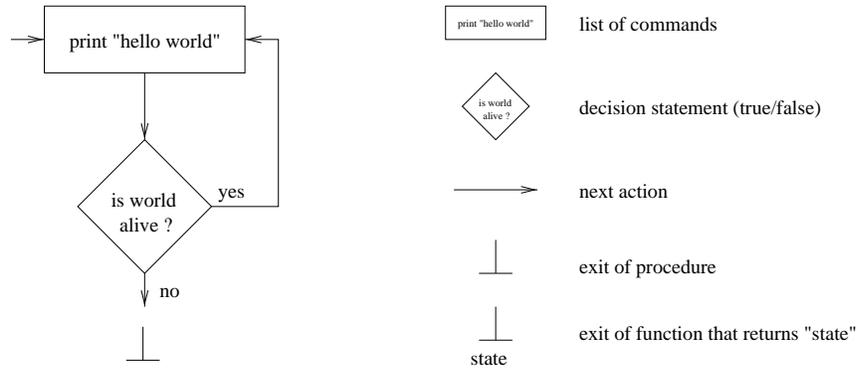

Fig. 3.2: *An example for a simple algorithm:* it starts with the list of commands at
the top of the figure and prints the string "hello world". Then it follows
the arrow to the decision-box. Now it tests for the state of the world, if
it is alive. If so, it repeats the output of the string. Otherwise, it stops.

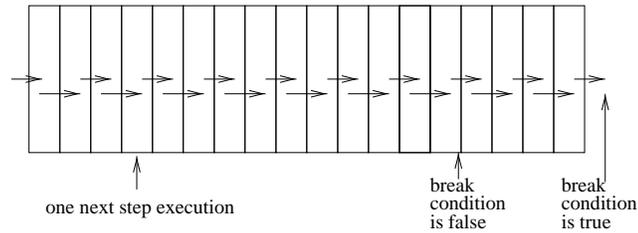

Fig. 3.3: *Next Step:* this is one execution of a core loop in a loop kernel of an
algorithm. Each small rectangular box represents one execution of a
next step in the algorithm class.

### 3.1.2 Next Step

Usually, a loop consists of the following:

- *body:* it contains the statements that have to be executed in each iteration

- *break condition:* if it changes its value, the iteration will stop

We will call the body of a core loop in a loop kernel of an algorithm the
**next step** of the algorithm (see figure 3.3). This is the most important and
complex part of an algorithm class; we have to examine it in detail.

### 3.1.3 Auxiliary Structures

Some algorithms use higher-order data structures such as stacks, lists, priority queues, or simple queues. Since we want to develop solutions that are language independent, we have to formulate appropriate language independent class definitions for these higher-order data structures.

**L.I.C.Definition 3.2** *interface* **stack***:*

1. Push($v$)

2. Pop()

3. Empty()

**L.I.C.Definition 3.3** *interface* **queue***:*

1. Append($v$)

2. Pop()

3. Empty()

**L.I.C.Definition 3.4** *interface* **priority queue***:*

1. Insert($v, i$)

2. ExtractMinimum()

3. DecreaseKey($v, i$)

4. Empty()

## 3.2 Depth First Search

**Definition 3.3** *A* **search** *in a graph is an ordering of a subset of a connected component of the graph that is rooted in a single node $s$.*

**Definition 3.4** *A* **depth first search** *is the following: start with a starting node. Visit each of the nodes adjacent to the current node calling the depth first search on each that has not been visited.*

Let $G$ be the graph of figure 3.4; we want to traverse it in depth first order. Starting at, say node "a", the algorithm inspects in every step a single edge. A possible depth first search for the sample graph $G$ is the following: $(0, 1), (1, 2), (1, 3), (3, 4), (4, 5)$ (where $(a, b)$ is an edge between $a$ and $b$). The graph search can be implemented as an algorithm class with the next step as in figure 3.5.

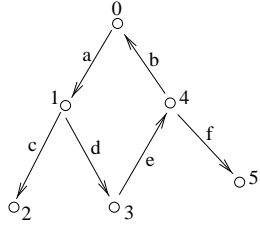

*Fig. 3.4:* The sample graph *G*.

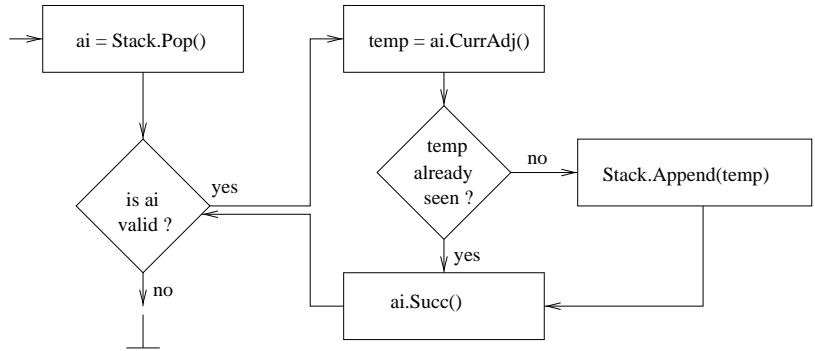

*Fig. 3.5:* simple dfs next step

Figure 3.1 shows an exemplary computation of a depth first search in the sample graph *G*. An adjacency iterator is popped from the stack. If it is valid, i.e. represents an edge, its current target node is inspected, otherwise we are done. If the current adjacent node — which is the target node of the represented edge — has not yet been seen, it will be appended to the stack. Then, the next incident edge will be considered. The "next"-step simply processes a whole node by checking if its adjacent nodes were already seen. In this case, the adjacent node will be appended to the stack.

The fixed node associated with the adjacency iterator on top of a stack is called the **current node** of the stack.

**Property 3.5** *If the stack is initialized with the adjacency iterator "s", then the list of current nodes before each execution of the* `next()`*-step describes a prefix of a depth first search in G.*

Basically this next step is exactly what is implemented in the current DFS implementation of the LEDA package. It equals the original DFS algorithm. □

Whenever a depth first search is needed, this alternative implementation might be used to introduce flexibility through the ability of exchanging data structures (different graph representations) or adding functionality (i.e. filter iterators or animation).

| step | action | content of stack | visited nodes |
|------|--------|------------------|---------------|
| 0 | adjacency iterator $(0,a)$ into stack | $(0,a)$ | 0 |
| 1 | pop adjacency iterator $(0,a)$ | - | 0 |
| 2 | $(0,a)$ valid? $\Rightarrow$ yes | - | 0 |
| 3 | current adjacency iterator of $(0,a)$ is $(1,c)$ | - | 0 |
| 4 | $(1,c)$ seen? $\Rightarrow$ no | - | 0,1 |
| 5 | append $(1,c)$ to stack | $(1,c)$ | 0,1 |
| 6 | advance$((0,a))=(0,-)$ | $(1,c)$ | 0,1 |
| 7 | $(0,-)$ valid? $\Rightarrow$ no | $(1,c)$ | 0,1 |
| 8 | pop $(1,c)$ | - | 0,1 |
| 9 | $(1,c)$ valid? $\Rightarrow$ yes | - | 0,1 |
| 10 | current adjacency iterator of $(1,c)$ is $(2,-)$ | - | 0,1 |
| 11 | $(2,-)$ seen? $\Rightarrow$ no | - | 0,1,2 |
| 12 | append $(2,-)$ to stack | $(2,-)$ | 0,1,2 |
| 13 | advance $((1,c))=(1,d)$ | $(2,-)$ | 0,1,2 |
| 14 | $(1,d)$ valid? $\Rightarrow$ yes | $(2,-)$ | 0,1,2 |
| 15 | current adjacency iterator of $(1,d)$ is $(3,e)$ | $(2,-)$ | 0,1,2 |
| 16 | $(3,e)$ seen? $\Rightarrow$ no | $(2,-)$ | 0,1,2,3 |
| 17 | append $(3,e)$ to stack | $(2,-)$, $(3,e)$ | 0,1,2,3 |
| 18 | advance $((1,d))=(1,-)$ | $(2,-)$, $(3,e)$ | 0,1,2,3 |
| 19 | $(1,-)$ valid? $\Rightarrow$ no | $(2,-)$, $(3,e)$ | 0,1,2,3 |
| 20 | pop $(3,e)$ | $(2,-)$ | 0,1,2,3 |
| 21 | $(3,e)$ valid? $\Rightarrow$ yes | $(2,-)$ | 0,1,2,3 |
| 22 | current adjacency iterator of $(3,e)$ is $(4,f)$ | $(2,-)$ | 0,1,2,3 |
| 23 | $(4,f)$ seen? $\Rightarrow$ no | $(2,-)$ | 0,1,2,3,4 |
| 24 | append $(4,f)$ to stack | $(2,-)$, $(4,f)$ | 0,1,2,3,4 |
| 25 | advance $((3,e))=(3,-)$ | $(2,-)$, $(4,f)$ | 0,1,2,3,4 |
| 26 | $(3,-)$ valid? $\Rightarrow$ no | $(2,-)$, $(4,f)$ | 0,1,2,3,4 |
| 27 | pop $(4,f)$ | $(2,-)$ | 0,1,2,3,4 |
| 28 | $(4,f)$ valid? $\Rightarrow$ yes | $(2,-)$ | 0,1,2,3,4 |
| 29 | current adjacency iterator of $(4,f)$ is $(5,-)$ | $(2,-)$ | 0,1,2,3,4 |
| 30 | $(5,-)$ seen? $\Rightarrow$ no | $(2,-)$ | 0,1,2,3,4,5 |
| 31 | append $(5,-)$ to stack | $(2,-)$,$(5,-)$ | 0,1,2,3,4,5 |
| 32 | advance $((4,f))=(4,b)$ | $(2,-)$,$(5,-)$ | 0,1,2,3,4,5 |
| 33 | $(4,b)$ valid? $\Rightarrow$ yes | $(2,-)$,$(5,-)$ | 0,1,2,3,4,5 |
| 34 | current adjacency iterator of $(4,b)$ is $(0,a)$ | $(2,-)$,$(5,-)$ | 0,1,2,3,4,5 |
| 35 | $(0,a)$ seen? $\Rightarrow$ yes | $(2,-)$,$(5,-)$ | 0,1,2,3,4,5 |
| 36 | advance $((4,b))=(4,-)$ | $(2,-)$,$(5,-)$ | 0,1,2,3,4,5 |
| 37 | $(4,-)$ valid? $\Rightarrow$ no | $(2,-)$,$(5,-)$ | 0,1,2,3,4,5 |
| 38 | pop $(5,-)$ | $(2,-)$ | 0,1,2,3,4,5 |
| 39 | $(5,-)$ valid? $\Rightarrow$ no | $(2,-)$ | 0,1,2,3,4,5 |
| 40 | pop $(2,-)$ | - | 0,1,2,3,4,5 |
| 41 | $(2,-)$ valid? $\Rightarrow$ no | - | 0,1,2,3,4,5 |

*Tab. 3.1:* Exemplary computation of a depth first search in the sample graph $G$ (see figure 3.4 ).

| line (see figure 3.1) | adjacency iterator |
|:---:|:---:|
| 0 | $(0, a)$ |
| 7 | $(1, c)$ |
| 19 | $(3, e)$ |
| 26 | $(4, f)$ |
| 37 | $(5, -)$ |
| 39 | $(2, -)$ |

*Tab. 3.2:* Observe that the current adjacent iterator before each execution of `next()` is representing a single node of a depth first tree. In this example, $0 - 1 - 3 - 4 - 5 - 2$ is a depth first search in the sample graph (see figure 3.4).

Actually, the shown implementation does not cover the registration of time stamps as needed in some versions of computation of strongly connected components in graphs[18].

In the next version, a depth first search is computed where each node is assigned two time stamps: the first one is set when the node is seen the first time; the last one is set when it is eventually removed from the stack. [CLR94] discusses a recursive approach to do this. For algorithm classes, we need the iterative approach.

At first, an informal description of the algorithm (see figure 3.6 for a complete overview):

- there is a stack that stores adjacency iterators

- there is a data accessor "mark" that yields `true` for a given iterator *it* if and only if `it.v` has been seen before

- an adjacency iterator is said to be **fresh** if and only if the current adjacent node has not yet been seen (`mark[it.curr_adj()] == false`) and *it* is valid.

- events for adjacency iterator "*it*":

    – if *it* is fresh then inspect the current adjacent node

---

[18] For example, in [CLR94] (pp. 488) time stamps were used for this.

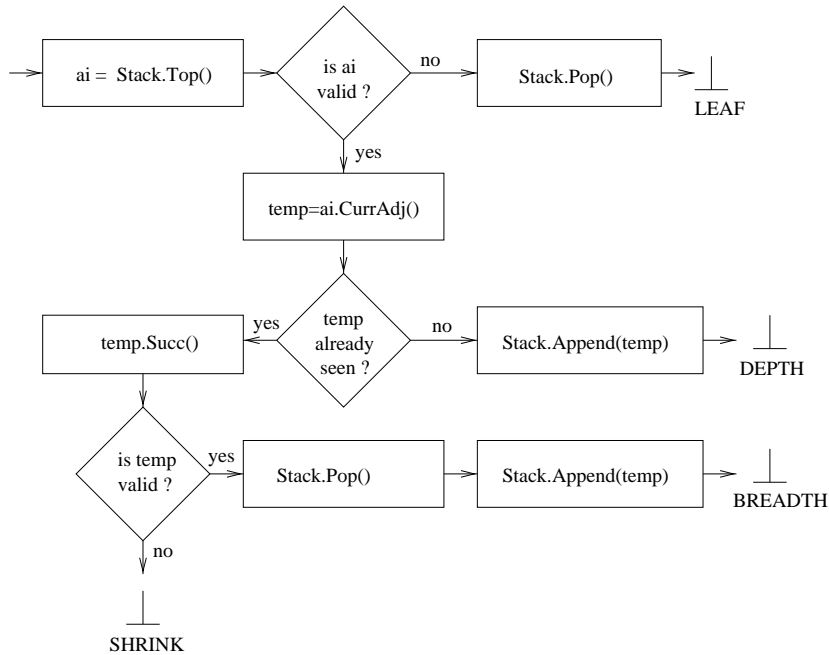

*Fig. 3.6:* complex dfs next step

- if *it* is not fresh but valid then advance the current incident edge to the next one; if no next one present, pop *it* from stack
- if *it* is not fresh and not valid then pop *it* from stack

This is slightly a different view on depth first search and the result of the algorithm is still the same.

This variant can be used to record time stamps for each node $v$ of the depth first tree when $v$ is first visited and last visited; the according adjacency iterator will be popped from the stack only if the adjacency list of its fixed node is consumed completely. The first time stamp can be recorded when it is seen first by the user. Whenever the user sees it again on top of the stack, a second time stamp will be recorded.

Since leaves are never fresh, they are seen only once. Therefore their timestamps must be equal. To do this, a state is returned, i.e. a description of the last action of the `next()`-step.

There are four cases:

1. `dfs_shrink`: an adjacency iterator was popped from the stack, i.e. the tree-walk returns in root-direction

2. `dfs_leaf`: same as `dfs_shrink`, but a leaf occurred

3. `dfs_grow_depth`: a new adjacency iterator was appended to the stack because it was detected as not seen before, i.e. the tree-walk goes in depth-direction

4. `dfs_grow_breadth`: the former current adjacency iterator was replaced by the successor iterator, i.e. the tree-walk goes in breadth-direction

Now suppose the algorithm next step returns `dfs_leaf`, we can handle this case for updating both time stamps at the same time.

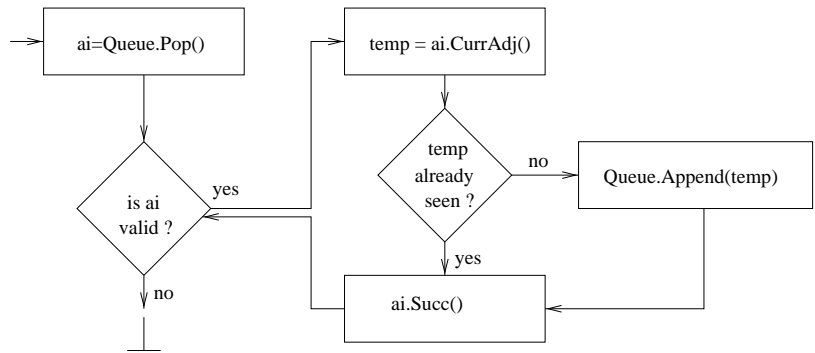

*Fig. 3.7:* bfs next step

## 3.3 Breadth First Search

**Definition 3.6** *A* **breadth first search** *is the following: start with a starting node. Visit each of the nodes adjacent to the current node. If a node has not yet been seen, append it to the queue $Q$. For each node in $Q$, call the breadth first search.*

The breadth first algorithm is completely analogous to the simple version of the depth first algorithm except that we do not use a stack for storing data but a queue. Let $G$ be the graph of figure 3.4, and we want to traverse it in a breadth first order. Starting at, say node "a", the algorithm inspects in every step a single edge.

A possible result for $G$ is $(0, 1), (0, 4), (1, 2), (1, 3), (4, 3), (4, 5)$.

The graph search can be implemented as an algorithm class with the next step as in figure 3.7.

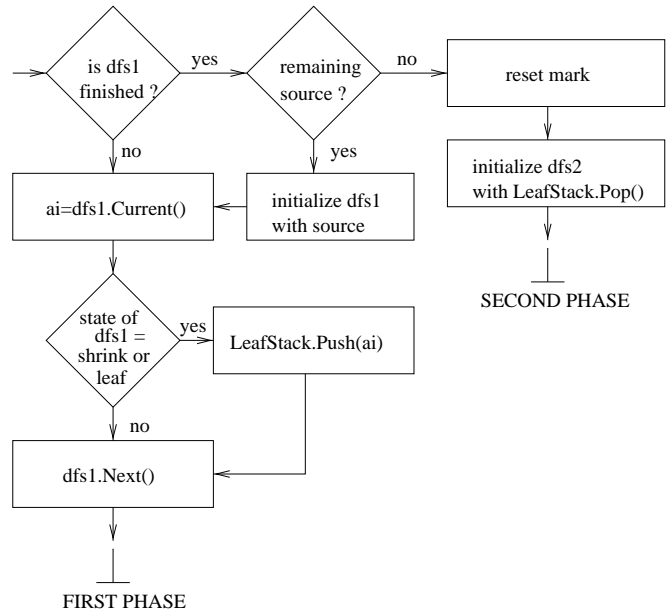

*Fig. 3.8:* In the first phase, a depth first forest of the graph will be computed.

## 3.4   Strongly Connected Components

**Definition 3.7** *$SC \subset V_G \times E_G$ is a* **strongly connected component** *for a graph $G = (V_G, E_G)$ if and only if $\forall v, w \in SC \Rightarrow \exists\, path\, P = (v_1, \ldots v_k) \wedge v = v_1 \wedge w = v_k$, i.e. $SC$ are the inclusion-maximal components of $G$.*

This algorithm works in two phases, where the first computes a depth first forest[19] (see figure 3.8), and the second the depth first trees of the transposed graph (see figure 3.9).

The **transpose** of a graph $G = (V, E)$ is the graph $G^T = (V, E^T)$, where $E^T = \{(v, u) \in V \times V : (u, v) \in E\}$. Thus, $G^T$ is $G$ with all its edges reversed.

This algorithm uses the depth first search implementation of section 3.2 and a stack $LeafStack$ to store all nodes which we have left in a search in the first phase of the algorithm.

*The first phase:* We maintain a linear node iterator $it$ that traverses the node set of the graph to attach every source node. If we have an empty stack we search for $it$ a new node which is not seen yet, and take it as the source of a new depth first tree. Each time we execute a next()-step of the dfs-algorithm, we know that a node has been left in a search; we notify that node in a stack of nodes: *LeafStack*. If there remains no new node, we reset *mark* for all nodes and initialize the depth first search of the second phase with the top node of *LeafStack*.

---

[19] The presented algorithm will be very similar to the algorithm in [CLR94] that computes strongly connected components using a depth first search algorithm.

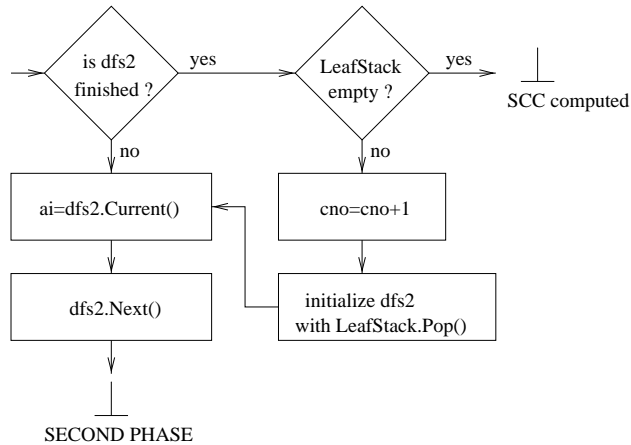

*Fig. 3.9:* In the second phase, depth first trees of the reversed graph are computed for each item of the stack.

*The second phase:* We use adjacency iterators for incoming edges to compute depth first trees of the reversed graph. The result will be that each depth first tree represents a strongly connected component.

### 3.4.1 Correctness

We will show that this algorithm functions correctly.
At first we show a property of the first phase:

**Property 3.8** *The first phase of the algorithm computes a depth first forest with finishing times $f[v]$ for each node $v \in V_G$ of a given graph $G = (V_G, E_G)$.*

*Proof:* Suppose that we can compute depth first trees. Then, if we have a mapping "mark" that yields *true* for every visited node, we can find new source nodes only by traversing the node set $V_G$ of $G = (V_G, E_G)$. Since "mark" is monotone[20], we can use a single node iterator to find the sources. If we start with a source node $s$, an adjacency iterator $it$ is initialized with $s$ and appended to the stack. Before a depth first next step is executed, the current adjacency iterator $ci$ is saved. If the return value is `dfs_shrink` or `dfs_leaf`, we know that this node will never be seen again, and $f[ci.v]$ can be set with a finishing counter variable (where $ci.v$ means the fixed node associated to $ci$). Since $f$ will only be set if the next step returns in direction to the root node, it is really a finishing time mapping for $V_G$. In the algorithm it is more efficient to use a stack of nodes that realize the finishing time stamps.                                                                    □

---

[20] I use the term "monotone" to describe that a node which is marked as seen will not be regarded as not seen later.

Since we will operate on the transpose of $G$, we have to show the following:

**Property 3.9** *Let $G_o$ be the graph described by an adjacency iterator for outgoing edges $ai_o$ and $G_i$ be the graph described by an adjacency iterator for incoming edges $ai_i$. Then it follows: $G_o = G_i^T$.*

*Proof:* The transpose of a graph is the graph where the edges are reversed, i.e. incoming edges of a node are outgoing edges and outgoing edges are incoming edges. Since both types of iterators are defined equally except in the direction of edges, $G_o$ is the transpose of $G_i$.                                  □

Now we need a statement concerning the second phase:

**Lemma 3.10** *The second phase of the algorithm computes depth first trees for nodes in order of decreasing $f[v]$ in $G^T$, the transpose of graph $G$.*

*Proof:* It is not a problem to reorder the node set in order of decreasing $f[v]$. If we used a stack in the first phase, we can simply process the stack by deleting the topmost element of it. It remains to show that with an adjacency iterator for incoming edges we can discover the transpose of $G$, but this follows from the definition of this iterator type.              □

Now we can show that strongly connected components will be computed correctly:

**Theorem 3.11** *If the linear node iterator is initialized with the first node of $G = (V_G, E_G)$, then each computed component is a strongly connected component.*

*Proof:* Since the node iterator traverses the node set $V_G$, it can find a new node that has not yet been seen for a new depth first tree, if there is still one remaining. With property 3.8 it follows that we have finishing time $f[v]$ for each node, and with lemma 3.10, we know that depth first trees in order of decreasing $f[v]$ in the transposed graph can be computed.

We can summarize the algorithm in the following steps:

1. call `DFS(G)` to compute finishing times $f(u)$ for each node $u$

2. compute $G^T$

3. call `DFS`($G^T$), but in the main loop of DFS, consider the nodes in order of decreasing $f(u)$ (as computed in line 1)

4. output the vertices of each tree in the depth-first forest of step 3 as a separate strongly connected component

Since this is exactly the algorithm which was proven in [CLR94] (pp. 489) as correct, it really computes strongly connected components. □

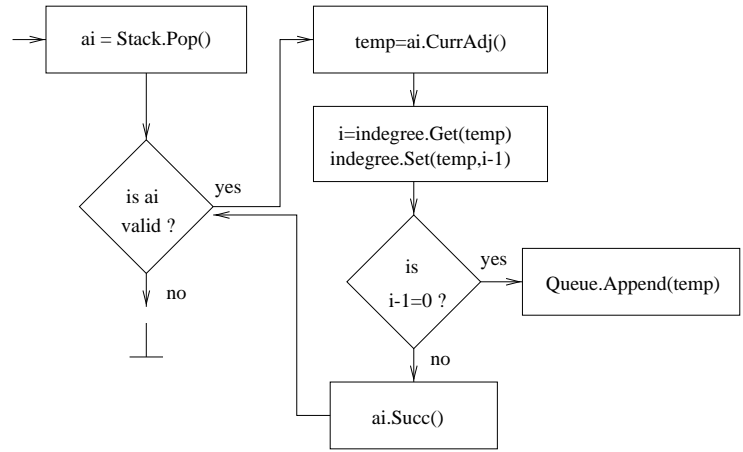

*Fig. 3.10:* An object of this class maintains an *internal queue*, which contains all
nodes (in form of adjacency iterators where the current node is equal
to the fixed node) that are not yet passed, but all its predecessors have
been passed. The implementation of the **next**-step here removes the first
element of the internal queue. Now, the in-degree of all nodes incident
to outgoing edges of that node will be decreased by 1. Every node
that becomes a source node in the current subgraph, i.e. this in-degree
counter is zero, will be inserted into the queue.

## 3.5 Topological Sorting

Topological sortings of graphs is implemented as an algorithm class which
generates objects that can traverse the graph in topological order by applying
the "next()" method.

**Definition 3.12** *A **topological sort** of a graph $G = (V, E)$ is a linear
ordering of all its vertices such that if $G$ contains an edge $(u, v)$, then $u$
appears before $v$ in the ordering.*

The algorithm computes topological sorts by maintaining a list of nodes with
zero in-degree. After deleting these nodes from the graph, other nodes will
have zero in-degree. By this, we will generate a topological ordering. In fact,
we will not explicitly delete the nodes from the graph, but emulate this by
decreasing an in-degree counter of affected nodes (see figure 3.10).

(1)   **for** each vertex $v \in V[G]$
(2)        **do** $distance[v] = \inf$
(3)   $d[s] = 0$
(4)   $PQ.\texttt{Insert}(s, 0)$
(5)   **while** $PQ \neq \emptyset$
(6)        $u = PQ.\texttt{ExtractMinimum}()$
(7)        **for** each vertex $v \in Adj[u]$
(8)            **if** $d[v] > d[u] + w(u, v)$
(9)                **then** $d[v] = d[u] + w(u, v)$
(10)                    **if** $v \in PQ$
(11)                        $PQ.\texttt{DecreaseKey}(v, d[v])$
(12)                    **else** $PQ.\texttt{Insert}(v, d[v])$

*Fig. 3.11:* Example of pseudo-code for the *Dijkstra* algorithm ($s$ is the source node and $d$ an array that stores distances for each node; $PQ$ is a priority queue implemented with Fibonacci heaps): it computes shortest paths in a graph $G$ with respect to nonnegative edge lengths, where the length of a path is the total sum of the lengths of its edges.

## 3.6 Dijkstra

Consider a graph with two special nodes, called the origin and destination. Associated with each edge is a nonnegative length. The objective is to find a shortest path between the origin and destination.

There is a very fast algorithm for this problem which runs in $\mathcal{O}(n \log n + m)$[21] (see figure 3.11). The essence is to explore outward from the origin, successively seeking the shortest paths to nodes in the network until the destination is reached.

**Definition 3.13** *A **shortest path** in a graph $G = (V_G, E_G)$ is a path $P = (v_1, \ldots, v_k)$ with the property that there is no path $P'$ connecting $v_1$ and $v_k$ with smaller edge lengths.*

The flexible implementation of Dijkstra makes it possible to cover the scenarios of table 3.3.

---

[21] See also in [CLR94] or [PS82], p128.

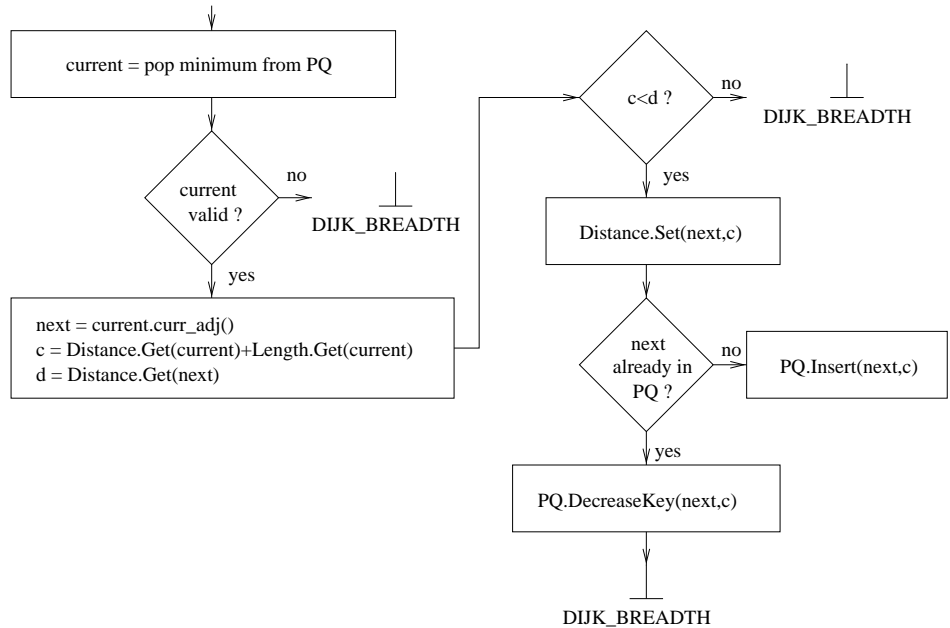

*Fig. 3.12:* Depth Phase of Dijkstra Next Step

### 3.6.1    Algorithm

The next step of this implementation of Dijkstra processes each time one single edge and not (like in other implementations) a complete node. This makes it possible to compute the predecessors of a shortest path tree simply by notifying the current edge of each current adjacency iterator before a next step as the predecessor of the current adjacent node.

In the design of the next step we discuss two different states: `DIJK_DEPTH` and `DIJK_BREADTH`. In the following we will refer to the different states as *depth phase* (see figure 3.12) and *breadth phase* (see figure 3.13). The algorithm starts by initializing the queue with the source node, where the distance of this node is set to zero. Then, it enters the depth phase.

*Depth Phase:* The current node with minimum distance is popped from the queue and its first adjacent node is explored. If this node improves the stored distance, it will be updated like this: if the node is not yet in the queue, it is inserted into the priority queue with the new distance. Otherwise only method `DecreaseKey` of the priority queue will be called to adjust the location of the node in amortized constant time. After, the algorithm continues in the breadth phase.

*Breadth Phase:* If the current adjacency iterator is valid, we consider the next adjacent node of the fixed node. If this node improves the stored dis-tance, it will be updated like this: if the node is not yet in the queue, it is inserted into the priority queue with the new distance. Otherwise only method `DecreaseKey` of the priority queue will be called to adjust the location of the node in amortized constant time. After, the algorithm continues in the breadth phase.

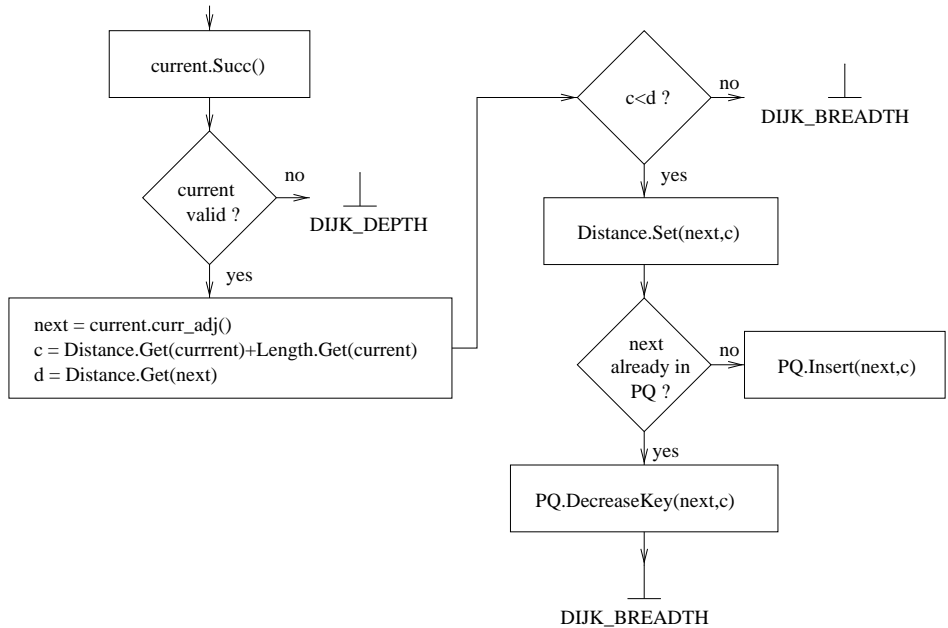

*Fig. 3.13:* Breadth Phase of Dijkstra Next Step

Otherwise, if the current adjacency iterator is not valid, i.e. has no adjacent successor, we try to consider a new node by entering the depth phase.

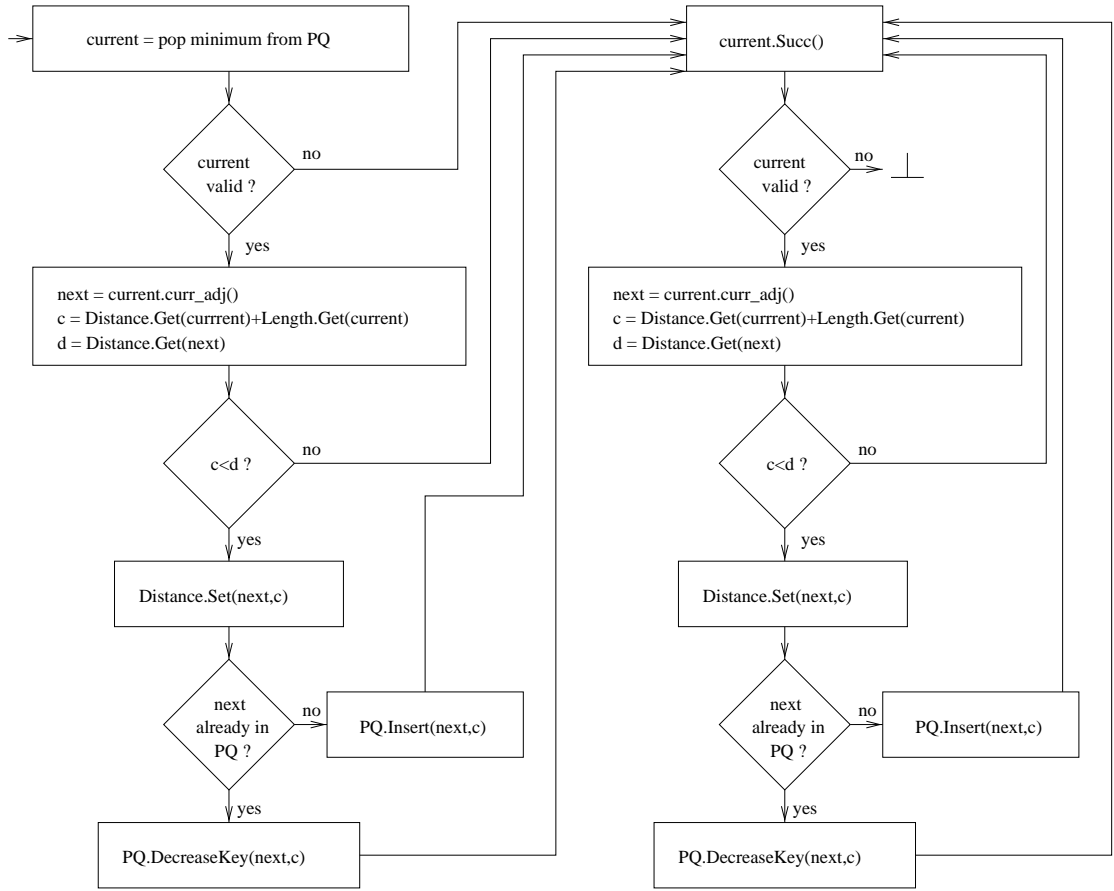

*Fig. 3.14:* New combined figure, made out of both parts of the next-step.

### 3.6.2 Correctness

**Property 3.14** *The variant "Dijkstra algorithm with two states" can be used in a way that is equivalent to the original Dijkstra algorithm.*

*Proof:* We try to compare the control flow graph with the pseudo-code in figure 3.11.

First we connect the given control flow graphs for the two states (in figure 3.14. Observe that we have to add the expression, which tests before each next step, if the priority queue is empty. In this case, the algorithm is finished.

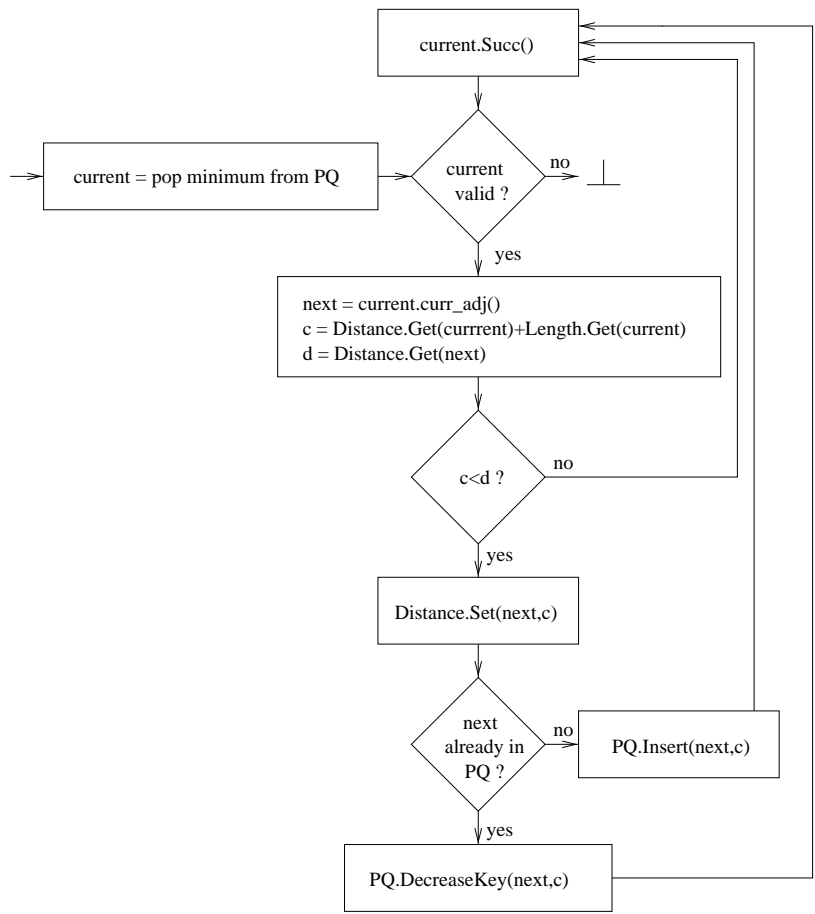

*Fig. 3.15:* Minimized graph of figure 3.14.

| Problem | Solution |
|---|---|
| compute a shortest path between two nodes $s$ and $t$ | execute `next()` until target node has been visited |
| *Euclidean case:* compute a shortest $(s, t)$-path in a bounded area in the plane | use a filter iterator and a predicate that checks edges for being inside the area and execute `next()` until target node has been visited |
| operation counting—how many operations? | use an observer adjacency iterator that counts the number of executions |
| on-line checking - e.g. are the edge lengths bounded by a certain constant? | can be done in data accessor for length that is checking the value in every call of function `get` |
| simultaneous graphical animation of the algorithm working | use observer iterators with an observer that has access to a special window in which all drawings take place |
| pre- Dijkstra - the algorithm can be interrupted and continued again | standard use of the class |
| pseudo-parallel execution of several instances of the Dijkstra algorithm (e.g. instances perform single iterations in a round-robin style) | use several instances of the algorithm |
| edge lengths and/or node distances are realized as template parameters of `GRAPH` instead of `edge_array` and `node_array` | use instances of `MemberAccessor` |
| shortest paths for all nodes of $R \subseteq V$, i.e. find $\forall v \in V$ a minimum distance $\min\{dist(r,v) : r \in R\}$ | use the algorithm class with $R$ as set of sources. |

*Tab. 3.3:* Different scenarios can be covered by the flexible implementation of Dijkstra.

$$current = PQ.\texttt{ExtractMinimum}()$$
**while** $current.\texttt{Valid}()$
$\quad\quad next = current.\texttt{CurrAdj}()$
$\quad\quad c = Distance.\texttt{Get}(current) + Length.\texttt{Get}(current)$
$\quad\quad d = Distance.\texttt{Get}(next)$
$\quad\quad$**if** $d > c$
$\quad\quad\quad$**then** $Distance.\texttt{Set}(next, c)$
$\quad\quad\quad\quad\quad$**if** $next \in PQ$
$\quad\quad\quad\quad\quad\quad PQ.\texttt{DecreaseKey}(next, c)$
$\quad\quad\quad\quad\quad$**else** $PQ.\texttt{Insert}(next, c)$
$\quad\quad current.\texttt{Succ}()$

*Fig. 3.16:* Conversion of the control flow graph of figure 3.15 into pseudo-code.

Now we encounter two equivalent parts and we can minimize the graph (in figure 3.15). We translate the minimized graph into pseudo-code and get figure 3.16.

We have to show that this part of pseudo-code does the same as the body of the loop of figure 3.11.

There are several observations:

$O_1$ Iterators can be removed and replaced by a forall-loop, i.e. an adjacency iterator can be replaced by a loop that traverses the adjacency list of the fixed node.

$O_2$ Data accessors can be replaced by arrays that have nodes or edges as domains:

1. $Distance.\texttt{Get}(it)$ and $Distance.\texttt{Set}(it, val)$ may be replaced by $d[it.v]$, where $it$ refers to $it.v$

2. $Length.\texttt{Get}(it)$ may be replaced by $w(it.e)$, where $it$ refers to $it.e$; alternatively, we can write $w(a, b)$ where $a$ is the source and $b$ the target node of $it.e$

$O_3$ *current* and *next* can be replaced by $u$ and $v$, respectively.

This leads to the version of figure 3.17, which is actually equal to the body of the loop of figure 3.11(5-12).

The remaining part is initialization and the break condition. Initialization is provided by the user and can be the same as in figure 3.11(1-4). The

$u = PQ.\texttt{ExtractMinimum}()$
**for** each vertex $v \in Adj[u]$
    **if**  $d[v] > d[u] + w(u, v)$
        **then** $d[v] = d[u] + w(u, v)$
            **if** $v \in PQ$
                $PQ.\texttt{DecreaseKey}(v, d[v])$
            **else** $PQ.\texttt{Insert}(v, d[v])$

*Fig. 3.17:* Application of the observations $O_1$, $O_2$ and $O_3$ to the pseudo-code of figure 3.16.

---

break condition in the converted algorithm can be made equal to the break condition of figure 3.11(5).

This completes the proof and we have shown that the algorithm class can be used in a way that is equivalent to the original Dijkstra algorithm.    □

This variant is an example of decomposing the smallest steps of an algorithm with the help of status flags. With this technique, any algorithm can be modified to have fine granulated basic steps.

By this design, we increase flexibility, since new functionality can be added around the next-steps from outside. For example, predecessor edges can be recorded from outside of the algorithm without modifying the internals of it [22].

---

[22] This is shown in the current implementation of the graph-iterator LEP — see algorithm `GIT_DIJKSTRA`

# 4. MATCHING ALGORITHM

This chapter describes a rather complex network algorithm, which computes a maximum cardinality matching of a given graph. The definitions and some parts of the algorithm refer to the discussion of a nonbipartite matching algorithm in [AMO93]. This discussion is very similar to that in [PS82] (pp. 218) and is based on the concept of flowers, introduced in [E65].

What is a matching? A **matching** $M \subseteq E$ of a graph $G = (V, E)$ is a subset of edges such that no two edges are incident to the same node. If an edge $e$ is in the matching $M$, it is called a **matched edge**, otherwise an **unmatched edge**. A node incident to a matched edge is a **matched node**, all other nodes are **unmatched nodes**. For each edge $e = (v, w) \in M$ we say, $v$ is matched to $w$ and $w$ is matched to $v$.

We will begin with a restricted version, namely bipartite matching, of the algorithm and will adapt it to fit the general case.

A simple application of bipartite matching, is an attempt to marry off as many adults in some village as possible. Marriages must be monogamous. No one should be forced to marry someone she or he dislikes. We will assume compatibility is symmetric, i.e., a pairing is compatible if neither party dislikes the other.

The edges in a bipartite graph for this problem represent the compatibilities. It is theoretically possible for everyone of the opposite sex. If everyone likes everyone else, then the graph is the complete $m \times n$ — bipartite graph (where $m$ is the number of men and $n$ the number of women). In this case, the maximum number of people that can be married is $2 \cdot min\{m, n\}$.

We will see an approach that computes a scenario where the number of couples is maximal. We start with some definitions.

## 4.1 Preliminaries

A path $P = (e_1, \ldots e_k)$ is an **alternating path** if each $e_i$ with odd $i$ is matched and even $i$ is unmatched. Such a path is called an **even alternating path** if it contains an even number of edges, an **odd alternating path** otherwise. An **alternating cycle** is an alternating path that is a cycle.

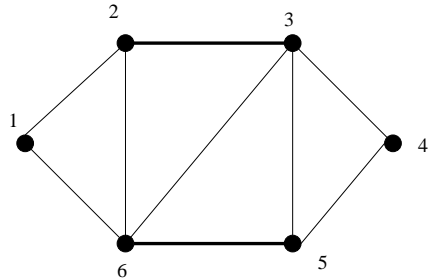

Fig. 4.1: Matching example. In this example, 1-2-3-6-5 is an even alternating path and 1-2-3-6-5-4 an odd one. 6-2-3-5-6 is an alternating cycle.

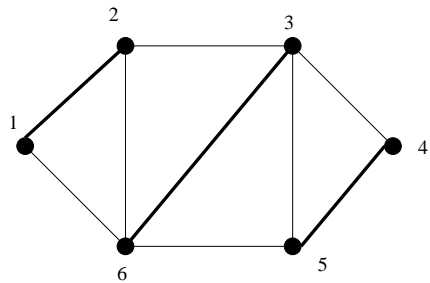

Fig. 4.2: Improved matching of figure 4.1 after processing the augmenting path 1-2-3-6-5-4.

We say, an odd alternating path with respect to a matching $M$ is an **augmenting path** if the first and last node in the path are unmatched (in the definition above, $k$ would be odd, as well). We call them augmenting paths, because after re-designating matched edges on the augmenting path as unmatched and unmatched edges as matched, we obtain another matching of increased cardinality $|M| + 1$ (see figures 4.1 and 4.2).

We now introduce the concept of symmetric difference because it will be quite useful later for the description of the algorithm:

Let $S_1$ and $S_2$ be two sets; the symmetric difference of these sets, denoted $S_1 \oplus S_2$, is the set $S_1 \oplus S_2 = (S_1 \cup S_2) - (S_1 \cap S_2)$. In other words, the symmetric difference are those elements that are members of one, but not both of $S_1$ and $S_2$.

**Property 4.1** *If $M$ is a matching and $P$ is an augmenting path with respect to $M$, then $M \oplus P$ is a matching of cardinality $|M| + 1$. Moreover, in the matching $M \oplus P$, all of the matched nodes in $M$ remain matched and two additional nodes, namely the first and last node of $P$, are matched.*

*Proof:* The symmetric difference here is a set-theoretic formulation to inter-change the matched and unmatched edges in $P$. We refer to the process of replacing $M$ by $M \oplus P$ as an **augmentation**. This operation yields a matching of cardinality $|M| + 1$, because all matched nodes remain matched. After the operation, the formerly unmatched first and last node of the augmenting path will be matched. Since the path will still be an alternating path, we have an improved matching with increased cardinality $|M| + 1$. The second conclusion follows from the definitions (see the beginning of this chapter). $\square$

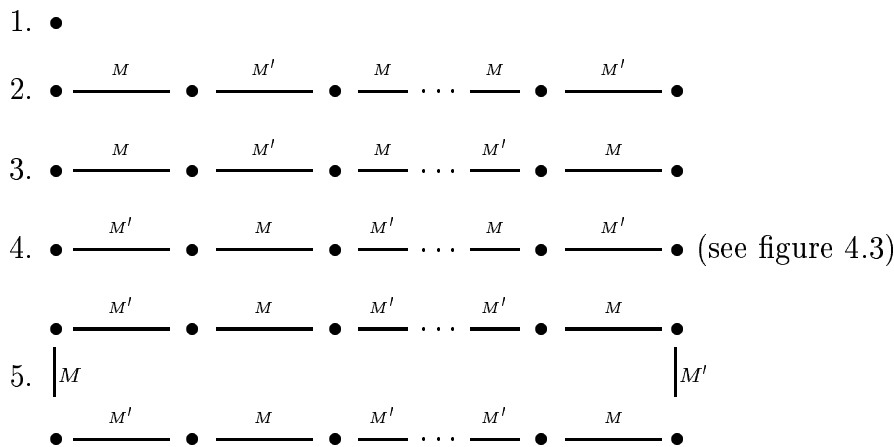

*Tab. 4.1:* Possible types of components formed by a symmetric difference of two matchings $M$ and $M'$. Since each subgraph $G'$ has only nodes with in-degree 0, 1 or 2 these are all possibilities of components. Each edge that is labeled with $M$ is matched in $M$, and every edge that is labeled with $M'$ is matched in $M'$.

**Property 4.2** *If $M$ and $M'$ are two matchings, their symmetric difference defines the subgraph $G' = (V, M \oplus M')$ with the property that every component is one of the five types shown in table 4.1.*

This property follows from the facts that in the subgraph $G'$ each node has degree 0, 1 or 2, and the only possible components with these node degrees are singleton nodes (as shown in table 4.1.1.), paths (as shown in table 4.1.2. to 4.), or even–length cycles (as shown in table 4.1,5.). $\square$

### 4.1.1 Augmenting Path Theorem

The algorithm will rely crucially on the following theorem:

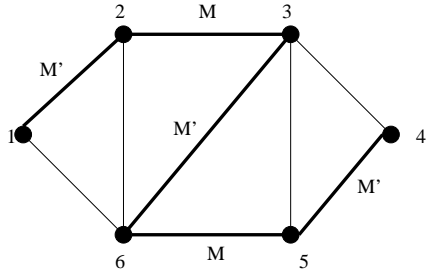

*Fig. 4.3:* The symmetric difference of the matchings of figure 4.1 and figure 4.2.

## Theorem 4.3 Augmenting Path Theorem.
*Let node $p$ be unmatched in a matching $M$ for a graph $G = (V, E)$. If $M$ contains no augmenting path that starts at node $p$ then node $p$ is unmatched in some maximum matching.*

*Proof.* We will show that we can compute such a maximum matching, where $p$ is unmatched. Let $M'$ be a maximum matching. If node $p$ is unmatched in $M'$ we are done. Assume $p$ is matched in $M'$ and consider the symmetric difference $E' = M \oplus M'$. Each component in $E'$ is one of the five types of table 4.1 (see property 4.2). Since node $p$ is unmatched in $M$, only possibilities 2. and 4. can fit here (in 4.1.2., $p$ will be on the right, in 4.1.4. $p$ will be on the left or on the right). Since there is no augmenting path starting at $p$, only possibility 2 remains and we have an even alternating path $P$ with node $p$ as starting node. Now we can augment $M'$ with the augmenting path $P$ and get a new matching $M'' = M' \oplus P$ where node $p$ is unmatched. □

**Corollary 4.4** *If a matching contains no augmenting paths then it is a maximum matching.*

*Proof:* If we have a matching $M$ that has no augmenting path we can successively delete all unmatched nodes for which there exists no augmenting path. We can do that without losing any possible augmenting path because each time we delete a node, we can still find a maximum matching, where that node is not matched (which is the same as not used in this context). Of course, after this procedure we have deleted all unmatched nodes. Since every node is now matched, we have a maximum matching.

Suppose we have a maximum matching and an augmenting path. Each augmentation increases the number of matched edges by one which is a contradiction since we already had a maximum matching. □

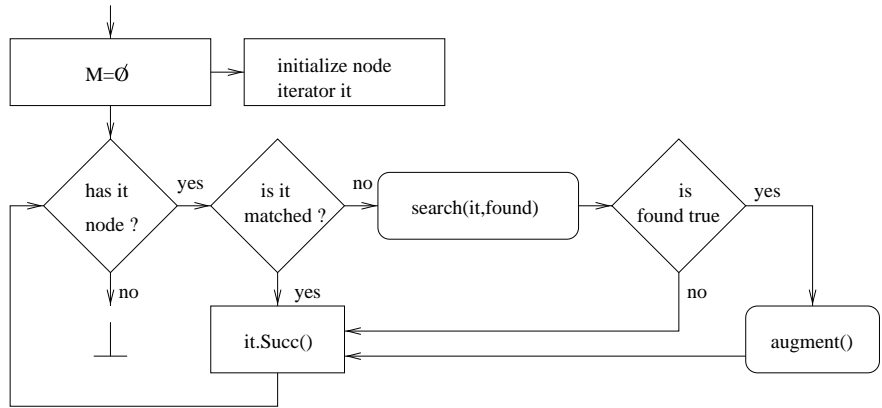

*Fig. 4.4:* **Bipartite matching algorithm:** first, we consider the empty matching
and initialize the node iterator *it* that traverses the node set. If *it* is
still referring to a node, we denote this node with *v*. If *v* is currently
unmatched, we found a possible starting node for an augmenting path and
start the search procedure appropriately. If we found an augmenting path
*P* (with $found = true$), we augment the matching *M* will be replaced by
$M \oplus P$. Otherwise, if *v* is matched or the search failed (with $found = false$), we move the node iterator *it* to the next node in the node set.

### 4.1.2 Bipartite Matching Algorithm

The augmenting path theorem suggests the following algorithm for comput-
ing a maximum cardinality matching: start with a feasible matching *M* (for
example the empty matching), try to find repeatedly an augmenting path *P*,
and replace *M* by $M \oplus P$. If there is no more path, we are done.

Each time we augment the matching, its cardinality increases by one. If the
algorithm terminates, we have a maximum matching according to corollary
4.4.

We try to find an augmenting path using a labeling technique which starts
at an unmatched node *p* and then uses a search algorithm to identify all
reachable nodes. If the algorithm finds an unmatched node, it has discov-
ered an augmenting path. If there is no such unmatched node, there is no
augmenting path starting at node *p*.

We will grow a search tree rooted at node *p* such that each path in the tree
from node *p* to another node is an alternating path. We refer to this tree as
an *alternating tree* and nodes in the tree are *labeled nodes* and the others are
*unlabeled*. The labeled nodes are of two types: *even* or *odd*. The root node is
labeled with even. Notice that whenever an unmatched node has as an odd label, the path joining the root node to this node is an augmenting path.

Figures 4.4, 4.5, 4.6 and 4.7 show the algorithm. We call this algorithm the *bipartite matching algorithm* because, as we will explain later, it will always establish a maximum matching in bipartite graphs (it may fail when applied to nonbipartite graphs):

**Lemma 4.5** *Let $G = (V, E)$ be a graph and $|V| = n$ and $|E| = m$. The bipartite matching algorithm runs in worst case time $\mathcal{O}(n \cdot m)$ for a given graph $G = (V, E)$.*

*Proof:* The algorithm executes the search and augment procedures at most $n$ times. The augment procedure clearly requires $\mathcal{O}(n)$ time. For each node $i$, the search procedure (see on the right of figure 4.5) performs one of the following two operations at most once: it executes *examine-even(i,found)*, or it executes *examine-odd(i,found)*. The latter operation requires constant time per execution. The former operation requires $\mathcal{O}(|Adj(i)|)$ (where $Adj(i)$ is the list of adjacent nodes of $i$), so a total of $\mathcal{O}(\sum_{v \in V} |Adj(i)|) = \mathcal{O}(m)$ time for all the nodes is needed.                                             □

Later, we will show that the algorithm identifies a matching correctly.

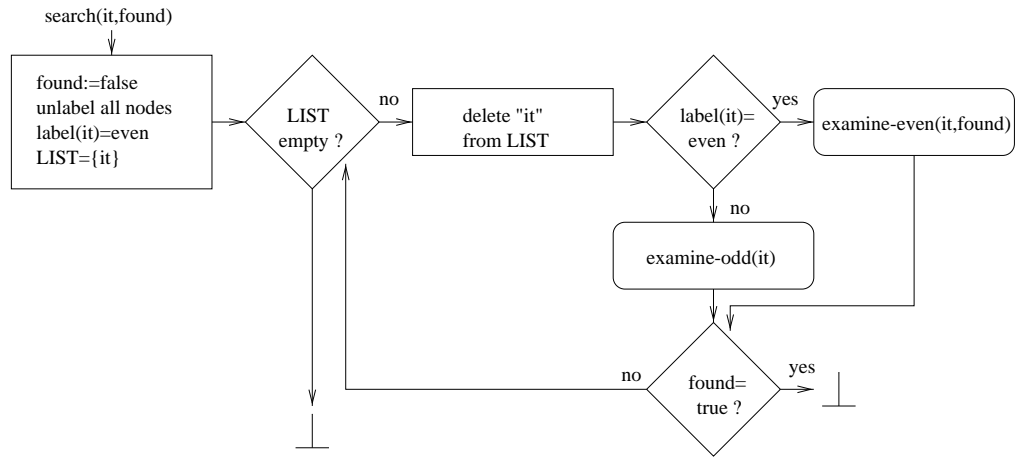

**Fig. 4.5:** *procedure search(i):* in each new search, all nodes are considered as unlabeled; it maintains a set, LIST, of labeled nodes and examines labeled nodes one by one. Depending on the label, it executes *examine-even* (see figure 4.6) or *examine-odd* (see figure 4.7). The search algorithm terminates when LIST becomes empty, or it has assigned an odd label to an unmatched node, thus discovering an augmenting path.

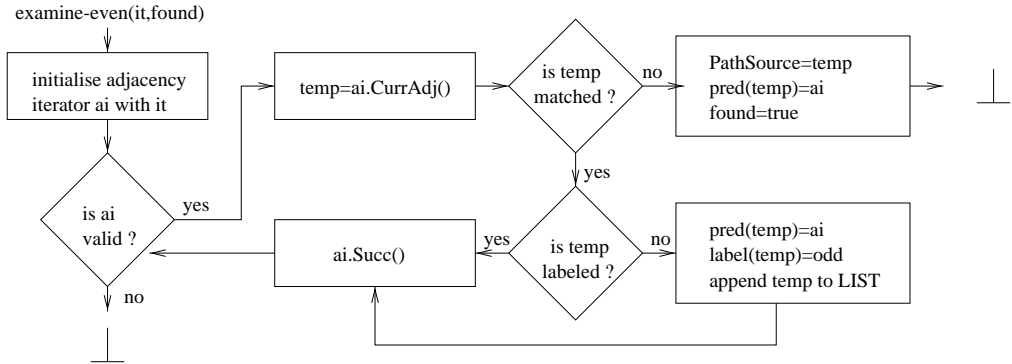

*Fig. 4.6: procedure examine-even(it,found):* first, we initialize an adjacency iterator *ai* with *it*, where *it* refers to node *i*. The algorithm traverses the adjacency list of *i* and assigns an odd label to each unlabeled target node *j* of the current incident edge. If we found an unmatched node, we have discovered an augmenting path; otherwise, we label this node with *odd* and add this node to LIST. Each time we store a new node *j* in LIST, we know that the predecessor node in the search tree is *i* and set *pred(j)* to *i*.

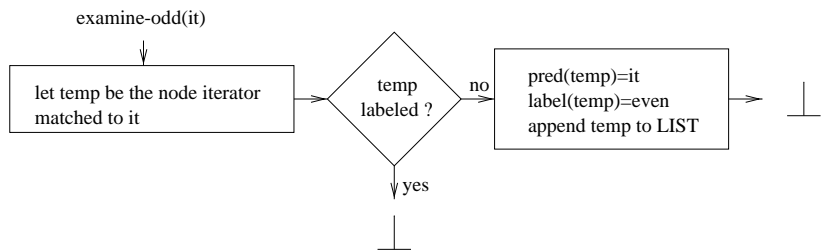

*Fig. 4.7: procedure examine-odd(it):* since *it* refers to a node *i*, we create a node iterator *temp* that refers to the node *j* that is matched to *i*. If *j* is not labeled yet with *even* or *odd*, we label it with *even* and append it to the LIST. Again, as in *examine-even*, each time we store a new node *j* in LIST, we know that the predecessor node in the search tree is *i* and set *pred(j)* to *i*.

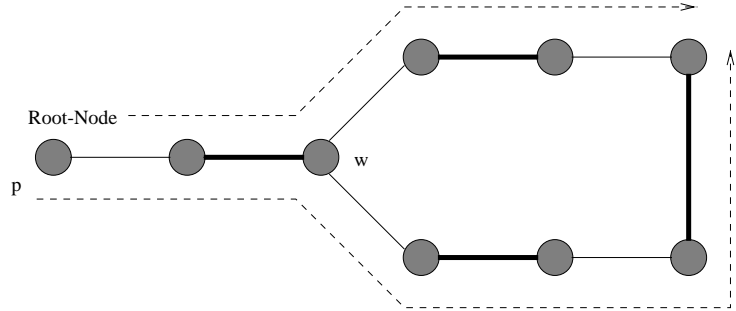

*Fig. 4.8:* Two distinct alternating paths from the root to some node in the blos-
som. The even alternating path terminates with a matched edge, and
the odd alternating path terminates with an unmatched edge.

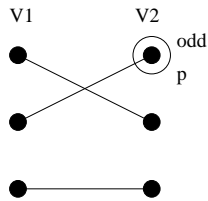

*Fig. 4.9:* We label nodes in $V_1$ with *even* and nodes in $V_2$ with *odd*. If the root node
$p$ is in $V_2$, we have to start with *odd* in the labelling process. Otherwise,
we have to start with *even*.

## 4.2   Unique Label Property

The presented bipartite matching algorithm does not work in the general
case, with nonbipartite graphs, because we might not be able to label the
nodes in the alternating tree consistently. The following property is the key
to why the labeling fails (and so the algorithm):

**Definition 4.6** *A graph G has the* **unique label property** *with respect to
a matching M and a root node p if the search procedure assigns a unique label
to every labeled node (in this case, even or odd) irrespective of the order in
which it examines the alternating tree.*

**Lemma 4.7** *A graph $G = (V, E)$ has the unique label property with respect
to a matching M if and only if it is bipartite.*

*Proof:* If a graph $G = (V, E)$ is bipartite we can partition $V$ into $V_1$ and $V_2$
so that every edge $(v, w) \in E$ has its endpoints in either $V_1$ or $V_2$. If the root
node is in $V_1$, we label all nodes in $V_1$ with even and the others with odd,
because any path alternates between nodes of both subsets. Similarly, for a root node in $V_2$, all nodes of $V_2$ are labeled odd and those of $V_1$ even (see figure 4.9).

If $G = (V, E)$ has the unique label property with respect to a matching $M$ we are able to assign a unique label to each node of $V$. Suppose we have two labels, e.g. *even* and *odd*, we may denote the set of "even" nodes with $V_1$ and the set of "odd" nodes with $V_2$. It follows that $G$ is bipartite. If $G$ is not connected, we can apply the idea to each connected component and the result is the same. □

We have already seen the bipartite matching algorithm which will be useful for the computation of maximum matchings in bipartite graphs:

**Property 4.8** *If a graph $G$ has the unique label property with respect to a matching $M$, we can find a maximum cardinality matching $M'$ with the bipartite matching algorithm.*

## 4.3 Contraction Iterators

For the matching algorithm, I used a contraction coordinator (see figure 2.8) that contracts a whole list. Every contraction level corresponds to a contracted blossom in the algorithm, and the algorithm has access to a list of nodes in a current blossom, which is necessary to recompute the matching while expanding a blossom. The contractor additionally stores in a given data structure all encountered edges, which are necessary for the re-computation of the matching, as well.

### Complexity of Contraction

**Lemma 4.9** *Contraction of $k$ nodes in a graph $G = (V, E)$ needs at most time $\mathcal{O}(|V| \cdot A_{worst})$, where $A_{worst}$ is worst case size of the adjacency lists. $N$ contractions need at most time $\mathcal{O}(|V| \cdot N + n^2)$. If a union-find data structure is used, we can improve the time bounds to $\mathcal{O}(|V| \cdot \alpha(|V|, N))$. Expansion can be implemented with the same complexity as contraction ($\alpha$ is the inverse ackermann-function).*

*Proof:* In every contraction, all contracted nodes that have to be updated are visited. Since we have at most $|V| - 2$ nodes in a formerly contracted blossom that is represented by a node in the current blossom (each blossom contains at least three nodes), it is dominated by $|V|$. Then, we have to construct the adjacency list of the resulting contracted node. For each node in the blossom we scan its adjacency list for the incident edges. The method requires $\mathcal{O}(|V|)$ plus time spent in scanning nodes in the adjacency list, which sums to $\mathcal{O}(n^2)$ over all contractions, because each node is part of a blossom at most once, so the algorithm will examine its adjacency list at most once. Expansion of a blossom requires scanning the created lists in the same manner like at contraction time, such that we have the same complexity.

Using a union-find data structure as described in [T83], we are able to improve the time bounds to $\mathcal{O}(m \cdot \alpha(m, n))$, where $m$ is the number of created blocks and $n$ is the number of other operations, which are in this example $N$ contractions. Since we have $|V|$ nodes, we have to create $|V|$ blocks. To realize the same time bounds for expansion we need to maintain the former states of the entities (nodes or again blossoms) which are contained in one blossom.                                                                               □

## 4.4   Nonbipartite Matching Algorithm

It is more difficult to find a maximum matching in nonbipartite graphs, since this property is violated by odd cycles. Suppose we enter an odd cycle in a certain node $k$ while examining the graph and try to follow the cycle in both directions. Then we can label each node in the cycle with both labels, even or odd (see figure 4.8). The idea is to eliminate all of these cycles in a certain way, and to do all augmentations slightly differently.

First of all, we want to define exactly what a blossom is and how this relates to alternating paths.

**Definition 4.10** *Let $G = (V, E)$ a graph, $M$ a matching and $p$ a root node. A **flower** is a subgraph $G' \subseteq G$ with two components: (1) a **stem** — an even (length) alternating path that begins at $p$ and ends at some $w \in V$ — and (2) a **blossom** — an odd (length) alternating cycle that begins and ends at $w$ and has no other node in common with the stem. We refer to node $w$ as the **base of the blossom**.*

Figure 4.8 shows an example of a flower (the first three nodes on the left form the stem, the right part is the blossom). Here are some properties of flowers which we will record for future reference:

**Property 4.11** *Flower properties.*
*(a) a stem spans $2l + 1$ nodes and contains $l$ matched edges for some $l \in \mathcal{N}$, $l \geq 0$.*
*(b) a blossom spans $2k + 1$ nodes and contains $k$ matched edges for $k \in \mathcal{N}$, $k \geq 1$. The matched edges match all nodes of the blossom except its base.*
*(c) the base of the blossom is an even node.*

*Proof: (a)* The stem is an even length alternating path (see definition 4.10) and the number of edges $t$ is even. So the number of nodes is $t + 1$, or with $l$ for the number of matched edges, $2l + 1$.

*(b)* Suppose the blossom contains $k$ matched edges, i.e. exactly $k + 1$ unmatched edges and $2k + 1$ nodes, because it is a cycle. Since the last edge of the stem is matched, the base node $w$ cannot be matched a node in the blossom. If any other node in the blossom is unmatched, there would be no alternating path

*(c)* When we number each node in the stem, every node with an odd number will be labeled even. For example, the root node, numbered with 1 is even, the second with 2 is odd. It follows that the base of the blossom which is numbered with $2l + 1$ is even. □

The algorithm tries to label some of the nodes even and some odd. However, every node in a blossom is qualified to receive any label; we may have two different alternating paths, contradicting each other in the blossom.

We would like to shrink those blossoms into a single, even labeled node. While the base node is the only one which is matched to a node outside of the blossom, we do not have to consider the other blossom nodes anymore (until we have discovered an augmenting path — see for example the shrinking process in figure 4.11). We will use contraction iterators from section 2.3.2. Whenever we encounter a blossom, we try to identify it and shrink it to a single node. The nonbipartite matching algorithm is a modified version of the bipartite matching algorithm with contraction iterators and blossom shrinking.

The modification is as follows:

1. Whenever we examine an even node $i$ and detect an adjacent even node $j$ (in the function examine-even) then we know we found a blossom (this is the only possibility if we start to search for augmenting paths at one single node). If we trace the predecessor indices of both nodes to identify the blossom $B$ we can shrink it to the base node of it. We do the same when we examine an odd node $i$ and detect again an odd node $j$ (in the function examine-odd).

2. If we found an augmenting path we augment the contracted graph (see figure 4.11) along the predecessors, starting at the unmatched node we have discovered in the search procedure.

3. Finally, before we begin to search new augmenting paths, we unshrink all blossoms and recompute the matching in them, which is done in a special way such that we get a correct matching (see for an example figure 4.12 and for a description lemma 4.14).

We come now to an algorithmic description of the modified algorithm: Figure 4.13 shows the nonbipartite cardinality matching algorithm. It executes the search-procedure of the bipartite matching algorithm and tries to find an augmenting path. Instead of the original examine procedures of the bipartite matching algorithm, it uses modified versions (see figures 4.14 and 4.15) which are able to detect blossoms. If one of them detects a blossom, it is contracted (see figure 4.16). After an augmentation, or if no augmenting path is found, it expands all blossoms and recomputes their matchings (see figure 4.17).

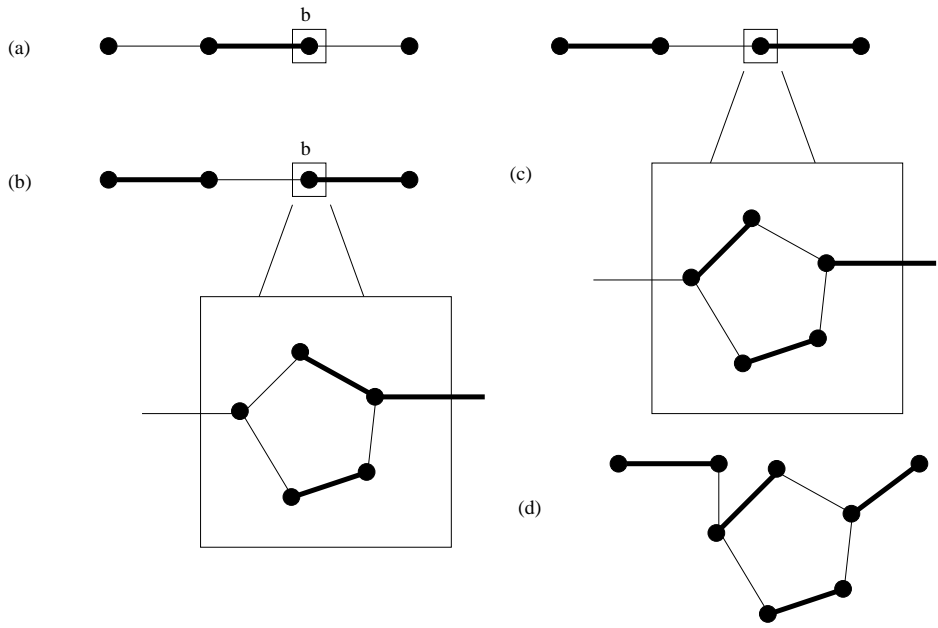

*Fig. 4.10:* This is an example for the behavior of the nonbipartite matching algorithm. In (a), we have already found an augmenting path and have already shrinked a disturbing blossom. We know how to augment this graph and get (b). Now, in a second phase, we unshrink the blossom and recompute the matching in it. Since the matching in the contracted graph (in the figure denoted with (b)) is in fact a matching, we know that only one distinct node in the blossom is matched to a node outside the blossom. We can identify this node in a doubly linked list of nodes. Both edges, incident to that node, cannot be matched in the blossom, but all others must be updated like in (c) such that we still have a correct matching (this is explained in the proof of Lemma 4.12) if we unshrink the blossom like in (d).

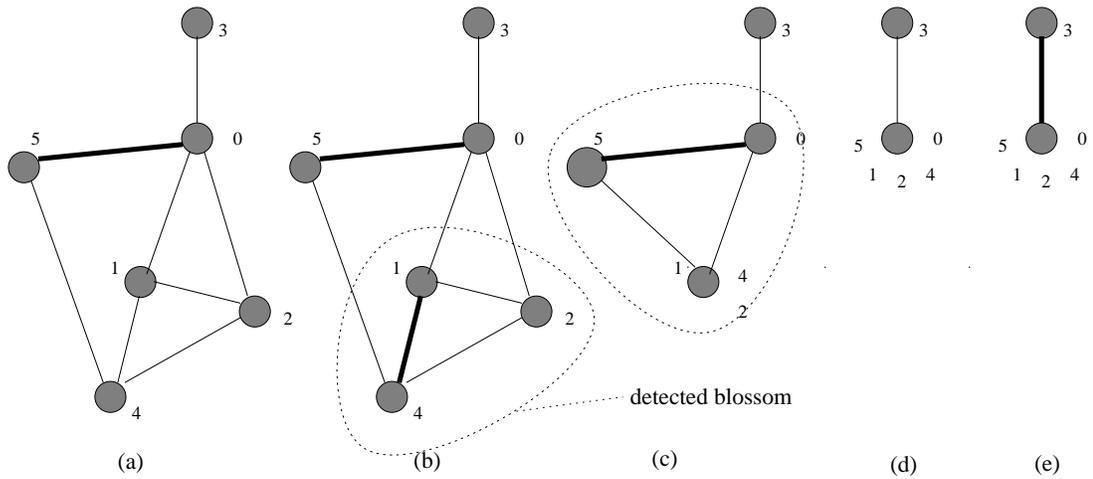

Fig. 4.11: *Example for the computation of a matching :* (a) augmenting path $(0, 5)$
(b) augmenting path $(1, 4)$ (c) blossom $2-1-4$ detected and contracted
(d) blossom $5-0-(2-1-4)$ detected and contracted (e) augmenting
path $(3, (5-0-(2-1-4)))$ found

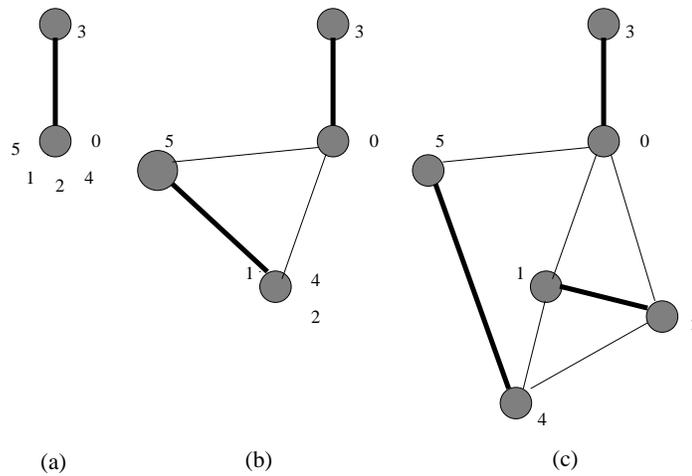

Fig. 4.12:  Example for unshrinking the blossoms: (a) matching after augmenta-
tion (b) blossom expanded to 5, 0 and $4-1-2$ and matching recomputed
(c) blossom expanded to 4, 1 and 2 and matching recomputed

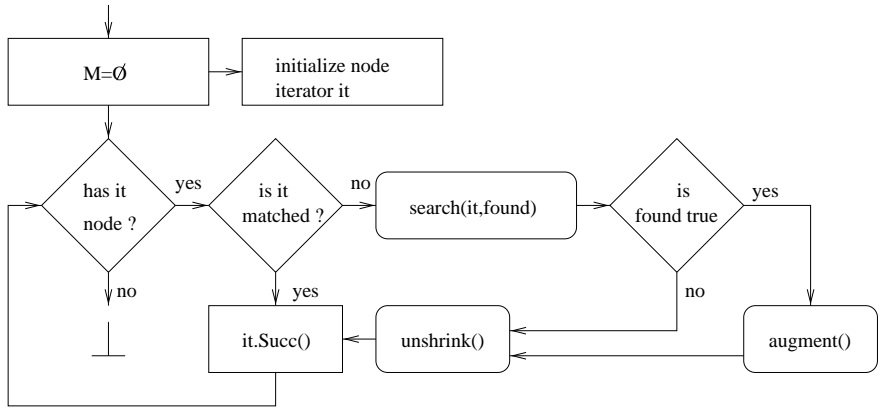

Fig. 4.13: **nonbipartite matching algorithm:** if the search algorithm discovers an augmenting path $P$, the matching $M$ will be replaced by $M \oplus P$. This is very similar to figure 4.4. After each search, all blossoms are unshrunken again.

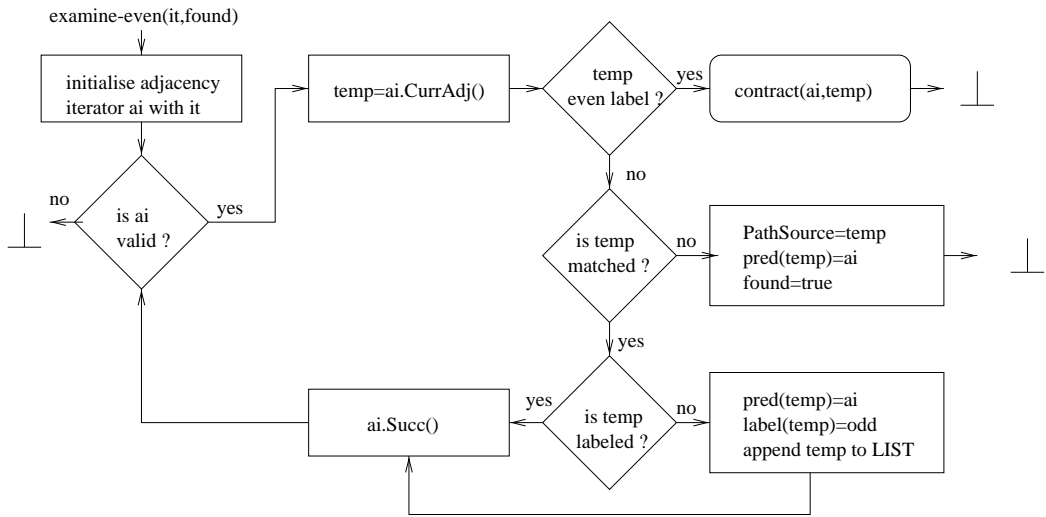

Fig. 4.14: *procedure examine-even(it,found):* if we encounter an even labeled node, we know that we have discovered a blossom. Then we call the contraction procedure that shrinks that blossom to the base node. Note that the rest of this function remains the same as in the original examine-even procedure of figure 4.6.

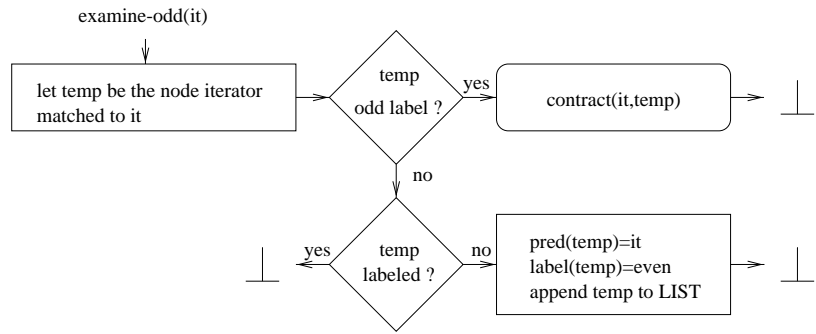

Fig. 4.15: *procedure examine-odd(it)*: if we encounter an odd labeled node, we know that we discovered a blossom. Then we call the contraction procedure that shrinks that blossom to the base node. Note that the rest of this function remains the same as in the original examine-even procedure of figure 4.7.

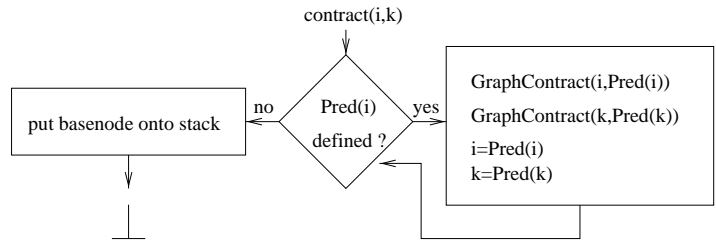

Fig. 4.16: *procedure contract(i,j)*: shrinks a blossom by tracing the predecessor entries until we reach the base node of the blossom. Each time we call a function that identifies two nodes (this is a function of the contraction iterator). Additionally, we maintain in a stack all encountered blossoms such that we can expand them later in reversed order. Since both parts of the blossom, the one path starting at $i$, the other starting at $k$ are of equal length (the search procedures grows a breadth first tree), we only have to test $pred(i)$ if it is defined.

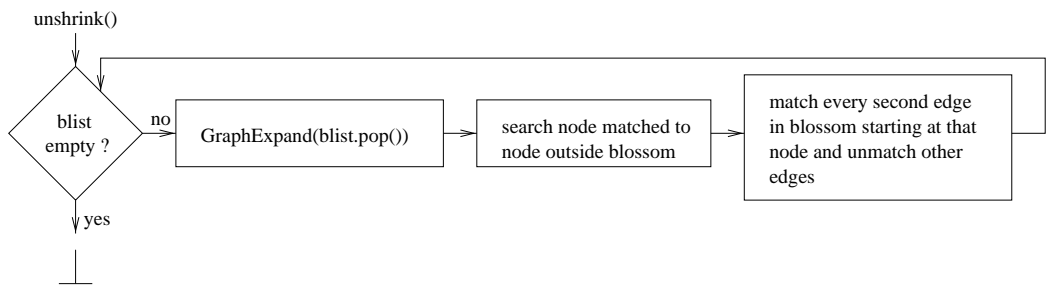

*Fig. 4.17: procedure unshrink():* we expand all blossoms in reversed order in which we created them by maintaining each of them in a stack *blist*. After we have expanded one blossom with *GraphExpand*, we recompute the matching for the subgraph defined by the blossom such that every node gets matched in the blossom, but not the node which is matched to a node outside the blossom (this one is the new base node of it). This can be done by matching every second edge (starting at the node that is matched outside the blossom) and unmatching the others.

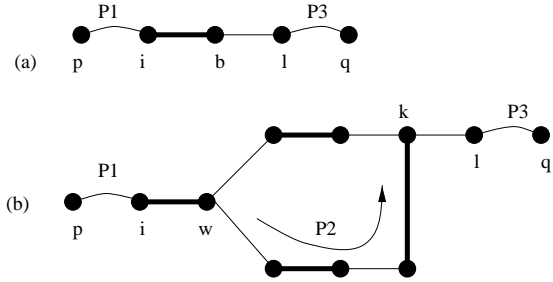

*Fig. 4.18:*  Identifying an augmenting path in the original graph: (a) augmenting
           path in contracted graph $G_C$ and (b) augmenting path in original graph
           $G$.

## 4.5   Correctness of the algorithm

We will see that the algorithm correctly finds a maximum matching. $G_C$ is
the graph that is represented by the contraction iterators, i.e. contracted
blossoms are seen as single nodes; $M_C$ is the matching according to $G_C$.

In the following, we will use character $b$ for naming blossoms that are repre-
sented by a node. If there are more blossoms, we use indices. $\hat{b}$ denotes the
nodes and edges of the blossom (only those from highest level).

First, we show that for each augmenting path in any contracted graph there
is an augmenting path in the original graph.

**Lemma 4.12** *Let $P_C$ be an augmenting path in the contracted graph $G_C$,
starting at root node $p$ with respect to the matching $M_C$. Then the original
graph $G$ contains an augmenting path starting at node $p$ with respect to the
original matching $M$.*

*Proof:* If the augmenting path $P_C$ does not contain blossom $\hat{b}$, it is also an
augmenting path in $G$ and the conclusion is valid. Next suppose that $b$ is an
interior node of $P_C$, i.e. $\hat{b}$ has a nonempty stem. In that case the augment-
ing path in $G_C$ will have the structure shown in figure 4.18(a). Recall from
property 4.11(c) that $b$ is an even node and that the alternating path from
node $p$ to $b$ ends with a matched edge. We can represent the augmenting
path in $G_C$ as $[P_1, (i, b), (b, l), P_3]$ ($P_3$ can also be empty). If we expand the
contracted blossom, we obtain for example the graph shown in figure 4.18(b).
Notice that node $l$ is incident to some node in the blossom, say node $k$. Since
every node $i$ in the blossom (except its base) is reachable from the root (or
from the base of the blossom) through two distinct alternating paths (one
with even length and one with odd length), $G$ contains an even alternat-
ing path from node $w$ (i.e., the base of the blossom) to node $k$ that ends with a matched edge. Let $P_2$ denote this path. Now observe that the path $[P1, (i, w), P_2, (k, l), P_3]$ is an augmenting path in the graph $G$. Whenever $b$ is the first node of the augmenting path, $p = w$ and the path $[P_2, (k, l), P_3]$ is an augmenting path in original graph $G$.

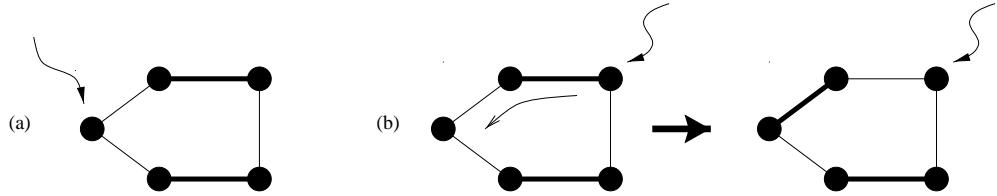

*Fig. 4.19:* Example for re-computation of the matching inside of a blossom for proving Lemma 4.13: (a) entering a blossom at the base node - nothing changes (b) entering a blossom somewhere else - the matching inside changes, the cardinality remains equal and the new base node is the entering node.

This result establishes the lemma whenever $P_C$ contains exactly one blossom. If $P_C$ contains several blossoms, we apply the lemma iteratively for all blossoms in the path. It may happen that after unshrinking one blossom several new blossoms appear, but there is a finite number of blossoms in total and the lemma is true for the general case. $\qquad\square$

The lemma shows that if we discover an augmenting path in the contracted graph, there is an augmenting path in the original graph. This lemma also shows that by contracting a blossom we do not add any augmenting paths beyond those that are contained in the original graph. It remains to show that the algorithm correctly computes a maximum matching, because it does not compute an augmenting path in the original graph.

**Lemma 4.13** *Let $\hat{b} = (V_b, E_b)$ be a contracted blossom in the contracted graph $G_C = (V_C, E_C)$ and let $M_C$ be the according matching for $G_C$ and $M_b$ the matching inside the blossom. Let $G = (V, E)$ be the graph after expansion of $\hat{b}$ and $M'$ be the according matching for $G$. Then, the algorithm expands $\hat{b}$ such that $|M'| = |M_C| + |M_b|$ and $M'$ is a matching.*

*Proof:* We consider two cases:

1. *$\hat{b}$ contains no blossoms.* Since $M_C$ is a matching, $\hat{b}$ is matched to one node $i \in V_C$. Inside $\hat{b}$, $i$ is matched in $M_C$ to one node, say $w \in V_b$. If $w$ is the base node of $b$, nothing changes (see figure 4.19(a)). If $w$ is matched in $M_b$ to another node $j \neq i$, the algorithm adapts the matching in the following way (see figure 4.19(b)): in a doubly linked list of edges (those of $E_b$) we follow the edges in direction of the matched edge $(w, j)$ and augment them until we reach the base of $\hat{b}$, i.e. $w$ is now the new base of this blossom with respect to the new matching $M_b'$. Since this is an odd path, the cardinality of this blossom remains unchanged.

Now, $|M_b| = |M_b'|$, $M' = M_C' \cup M_b'$ and $M' = M_C + M_b$. Moreover, $M'$ is a correct matching after expanding blossom $\hat{b}$.

2. $\hat{b}$ *contains* $t > 0$ *blossoms*. After expansion in case 1, the blossom has produced several blossoms, but the idea of case 1 can be applied recursively to each of them such that we still have the desired property.

The *deepest base node* of $b_1$ is inductively like this: for a single blossom $b_1$ it is the base node of $b_1$; if the base node of $b_i$ is a blossom $b_{i+1}$, it is the base node of $b_{i+1}$ ($i$ means in $b_i$ the nesting depth in the blossom hierachy).

**Lemma 4.14** *If the search procedure found an augmenting path $P_C$ in the contracted graph $G_C$ starting at root node $p$ with respect to the matching $M_C$, then the augmenting procedure will augment the original matching $M$ and yield a new matching $M'$ with the following properties:*

1. *$|M'| = |M| + 1$*

2. *all nodes that are matched in $M$ are matched in $M'$*

3. *start and end node (or deepest base node of the referring blossom, respectively) of $P_C$ are matched in $M'$*

*Proof:* In the first phase, the algorithm augments the matching $M_C$ with the augmenting path $P_C$ and yields a new matching $M_C'$. In $M_C'$, start and end node of $P_C$ are matched and all nodes that were matched in $M_C$ are still matched in $M_C'$. No other nodes are matched. It follows that $|M_C'| = |M_C| + 1$ and all nodes that were matched in $M_C$ are now matched in $M_C'$. Start and end node (or base nodes of the blossoms) of $P_C$ are matched.

In the second phase, all blossoms are expanded. For each of the blossoms $b_1, b_2, \ldots b_z$ in the augmenting path, we apply lemma 4.13 and get the matching $M' = M_C' \cup M_{b_1}' \ldots \cup M_{\hat{b}_z}'$. The number of matched edges remains equal and the nodes that were matched in $b_i$ with respect to $M_{b_i}$ remains matched in $b_i$ with respect to $M_{b_i}'$, so $M_{b_i} = M_{b_i}'$. If the start node (end node) of $P_C$ refers to a blossom, lemma 4.13 ensures that the deepest base node becomes matched.

$\square$

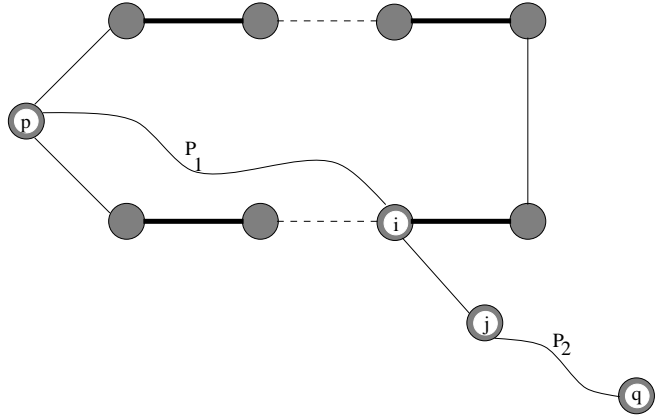

Fig. 4.20: *Proving case 1 of Lemma 4.15:* augmenting path in original graph *G*.

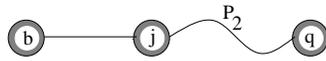

Fig. 4.21: *Proving case 1 of Lemma 4.15:* augmenting path in contracted graph *G_C*.

We now need to prove the converse result: If $G$ contains an augmenting path in $G$ from node $p$ to some node $q$ with respect to the matching $M$, then $G_C$ also contains an augmenting path from node $p$ (or the node representing the blossom which contains node $p$) to node $q$ (or the node representing $q$) with respect to the matching $M_C$. This result will show that by contracting nodes of a blossom $\hat{b}$, we do not miss any augmenting paths from the original graph.

**Lemma 4.15** *If $G$ contains an augmenting path from node $p$ to node $q$ with respect to a matching $M$, then $G_C$ contains an augmenting path from node $p$ (or the node representing the blossom which contains node $p$) to node $q$ with respect to the matching $M_C$.*

*Proof:* Suppose that $G$ contains an augmenting path $P$ from node $p$ to node $q$ with respect to a matching $M$ and that nodes $p$ and $q$ are the only unmatched nodes in $G$. We incur no loss of generality in making this assumption since nodes $p$ and $q$ are the only unmatched nodes that appear in $P$, so this path remains an augmenting path even if we delete the remaining unmatched nodes. If the path $P$ has no node in common with the nodes in the blossom $\hat{b}$, we have nothing to prove because $P$ is also an augmenting path in the contracted graph. When $P$ has some nodes in common with the blossom $\hat{b}$, we consider two cases:

1. *The blossom has an empty stem.*

   Node $p$ is the base of the blossom. Let node $i$ be the last node of the path $P$ that lies in the blossom. Path $P$ has the form $[P_1, (i, j), P_2]$ for some node $j$ and some unmatched edge $(i, j)$ (see figure 4.20).

   Note that the path $P_1$ might have some edges in common with the blossom. Now notice that $[(b, j), P_2]$ is an augmenting path in the contracted graph (while $b$ represents the blossom containing $p$) and we have established the desired conclusion (see figure 4.21).

2. *The blossom $\hat{b}$ has a nonempty stem.*

   Let $P_3$ denote the even alternating path from node $p$ to the base $w$ of the blossom and consider the matching $M' = M \oplus P_3$. In the matching $M'$, node $p$ is matched and node $w$ is unmatched. Moreover, since the matchings $M$ and $M'$ have the same cardinality, $M$ is not a maximum matching if and only if $M'$ is not a maximum matching. By assumption, $G$ contains an augmenting path with respect to $M$. Therefore, $G$ must also contain an augmenting path with respect to $M'$. But with respect to the matching $M'$, nodes $w$ and $q$ are the only unmatched nodes in $G$, so the graph must contain an augmenting path between these two nodes.

   Now let $M'_C$ denote the matching in the contracted graph $G_C$ corresponding to the matching $M'$ in the graph $G$. Note that $M_C$ might be different than $M'_C$. In $M'$, the blossom $\hat{b}$ has an empty stem and the analysis of the first case implies that the graph contains an augmenting path after we contract the nodes of the blossom. Consequently, $G_C$ contains an augmenting path with respect to the matching $M'_C$. But since $M'_C$ and $M_C$ have the same cardinality, $G_C$ must also contain an augmenting path with respect to the matching $M_C$. This conclusion completes the proof of the lemma.

   $\square$

The preceding two lemmas show that the contracted graph contains an augmenting path starting at node $p$ if and only if the original graph contains one. Therefore, by performing contractions we do not create new augmenting paths containing node $p$ nor do we miss any. As a consequence, the nonbipartite matching algorithm correctly computes a maximum matching in the graph.

## 4.6 Complexity of the Nonbipartite Matching Algorithm

**Theorem 4.16** *The nonbipartite matching algorithm identifies a maximum matching in a graph in worst case time $\mathcal{O}(n^3)$.*

*Proof:* Each search is invoked $n$ times and performs contractions and expansions in $\mathcal{O}(n^2)$ time (Lemma 4.18 and Lemma 4.19). It needs without contractions and expansions at most $\mathcal{O}(n \cdot m)$ time (Lemma 4.5). This sums to $T = n \cdot \mathcal{O}(n^2) + \mathcal{O}(n \cdot m) = \mathcal{O}(n^3)$. $\qquad\square$

**Lemma 4.17** *The search procedure performs at most $n/2$ contractions.*

*Proof:* A blossom contains at least three nodes. A contraction reduces the number of visible nodes by two or more. Since we have $n$ nodes, we can perform at most $n/2$ contractions in each search. $\qquad\square$

**Lemma 4.18** *All contractions in one search need at most time $\mathcal{O}(n^2)$ in total.*

*Proof:* Suppose we contract all nodes in the graph $G = (V, E)$ and each blossom contains 3 nodes. Only one blossom contains nodes that are not contracted. Every other blossom contains one contracted node and two uncontracted ones. The runtime sums to $T_C = \sum_{i=1}^{n/2} (i + 2) \cdot A_i$, using lemma 4.9 ($A_i$ is length of an adjacency list). Since the time spent for scanning the adjacency lists is at most $\mathcal{O}(n^2)$ (see Lemma 4.9), we have

$$
\begin{aligned}
T_C &\leq \sum_{i=1}^{n/2} i + 2 + n^2 \\
&\leq \frac{(n/2) \cdot ((n/2) - 1)}{2} + 2 + n^2 \\
&= O(n^2)
\end{aligned}
$$

$\qquad\square$

**Lemma 4.19** *All expansions need at most time $\mathcal{O}(n^2)$.*

The complexity discussion equals lemma 4.18, because expansion can be implemented as the reversed process of contraction. $\qquad\square$

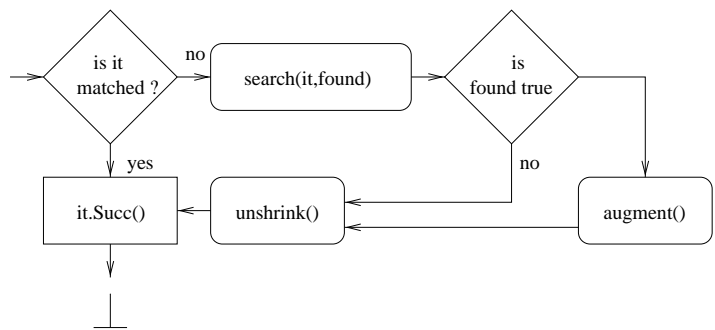

*Fig. 4.22: Nonbipartite Matching Algorithm: next step.*

## 4.7  Loop Kernel Version

When we identify the outer loop as the core loop of the algorithm, we can implement a next step in the way as described in figure 4.22. Together with the procedure definitions from the figures 4.5, 4.14 - 4.17 we may implement an algorithm class that computes a maximum cardinality matching in non-bipartite graphs, in which each next step searches a new augmenting path and if it succeeds, it applies the path to the current matching to maximize it.

This chapter demonstrated how a complex algorithm can be rewritten using iterators and data accessors while some parts of the algorithm are hidden in the requirements of iterators, i.e. contraction handling is done completely inside of the iterators.

Once contraction iterators are provided, it will be simple to implement the algorithm in a language such as `C++`.

The idea behind the design is essentially identifying a part of an algorithm that is only "helping" to create the result: here, contraction is used to "help" finding an augmenting path in a graph with respect to a current matching. Since it is not a sub-algorithm that can be separated from the matching algorithm in a procedural sense, we try to separate the "helping part" from the algorithm by "enhancing" the functionality of underlying data structures. Here, I did not try to provide an implementation that improves the runtime of previous implementations, but to show how certain parts of an algorithm realization can be hidden in an elegant way. Therefore, we develop contraction iterators that provide blossom based contraction. In the algorithm, we used these iterators, but did not assume anything about the internal realization of the contraction iterators. Thus, two goals were achieved:

   1. *flexibility:* the realization of contraction can be changed without mod-

ifying the matching algorithm itself

2. *design aspect:* algorithm and contraction are divided into independent parts, and are therefore improved in terms of object oriented design

Although there are more efficient implementations, this version of the nonbipartite matching algorithm can serve as a model, how a complex algorithm can be analyzed and implemented in a more object–oriented way.

# 5. IMPLEMENTATION IN C++

This chapter shows how the ideas of the previous chapters can be implemented efficiently (see [S91] for a reference on C++).

## 5.1  Template Feature

A **template** defines a family of types or functions. A template declaration may look like this, where `TList` is a list of template parameters, separated by a comma and each of which is e.g. a class type, introduced with the class-keyword. `declaration` defines a function, a class, or a static data member.

```
template < TList > declaration
```

This is an example of a template class `Test`, which can be instantiated with a given type.

```
template < class T >
class Test {
  T _t;
  Test(T t) : _t(t) { }
};
```

For example, we may instantiate the template class with `int` as the template parameter, and we call the constructor of that class with a given integer value 5 to construct an object $k$ of type `Test<int>`:

```
Test<int> k(5);
```

A second possibility is to use template functions, i.e. `declaration` will be a function declaration. A declaration is correct if the function declaration is correct and the list of arguments in the parameter list of the function determines in a unique way the list of template arguments. Therefore, we do not have to explicitly provide a template parameter when calling the function `print<int>(int)`. So the compiler knows implicitly the list of template arguments and is able to instantiate the function.

```
template < class T >
void print(T t) {
   /* do something */
}
```

It can be used like this:

```
int k(5);
double l(6.4);
print(k);
print(l);
```

**Member templates** are simply templates that are members. For example, the following template class `Pair` contains a template as a class member:

```
template <class A, class B> struct Pair {
  A _a;
  B _b;
  Pair(A a, B b) : _a(a) , _b(b) {}
  template <class T1,class T2> Pair(Pair<T1,T2> p) :
    _a(p._a), _b(p._b) { }
};
```

Now we can use this class to construct a pair from a second pair of different types, i.e. we construct an object of class `Pair<long,double>` from an object of class `Pair<int,float>` (conversion from `C` to `A` and `D` to `B` is assumed to be supported).

```
Pair<int,float>   a(10,10.5);
Pair<long,double> b(a);
```

Without member templates, we could only declare a constructor with fixed argument types such as `Pair(Pair<int,float> p)` or we could use the template arguments of class `Pair` itself such as `Pair(Pair<A,B> p)`. Unfortunately, most of the common compilers are not able to process member templates, but since this feature is covered by the ANSI `C++` standard, it will be supported in the future.

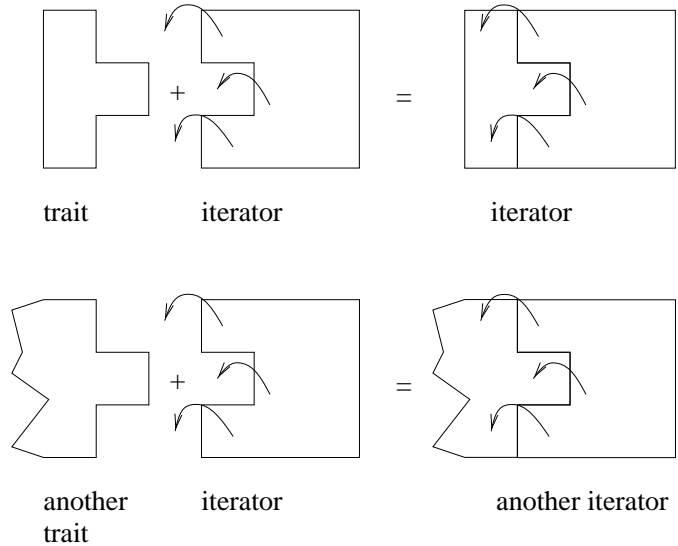

*Fig. 5.1:* The traits-technique make it possible to divide the implementation of an iterator into two parts: the first is specific for the underlying graph representation and the other is specific for the iterator. For example, suppose "iterator" is a node iterator, "trait" is a traits class for LEDA types (`node`, `graph`) and "another trait" is a traits class for XYZ types (from a different library, say XYZ). Then we have a node iterator for LEDA types and a node iterator for XYZ types without rewriting the code for iterators.

## 5.2 Traits Classes

With **traits classes** we can make a class polymorphic without introducing large performance loss. Dynamic polymorphism using inheritance normally uses runtime method binding, which is not necessary here.

A **traits class** is a class that is suitable as an "input class" for another class and contains type definitions and static methods. The parameterized class defines "by usage" a convention for the names of these types and the signatures of the static functions. When designing a traits class, one has to adhere to this convention.

Iterators are implemented using traits classes as template parameters. Iterators use only the types and static functions defined in the traits class while the traits class invokes the correct method of the underlying graph structure. If such a method is not defined by the graph structure, the traits class has to provide the desired functionality, itself. We refer to a class, which uses a traits class as a *master class.*

The following example will explain the structure of traits-classes:

First a master class is defined, which demonstrates the usage of a traits class. This class requires a static method "print":

```
template<class traits>
class TestTraits {
public:
  typedef  traits::value_type value_type;
private:
  value_type _f;
public:
  TestTraits(value_type f) : _f(f) {
    traits::print(_f); }
};
```

And now follows an example of a traits-class:

```
class traits1 {
public:
  typedef  int value_type;
  static void print(value_type f) { cout << f; }
};
```

By instantiating master class `TestTraits` with class `traits1`, we can give class `TestTraits` functionality:

```
TestTraits<traits1> K(5);
```

This prints "5" to the standard output[23].

To change class `TestTraits` functionality we can write a different class `traits2`:

```
class traits2 {
public:
  typedef  int value_type;
  static void print(value_type f) {
    while(f>0) cout << f--; }
};
```

---

[23] A short explanation: `TestTraits<traits1>` is a type produced by instantiating class `TestTraits<>` with `traits1`. `K(5)` will be an object of this type, initialized with 5, i.e. the constructor of that class will be invoked with 5.

This produces "54321" and is an example for exchanging the functionality in an elegant way without violating efficiency because the function `print` may be bound in both cases statically by the compiler.

Traits classes must always have a defined structure and functionality to avoid compile and runtime errors. To do this we discuss two ideas:

- *Syntactic Requirements:* today, there is no language support in `C++` for template requirements. Therefore, we must specify all requirements in the documentation of the code.

  E.g., let `A` and `B` be two master classes and `C` be traits classes fitting into the requirements of class `A`. Suppose that `B` has the requirements of `A` and an additional one that `C` does not meet. Then, `C` will only fit for `A` but not for `B`. The compiler will not know if `C` fits for `B` before using `C` as a template for `B`. The compiler reports an error-message if `B` uses something not present in `C`. It will not report that requirements are violated and might make debugging more difficult. It gets problematic, if the current implementation of `B` only use the requirements of `A`, because there is no such error-message although the requirements are not met by `C`. With a requirement documentation, we are able to check the list of requirements at programming time to avoid these situations.

- *Semantic Requirements:* the master class is correct if at least the traits classes are implemented correctly according to the requirement specification. Therefore, the documentation must describe the behaviour, and the pre- and postconditions of the static methods. Additionally, the requirements for the type definitions must be listed.

## 5.3   Iterators

The iterators can be implemented as concrete classes, for example like the following:

```
class node_iterator {
  node pointer;
  graph _G;
public:
  node_iterator(graph G, node n) : pointer(n), _G(G) { }
  node_iterator operator++() {
    pointer=_G.succ_node(pointer);
    return *this; }
  bool valid() { return pointer==nil; }
```

```
        node get_node() { return pointer; }
    };
```

In this example, `creation()` is represented by `node_iterator(graph G, node n)` and `succ()` by `operator++()`. Unfortunately, the iterator class is *only* suitable for graphs of type `graph` and nodes of type `node`. Moreover, iterators with adapted functionality (for example iterators that see only a subset of nodes) would have to be re-implemented.

These iterators may be implemented independently of the underlying data representation using the traits technique or inheritance (provided for example by `C++` and Java).

Now we will see how adjacency iterators for outgoing nodes are implemented in `C++` using the traits class mechanism. At first a part of the traits-class[24] already specialized for the use with LEDA:

```
    struct OutNodeEdgeTrait {
        typedef edge        T_Edge;
        typedef node        T_Node;
        typedef graph       T_Graph;
        static T_Node node_null()   { return nil; }
        static bool is_node_null(T_Node v) { return v==nil; }
        static void forward(T_Graph G, T_Node&, T_Edge& e) {
          e=G.adj_succ(e); }
        static T_Node first_node(T_Graph G) {
          return G.first_node(); }
        static T_Edge first_incident_edge(T_Graph G, T_Node v) {
          if (is_node_null(v)) return edge_null();
          return G.first_adj_edge(v); }
        static void curr_adj(T_Node& n, T_Edge& e, T_Graph G) {
          if (is_edge_null(e)) return;
          n=target(e);
          e=G.first_adj_edge(n); }
    ...
    };
```

There are several type definitions, which provide uniform access to the node, edge or graph type. Then there is a list of static methods, which perform basic graph operations. For example, `first_node()` gives the first node in the set of (ordered) nodes of the given graph $G$. If nodes are not pointers, there must be an alternative implementation for testing if the current node is valid or not. This is done with the functions `node_null()` and `is_node_null(v)`, where the latter tests if $v$ is invalid. The first incident edge in the list of incident edges will be returned by function `first_incident_edge(G,v)`. `curr_adj(n,e,G)` computes the successor adjacency iterator in direction of the current adjacent node.

---

[24] The prefix `T_` in `T_Node`, `T_Edge` and `T_Graph` means "this is a traits type". In the following only constant references to the underlying graph structure are exchanged, but not the complete graph (i.e. in the code, we use `T_Graph` instead of `const T_Graph&`).

The traits class is the only bridge to the underlying graph representation. If one has an alternative representation, only this bridge has to be modified. Currently there are 19 functions that have to be implemented in a traits class to provide the correct interface for bi-directional iterators, i.e. successor and predecessor iterators can be computed. If an algorithm only requires forward iterators and there are no graph updates, only 13 functions have to be implemented.

`C++` provides no method to assure the correct use of types and functions in traits classes that are used as template parameter classes. The only restriction lies in the use of correct names. If a static function or type definition is never used, it cannot be guaranteed that all required functions exist. This is especially necessary for libraries since it should be expected that any detail could be used.

In practice, this means that traits-based iterators only work correctly as long as they use traits classes that are correctly implemented according to the traits specification of the iterator class.

Since we now have a traits class, we want to write concrete iterators. In method `operator++()`, the static function `forward` of the traits-class is invoked. Other method implementations follow in the same style. The result are iterators that are independent of concrete graph representations.

```
template<class traits>
class AdjIt {
public:
   typedef traits::T_Edge        edgetype;
   typedef traits::T_Node        nodetype;
   typedef traits::T_Graph       graphtype;
private:
   nodetype                           _n;
   edgetype                           _e;
   graphtype&                         _G;
   typedef AdjIt<traits> self;
public:
```

```
AdjIt(traits::T_Graph& G) : _G(G) ,_n(traits::first_node(G)) {
  traits::assign(_e,traits::first_incident_edge(G,_n)); }
void update(traits::T_Node n) {
  traits::assign(_n,n);
  traits::assign(_e,traits::first_incident_edge(_G,_n)); }
bool has_node() const{ return !traits::is_node_null(_n); }
bool valid() const { return !traits::is_edge_null(_e); }
traits::T_Edge  get_edge() const { return _e; }
traits::T_Node  get_node() const { return _n; }
self curr_adj() const {
  self temp(*this);
  traits::curr_adj(temp._n, temp._e, temp._G);
  return temp; }
self& operator++ () {
  traits::forward(_G,_n,_e);
  return *this; }
 ...
};
```

## 5.4   Data Accessors

Data accessors are implemented as instances of a data accessor class for
which there are two templatized functions with standardized signature, which
controls the data access:

- the value associated to an item, which an iterator refers to can be
  retrieved with the following function template: `T get(DA da, Iter it)`, where *da* is the current data accessor and *it* the iterator.

- the value associated to an item, which an iterator refers to might be
  changed if it makes sense with the following function template `void set(DA da, Iter it, T value)` where *da* and *it* are the current data
  accessor and iterator.

Several data accessors use a so–called object accessor (see language indepen-
dent class definition 2.29) to access the item that the iterator refers to (for
an example see language independent class definition 2.32).
An object accessor is similar to a data accessor in that it gives flexible access
to data, but it accesses data in a less abstract way. Informally, it assumes
that the given value has certain member functions that return the right value.

More specifically, an object accessor of type `OA` is required to come with a function `item get_object(OA oa, Iter it);` where `item` is the item that is appropriate e.g. for the handler object of a handler accessor (see table 5.1). Since we are using iterators for graphs, we are only interested in item types like node or edge and therefore we require for iterators, which refer to nodes to have the type definition `Iter::nodetype` and the member function `Iter::nodetype Iter::get_node() const;`. Consequently, iterators that refer to edges need to have the type definition `Iter::edgetype` and member function `Iter::edgetype Iter::get_edge() const;`.

For example, adjacency iterators must provide both possibilities because they always refer to a fixed node and an incident edge.

All implemented graph iterators[25] provide the necessary member functions, so we can develop two object accessors, a node and an edge accessor. With member template capable compilers, we might write this:

```
class NodeAccessor{
  template<class Iter>
  node get_object(Iter it) {
    return it.get_node(); }
};
class EdgeAccessor{
  template<class Iter>
  node get_object(Iter it) {
    return it.get_edge(); }
};
```

Unfortunately, member templates are not provided by recent compilers[26] and we have to write a workaround such that we can use these object accessors with all compilers:

```
struct NodeAccessor { } nodeacc;
  // object accessor of type NodeAccessor
struct EdgeAccessor { } edgeacc;
  // object accessor of type EdgeAccessor

template<class Iter>
```

---

[25] See graphiterator LEP or the graphiterator part in LEDA.

[26] Several compiler manufacturers try to implement member templates, but this seems to be very complicated. Only few compilers are able to this, for example `KCC` from company Kai Associates. The maybe most popular compiler, `gcc` is not able to handle member templates (version 2.7.2).

```
node get_object(NodeAccessor, Iter it) {
  return it.get_node(); }

template<class Iter>
edge get_object(EdgeAccessor, Iter it) {'
  return it.get_edge(); }
```

A possible implementation is a handler accessor that hides the internal use
of handler objects that provide `operator[](node)` or `operator[](edge)`:

```
template<class T, class Handler, class StructAccessor>
class HandlerAccessor {
public:
  typedef T value_type;
private:
  Handler&  i_handler;
  StructAccessor i_sa;
public:
  HandlerAccessor(Handler& handler, StructAccessor sa)
    : i_handler(handler), i_sa(sa) { }
  StructAccessor internal_sa()    { return i_sa; }
  Handler& internal_map()  { return i_handler; }
};

template<class Iter, class Handler, class StructAccessor, class T>
T get(HandlerAccessor<T,Handler, StructAccessor> da,
  Iter it)  {
  return da.internal_map()[get_object(da.internal_sa(),it)]; }

template<class Iter, class Handler, class StructAccessor, class T>
void set(HandlerAccessor<T,Handler, StructAccessor>  da,
  Iter it, T value)  {
  da.internal_map()[get_object(da.internal_sa(),it)]=value; }
```

A special case of object accessor is a *struct accessor*. Often different node
(or edge) parameters are organized as one struct. The method get_object
of a struct accessor returns this struct and allows data accessors to access
the individual members of the struct.
Now follows the simpler member template version of this class:

```
template<class T, class Handler, class StructAccessor>
```

```
class HandlerAccessor {
public:
  typedef T value_type;
private:
  Handler&  i_handler;
  StructAccessor i_sa;
public:
  HandlerAccessor(Handler& handler, StructAccessor sa)
    : i_handler(handler), i_sa(sa) { }

  template<class Iter>
  T get(Iter it) {
      return i_handler[get_object(i_sa,it)]; }

  template<class Iter>
  void set(Iter it, T value) {
      i_handler[get_object(i_sa,it)]=value; }
};
```

In the following we will assume that for data accessors, appropriate wrapper classes were defined, which add a top and bottom value to the set of member variables. For example, a data accessor that refers to a boolean type attribute may have `true` and `false` as top and bottom value. A data accessor that refers to an integer type attribute may have `MAX_INT` and 0 as top and bottom value.

A possible wrapper class may look like this:

```
template <class DA>
class TopBot_Integer {
public:
  typedef DA::value_type value_type;
  value_type value_null, value_max;
private:
  DA&  i_da;
  TopBot_Integer(DA& da) : i_da(da), value_null(0),
    value_max(MAX_INT) { }
  DA& internal_da() { return i_da; }
};

template<class DA>
DA::value_type get(TopBot_Integer<DA> da, Iter it) {
  return get(da.internal_da(),it); }
```

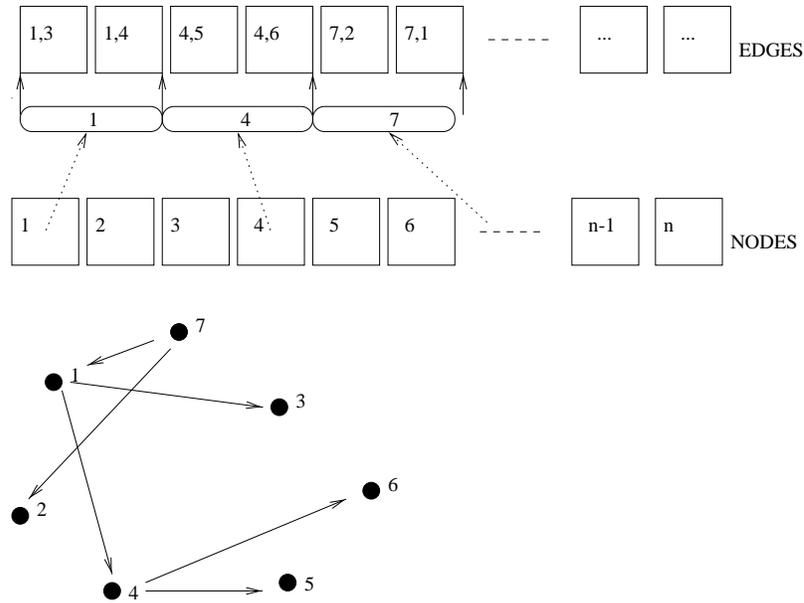

*Fig. 5.2:*   The alternative graph structure **fgraph** stores nodes and edges in a
            different way than **graph** (from LEDA) and is optimized in space re-
            quirements and access to the list of outgoing edges for each node. Each
            node knows the interval in the array of edges with incident edges and the
            edges are grouped according to the source node.

For example, if a Dijkstra algorithm uses a data accessor for distance values,
it uses both member values of the data accessor **value_null** and **value_max**.

Table 5.1 shows several types of data accessors, where the basic requirements
of a data accessor are explained.

## 5.5   Adaption to Other Graph Representations

Suppose we know at a certain stage of program execution that a data struc-
ture has a certain stable feature, e.g. the graph is constant. Then we might
convert the LEDA graph structure into a faster constant version on which
we run the same algorithms. With the generic algorithms presented here and
the possibility of adapting traits classes the effort is very little.

### 5.5.1   Fast Graph Data Structure

The traits class (incomplete in the sense that functions with void function-
ality are skipped) for adjacency iterators for outgoing edges and a special

| description | definition | value access |
|---|---|---|
| `DataAccessor` | • `da.T` is a type | $da.$**Get**$(it)$ returns the associated value for the iterator $it$ which is of type `da.T` and $da.$**Set**$(it, val)$ sets the associated value to $val$ |
| `ObjectAccessor` | • `da` is a `DataAccessor`<br>• `da.object` is a pointer to member function of class of $it$ | $da.$**Get**$(it)$ returns $it.(*da.object)()$ |
| `HandlerAccessor` | • `da` is a `DataAccessor`<br>• `da.oa` is an `ObjectAccessor`<br>• `da.handler` is a handler object | $da.$**Get**$(it)$ returns $da.handler(da.oa(it))$ and $da.$**Set**$(it, val)$ sets $da.handler(da.oa(it))$ to $val$ |
| `MemberAccessor` | • `da` is a `DataAccessor`<br>• `da.oa` is an `ObjectAccessor`<br>• `da.ptr` is a pointer to member value<br>• `da.oa[it]` yields a structure | $da.$**Get**$(it)$ returns $da.oa(it).(*da.ptr)$ and $da.$**Set**$(it, val)$ sets $da.oa(it).(*da.ptr)$ to $val$ |
| `MethodAccessor` | • `da` is a `DataAccessor`<br>• `da.oa` is an `ObjectAccessor`<br>• `da.ptr` is a pointer to member function<br>• `da.oa[it]` yields a structure | $da.$**Get**$(it)$ returns $da.oa(it).(*da.ptr)()$ |
| `ConstantAccessor` | • `da` is a `DataAccessor`<br>• `da.ca` is a value of type `da.T` | $da.$**Get**$(it)$ returns $da.ca$ |
| `CalcAccessor` | • `da` is a `DataAccessor`<br>• `da.s`, `da.t` are `DataAccessors`<br>• `da.calc` is a calculator object: there has to be a function `calculate( da.calc, result, da.s, da.t )` where `result` is of type `da.T` | $da.$**Get**$(it)$ returns the result of **calculate**$(da.calc, result, da.s, da.t)$ |

*Tab. 5.1:* these are the different types of data accessors; $da$ ($it$) represents a current data accessor (iterator)

graph representation `fgraph`:

```
struct fNodeEdgeTrait {
  typedef fedge* T_Edge;
  typedef fnode* T_Node;
  typedef fgraph T_Graph;
  static T_Node node_null()  { return nil; }
  static T_Edge edge_null() { return nil; }
  static bool is_node_null(T_Node v) { return v==nil; }
  static bool is_edge_null(T_Edge e) { return e==nil; }
  static void forward(T_Graph G, T_Node&, T_Edge& e) {
    e=G.succ_adj(e); }
  static T_Edge first_incident_edge(T_Graph G, T_Node v) {
    return G.first_adj_edge(v); }
  static void assign(T_Node& n1, T_Node n2) { n1=n2; }
  static void assign(T_Edge& e1, T_Edge e2) { e1=e2; }
  static bool is_equal(T_Edge e1, T_Edge e2) { return e1==e2; }
  static bool is_equal(T_Node n1, T_Node n2) {  return n1==n2;   }
  static void curr_adj(T_Node& n, T_Edge& e, T_Graph G) {
    if (is_edge_null(e)) return;
    n=e->target();
    e=G.first_adj_edge(n); }
  static T_Node first_node(T_Graph G) {
    return G.first_node(); }
};
```

After converting a LEDA graph to this fast graph structure, shortest path
runs much faster (see figure 5.3). In practice, it might happen that a very
large and inefficient graph structure is (implicitly) given - then conversion
can be too time consuming (for instance, if the algorithm runs faster than
the conversion would be). If possible, adapting the traits classes is a better
solution and may result in better performance compared to conversion.

### 5.5.2   Implicit Definition: Complete Graphs

Another possibility is that a mathematical description of the graph is avail-
able but the explicit realization of it as a LEDA graph with nodes and edges
would be too space consuming. The solution is again modifying the traits
classes.

A complete graph needs a lot of memory (space $\mathcal{O}(n+m) = \mathcal{O}(n^2)$, $n$ number
of nodes, $m$ number of edges). It is possible to reduce the space bounds for complete graphs to linear size since we only need the nodes as materialized entities.

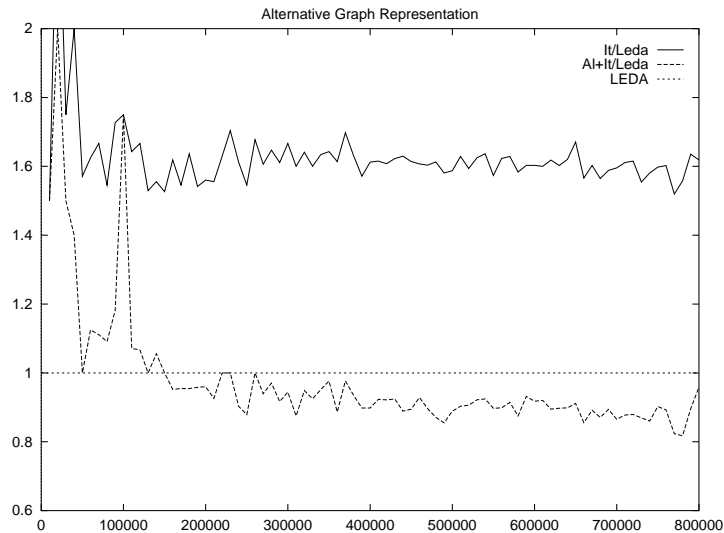

*Fig. 5.3:* Runtime comparison of different versions of Dijkstra (LEDA and iterator version and the alternative version with fast graphs ) on randomly generated graphs with 1.000 nodes and edges from 10.000 to 800.000). The graphs are generated by choosing arbitrarily three nodes and connecting them with edges until the number of edges is reached; no parallel edges are generated.

### Traits Class

This is the appropriate traits-class [27]:

```
class MathNodeEdgeTrait {
  typedef pair<int,int> T_Edge;
  typedef int T_Node;
  typedef int T_Graph;
public:
  static T_Node node_null() { return -1; }
  static bool is_node_null(const T_Node& v) { return v==-1; }
  static T_Edge edge_null() {
    return make_pair(-1,-1); }
```

---

[27] This traits class is used in an example program that computes shortest paths for complete LEDA graphs, where we use the reduced representation.

```
static bool is_edge_null(const T_Edge& v) {
  return v.second==-1; }
static T_Edge new_edge(T_Graph& G, T_Node v1, T_Node v2) {
  return make_pair(-1,-1); }
static void del_edge(T_Graph& G, T_Edge e) { }
static void forward(T_Graph G, T_Node&, T_Edge& e) {
  if (e.first==-1) return;
  if (e.second<0 || G-1==e.second) e.second=-1;
  else ++e.second; }
static void backward(T_Graph G, T_Node&, T_Edge& e) {
  if (e.first==-1) return;
  if (e.second<0) e.second=-1;
  else --e.second; }
static T_Edge first_incident_edge(T_Graph G, T_Node v) {
  return make_pair(v,0); }
static void assign(T_Node& n1, const T_Node& n2) { n1=n2; }
static void assign(T_Edge& e1, const T_Edge& e2) { e1=e2; }
static bool is_equal(const T_Edge& e1, const T_Edge& e2) {
  return ( e1==e2 || (e1.second==e2.second && e1.second==-1)); }
static void curr_adj(T_Node& n, T_Edge& e, T_Graph G) {
  if (is_edge_null(e)) return;
  n=e.second;
  e=make_pair(n,0); }
static T_Node first_node(T_Graph G) { return 0; }
static T_Node last_node(T_Graph G) { return G-1; }
static T_Edge first_edge(T_Graph G) {
  return make_pair(0,0); }
static T_Edge last_edge(T_Graph G) {
  return make_pair(G-1,G-1); }
static T_Edge last_incident_edge(T_Graph G, T_Node v) {
  return make_pair(v,G-1); }
static bool is_equal(T_Node n1, T_Node n2) {
  return n1==n2;    }
};
```

## Data Handling

If we cannot assume anything about the constellation of data, we have to
store them explicitly into a container (a list, an array or indexed table in a
database; space $\mathcal{O}(n)$ for nodes or $\mathcal{O}(m) = \mathcal{O}(n^2)$ for edges).
Sometimes, values can be computed online from other values. If, for example,

edge lengths can be computed from node locations, we can improve the overall space bounds to $\mathcal{O}(n)$, since edge lengths will take constant size.

There is a special data accessor (see table 5.1) that is able to calculate values out of some values associated to the source and target nodes of an edge: `CalcAccessor`. Unfortunately, the use is not easy, but not as difficult as rewriting a specialized version of all possible algorithms for on-line calculations of attributes.

To compute the length of an edge the program calculates the absolute value of the difference of the indices associated to the nodes (the formula to compute the length is only exemplary).

**concrete realization:** we will need the following ($ai$ is an adjacency iterator):

1. two appropriate object accessors

2. $ai$ refers to node $v$ and edge $e$: we need two data accessors that return the location of the source and the target node of $e$

3. a function that calculates the formula

4. the edge length calculation data accessor

(1.) Since we have designed new traits-classes, we also must define appropriate object accessors. If we use the same technique as described in section 5.4 for object accessors, we can write the following:

```
struct MathNodeAcc { };
struct MathEdgeAcc { };

template<class Iter>
int get_object(MathNodeAcc, Iter it) {
  return it.get_node(); }
template<class Iter>
int get_object(MathEdgeAcc, Iter it) {
  return it.get_edge(); }

MathNodeAcc mathgraph_node;
MathEdgeAcc mathgraph_edge;
```

(2.) Then we need two data accessors that return the location of the source and the target node: `SourceAcc` and `TargetAcc`. Let $ai$ be an adjacency iterator with fixed node $v$ and incident edge $e$. `SourceAcc` accesses $v$ and returns the index of that node. `TargetAcc` accesses $e$ and returns the index of the target node. This will be used for the exemplary formula.

```
struct transfer_s {
  int operator[](int k) const {
    return k; }
};
struct transfer_t {
  int operator[](pair<int,int> P) const {
    return P.second; }
};
transfer_s ts;
transfer_t tt;
typedef HandlerAccessor<int, transfer_s, MathNodeAcc>
  SourceAcc;
typedef HandlerAccessor<int, transfer_t, MathEdgeAcc>
  TargetAcc;
```

(3.) The correct calculation function must be selected. To ensure this at compile-time, we use the technique from above: create a type with void functionality and use it for overloading a function. An object of that type will be stored in the data accessor class and assures that the compiler will use the following function:

```
struct LengthCalc { } lengthcalc;
void calculate(LengthCalc, int& result, int s, int t) {
  result=abs(s-t); }
```

(4.) Now we can create the data accessor that computes the length of an edge online.

```
SourceAcc SA(ts,mathgraph_node);
TargetAcc TA(tt,mathgraph_edge);
typedef CalcAccessor<int,SourceAcc,TargetAcc,LengthCalc> Length;
Length length_da(SA,TA,lengthcalc);
```

With this complex data accessor we reduced the space bounds from $\mathcal{O}(n+m)$ to $\mathcal{O}(n)$. In complete graphs this means $\mathcal{O}(n^2)$ to $\mathcal{O}(n)$. The other attributes needed for Dijkstra, distance and maybe the predecessor entry, are stored in $\mathcal{O}(n)$ space.

In summary, Dijkstra can be implemented for complete graphs with time bounds $\mathcal{O}(m \cdot T(n,m) + n \cdot T'(n))$ where $T(n,m)$ is the complexity of a single length calculation and $T'(n)$ the (possibly amortized) complexity of a single queue update (for the priority queues in LEDA, $T'(n) = \mathcal{O}(\log n)$). The space requirements are $\mathcal{O}(n)$, if we can represent the length of edges as describes above.

## 5.6   Algorithm Classes

Algorithm classes are implemented in `C++` straightforward to their LIC definition (LICD: section 2.3, algorithms: chapter 3): algorithms as objects and the ability to "advance" an algorithm will be implemented by writing a class, which knows a certain state that can be advanced by applying a method with name "next()" to an object of this class.

```
class Algorithm {
  int state, endstate;
public:
  Algorithm(int max) : endstate(max), state(0) { }
  void next() { state++; }
  bool finished() { return state>=endstate; }
};
```

With this class `Algorithm` we can easily instantiate an algorithm object:

```
Algorithm alg(5);
while (!alg.finished()) alg.next();
```

This small piece of code creates an algorithm object and invokes "next()" until it has reached an end state.

An advantage of this design is that we can write basic algorithms, which can be used in a standardized way and if needed, inspection of internal states and variables can be provided without writing complex code. Additionally, it makes it possible to write persistent algorithms, if the member variables are persistent.

Actually, those algorithms are quite more flexible than ordinary written algorithm functions:

```
template<class Alg>
class OutputAlg {
  Alg alg;
public:
  OutputAlg(int m) : alg(m) {
    cout << "max state: " << m << endl; }
  void next() {
    cout << "old state: " << alg.state;
    alg.next();
    cout << " new state: " << alg.state << endl; }
  bool finished() { return alg.finished(); }
};
```

This wrapper algorithm can be used like this:

```
OutputAlg<Algorithm> alg(5);
while (!alg.finished()) alg.next();
```

In addition to the algorithm mentioned earlier this wrapper writes the internal states to the standard output.

This is as efficient as rewriting the "`Algorithm`"-class with an output mechanism, but provides more flexibility.

## 5.7   Algorithms in `C++`

Now follows the next step implementations of the five basic algorithms breadth first search, depth first search, toplogical sorting, computation of the strongly connected components and dijkstra shortest path computation. The concrete classes can be found in the actual implementation, which is the Graphiterator Leda Extension Package [GitLep]. All pre- and post-processing operations such as initialization of data structures are missing. In this section we show that the figures  3.7,  3.5,  3.6,  3.12,  3.13,  3.8,  3.9 and  3.10, which describe the algorithms can be implemented in a short way, and is simple as well (if you are used to iterators and data accessors). In each of the following sub-sections, the source code for the `C++`-implementation of the according next step will be shown, and a short explanation tries to connect the source code with the pictures (in figures  3.5, 3.6, 3.7, 3.8, 3.9, 3.13 and  3.10).

### 5.7.1   Breadth First Search

In the class `GIT_BFS` (see [GitLep]), the following member variables are maintained: a queue $q$, parameterized with adjacency iterators and a data accessor $ma$, which holds a boolean attribute. In each step, we look at the adjacency iterator from the top of the queue and test each adjacent node if we have already seen it.

In the positive case, we append the new adjacency iterator to the queue. In each next step we will see every node of the graph in breadth first order, beginning at a certain node[28] (while the queue was initialized in the beginning with the certain node).

---

[28] see also figure  3.7

```
AdjIt ai(q.pop());
while (ai.valid()) {
  AdjIt temp=ai.curr_adj();
  if (get(ma,temp)!=true) {
      set(ma,temp,true);
      q.append(temp); }
  ++ai;
}
```

### 5.7.2  Depth First Search

The simple version of depth first search is basically identical with breadth first search, but differs in that it uses a stack *st* instead of a queue (see also figure 3.5):

```
AdjIt ai(st.pop());
while (ai.valid()) {
  AdjIt temp=ai.curr_adj();
  if (get(ma,temp)!=true) {
      set(ma,temp,true);
      st.push(temp); }
  ++ai;
}
```

The more complex version of dfs tries to do the nested loop of the code example above in the main loop. The size of the stack will only decrease if we have explored the complete neighborhood of the fixed node of the current adjacency iterator on top of the stack (see also figure 3.6):

```
AdjIt ai(st.top());
if (ai.valid()) {
  AdjIt temp=ai.curr_adj();
  if (!get(ma,temp)) {
    st.push(temp);
    set(ma,temp,true);
    return dfs_grow_depth;
  } else {
    ++ai;
    st.pop();
    if (ai.valid()) {
      st.push(ai);
      return dfs_grow_breadth; }
    else return dfs_shrink;   }
}
else {
  st.pop();
  return dfs_leaf; }
```

### 5.7.3  Dijkstra

Dijkstra is rather complex compared to the previous algorithms.

Here, we maintain a variable `current` that holds the current adjacency iterator, since we process in each step one incident edge of a fixed node. Suppose, the current adjacency iterator is valid, i.e. it refers to an incident edge of the fixed node, we can move to the next incident edge and test if we can improve the distance value (which is here of type `disttype`) of the current adjacent node. In the other case, if the current adjacency iterator is invalid, we try get a new iterator from the priority queue and test again if we can improve the distance value (see also figure 3.13):

```
        if (!current.valid()) {
          current=pq.find_minimum();
          pq.delete_minimum();
          set(QI,current,QI.value_null);
          if (current.valid()) {
            valid=true;
            AdjIt current_adj=current.curr_adj();
            disttype c(get(distance,current)+get(length,current));
            disttype d(get(distance,curr_adj));
            if (c<d) {
              if (d<distance.value_max)
                pq.decrease_priority(get(QI,current_adj),c);
              else
                set(QI,current_adj,pq.insert(c,current_adj));
              set(distance, current_adj, c);
            }
          }
        } else {
          ++current;
          if (current.valid()) {
            AdjIt current_adj=current.curr_adj();
            disttype c(get(distance,current)+get(length,current));
            disttype d(get(distance,curr_adj));
            if (c<d) {
              if (d<distance.value_max)
                pq.decrease_priority(get(QI,curr_adj),c);
              else
                set(QI,current_adj,pq.insert(c,curr_adj));
              set(distance, current_adj, c);
            }
          }
        }
```

### 5.7.4 Topological Sorting

Here, the algorithm maintains a queue of iterators referring to nodes with zero in-degree, i.e. nodes that have no incoming edges. In each next step, we get one iterator from the stack and think of it as deleting the node from the graph, which is what we actually do not do. Instead, we decrease the number of incoming edges in each outgoing edge, which is stored in the data accessor *indeg*. For that purpose, the algorithms maintains an attribute for each node that stores the number of nodes that has zero in-degree in the current virtual graph (i.e. $get(indeg, it) == 0$).

If all nodes are processed, we saw all nodes in topological order outside of the next step.

```
OutAdjIt ai=zl.pop();
while (ai.valid()) {
  OutAdjIt temp=ai.curr_adj();
  indegreetype tempval=get(indeg,temp)-1;
  set(indeg,temp,tempval);
  if (tempval == indeg.value_null)
    zl.append(temp);
  ++ai;
}
```

### 5.7.5  Strongly Connected Components

The class that computes strongly connected components maintains two depth first search algorithm objects (`dfs1` and `dfs2`). Both share the same data accessor for marking nodes (`ma`), and there are two adjacency iterators, one for `dfs1` that traverses the outgoing edges of a fixed node (`oai`) and one for `dfs2` that traverses the incoming edges of a fixed node (`iai`). Additionally we have a counter variable for components that is increased every time we see a new component.

During the first phase, we compute a depth first forest according to the outgoing edges of a fixed node. Suppose the algorithm `dfs1` is not finished yet, then `oai` refers to the current adjacency iterator. Every time we leave a node in the depth first algorithm, we put it onto the node stack *LeafStack*, since we can see later on the nodes, which were last left only by processing the stack. If we have completed a depth first search, the global node iterator `it` will move to the successor node that has not been seen before. Then, there are two cases:

1. *there is a successor node:* we then can initialize a new depth first search

2. *there is no successor node anymore:* then we have seen all possible nodes and we enter the second phase. Before we continue, we refresh data accessor `mark` because we want to reuse it in the second phase. The adjacency iterator for incoming nodes will now refer to a node from the node stack built during the first phase. Since we explore for the first time a new component, we initialize the component number to 0.

The depth first search algorithm for incoming edges will be used in the following, because the transposed graph has to be inspected and it will be initialized with an iterator that refers to the current node from the node stack *LeafStack*. Then it returns with `NEXT2`, because in the next stage, the next step will enter the second phase.

```
if (dfs1.finished()) {
   it++;
   while (it.valid() && get(ma,it)) it++;
   if (it.valid()) {
     oai.update(it);
     if (!get(ma,oai)) dfs1.init(oai); }
   else {
     it.reset();
     while (it.valid()) set(ma,it++,false);
     iai.update(V.pop());
     if (!get(ma,iai)) {
        component=0;
        dfs2.init(iai); }
     return NEXT2;
   }
}
oai=dfs1.current();
int return_state=dfs.next();
if (return_state==dfs_shrink || return_state==dfs_leaf)
   V.push(_oai.get_node());
return NEXT1;
```

In this phase, we process all depth first search trees, which can be visited starting at all nodes from the node stack *LeafStack* . Suppose that the current depth first search is not finished, then we execute the next step of the depth first search. Observe that the component number will now describe, in which component the current fixed node of the current dfs is located. `next_unseen()` will execute the simple next step of dfs (see figure 3.5). In the other case, if we have finished a depth first search, we look into the stack to get a new node that has not been seen before by a dfs. If we found one, we increase the component number since we entered a new component and initialize the new search with the node from the stack. If there is no more node, we have found all components and the algorithm is finished.

```
         if (dfs2.finished()) {
           while (!V.empty() ) {
             iai.update(V.pop());
             if (!get(ma,iai)) break;
           }
           if (V.empty() || !iai.has_node()) return NEXT_DONE;
           component++;
           dfs2.init(iai);
         }
         if (!dfs2.finished())
           dfs2.next_unseen();
         return NEXT2;
```

## 5.8   Relation to LEDA

As the use of templatized code is a very difficult task, all classes were special-
ized for the LEDA graph class. They were integrated to the LEDA software
tree since version 3.6.

To see how easy it is to use iterators in own algorithms, one may look on the
excerpt from the LEDA manual pages (see figure 7.3.2).

It is now easy to use iterators. There are wrapper classes for all possibilities
of data storage in LEDA that simplify the access in form of data accessors.
For example, for a given `node_array<bool>` we can create an instance of
`node_array_da<bool>` (see figure 7.3.3) and initialize it with the given node
array. The result will be a data accessor.

To summarize, here is the concrete C++-code:

```
    graph G;
    node_array<bool> _mark(G,false);
    OutAdjIt ai(G);
    node_array_da<bool> mark(_mark);
```

With these data structures we can now write a simple breadth first search
algorithm. At first we need a simple queue that can be templatized with the
element type `OutAdjIt`. Iterators can be updated using method `update` and
`curr_adj` computes the successor iterator in direction of the current incident
edge, i.e. the new fixed node would be the current adjacent node.

```
    queue<OutAdjIt> Q;
    OutAdjIt temp(ai);
    Q.clear();
    Q.append(ai);
```

```
set(mark,ai,true);
while (!Q.empty()) {
  ai.update(Q.pop());
  while (ai.valid()) {
    temp=ai.curr_adj();
    if (get(mark,temp)!=true) {
      set(mark,temp,true);
      Q.append(temp); }
    ++ai;
  }
}
```

In the graphiterator LEDA extension package there are not only the specialized versions of the data structures but also the templatized ones, which can be adapted to other graph representations. Additionally, all graph algorithms are implemented in generic fashion and may be reused in own applications. A nice observation is that algorithm classes are quite similar to iterator objects although they have not the same interface. It is easy (as shown in the package) to write a wrapper class that modifies the interface. For example, you can use an iterator that traverses the strongly connected components and can give you the component number of the current component.

# 6. CONCLUSION

In the following, I will conclude the results and experiences of the approach of using iterators in graph algorithms.

At first it is interesting to examine the overhead caused by the approach. Time and space costs may be described in $\mathcal{O}$-notation. If we compare two alternative approaches with the same complexity, we may be interested in constant factors. The overhead induced by using iterators, data accessors, and algorithm classes is a constant factor and is — if implemented in `C++` using traits classes — quite small. Additionally, algorithms that are time-critical (for example in real-time applications) maybe inspected even before the algorithm has been terminated and this might reduce the run–time.
For example, pipelining can be used without changing the algorithm or introducing overhead. Algorithms are controllable like a process without expensive and difficult interprocess solutions. Thus, resources can be managed optimally. If an underlying data structure is exchanged by a more efficient one (in a specific situation), then time and space bounds may be improved.

A large set of tests on different algorithms (all presented in this paper) has been tested with different kinds of generated graphs (heavy cyclic graphs, acyclic graphs, planar graphs, grid graphs, random graphs, triangulated planar graphs). The overall overhead, which is smaller than 100% can be quite acceptable, especially, if we consider the Dijkstra shortest path computations, which are between 60% and 80%. The Dijkstra shortest path computation on a selection of "real data"- graphs (train stations and connections in France, Germany, Switzerland and Austria) showed that the overhead was smaller than 70%. In near future, the overhead may even be smaller in situations where the run-time flexibility of dynamic binding is not required — all function calls could be inlined and optimized as if they were hand-coded. Hence, there is hope that progress in compiler technology will reduce the overhead even further.

It might be interesting to see how the design influences run-time performance compared to the standard implementation of Dijkstra in the LEDA package (see figure 7.1). To evaluate the overhead on real-world data, I solved the all-pairs shortest-path problem (in this case actually $n$ times Dijkstra, where $n$ is the number of nodes in the graph) on several train graphs (Germany, local trains in Germany, Austria, France, Switzerland and all of Europe). The overhead ranges from 48 % to 68 %.

This design is not only *efficient*, but also *easy to use*, because there are different implementations of data structures and algorithms for users with different experience in flexible algorithm programming. Additionally, there are LEDA manual pages, which have a practice-oriented layout. Sample algorithms are provided to serve as a pattern for implementing new ones.

Data structures may be improved in terms of *reliability* by adding consistency checks to data structures or creating wrapper classes for algorithms that do consistency checking. For example, safe iterators are provided as a demonstration on how to implement safety features in iterators.
Graphs may be represented by tables in a database and provide iterators as wrappers for cursors. Then every algorithm that is designed for the use of iterators may benefit from the databases *robustness*.

The LICD concept (see section 2.3) allows us to implement the data structures in object-oriented languages such as `C++`. Algorithms designed for these data structures can be adapted very easily since only language-specific syntactic differences have to be considered.
Since algorithms and data representations are decoupled, the algorithms can be used with any representation. So all data structures and algorithms are *portable* to various languages and can be used with different data representations.

The concepts presented here provide a high degree of *flexibility*: iterators may be extended with new functionality (e.g. filter predicates), data accessors may be changed in the way they retrieve information (e.g. LEDA edge array $\Rightarrow$ online calculation of Euclidean lengths), and algorithm classes may be specialized for a specific domain (e.g. Dijkstra shortest path between two locations, where computation runs simultaneously from both locations).

All of these features motivate the belief that this design is really reusable. A remaining question is — how big is the practical effort for designing these structures and substructures? In fact, designing algorithms for iterators and data accessors is more complicated than for more convenient data structures, e.g. the LEDA graph. Fortunately, this increased overhead will be rewarded in future — if code has to be reused in some other situation, because the existing algorithm code can be reused without change.

# 7. APPENDIX

## 7.1 Syntatic and Semantic Description

The following definitions are used in a language independent class definition $L = (S, M, F, I)$. I will use $\|x\|$ as the semantic value $x$.

- $r \in RExpr$: then

  - $r = \begin{cases} r_1 & \text{if } e_1 \\ r_2 & \text{if } e_2 \\ & \vdots \\ r_{k-1} & \text{if } e_{k_1} \\ r_k & \text{otherwise} \end{cases}$

    where $r_1, \ldots r_k \in ArithExpr$ and $e_1, \ldots e_{k-1} \in BoolExpr$

    $\|r\| = \begin{cases} \|r_i\| & \text{if } \exists 1 \leq i \leq k-1 : \|e_i\| = \texttt{true} \\ \|r_k\| & \text{otherwise} \end{cases}$

  - $r = r' \in ArithExpr$

    $\|r\| = \|r'\|$

- $r \in ArithExpr$: then

  - $r = (e\ a\ \ r')$ where $e \in Entity$, $a \in ArithOp$ and $r' \in ArithExpr$

    $\|r\| = \begin{cases} \|e\| + \|r'\| & \text{if } a = "+" \\ \|e\| - \|r'\| & \text{if } a = "-" \end{cases}$

  - $r = r' \in SetExpr$

    $\|r\| = \|r'\|$

  - $r = r' \in BoolExpr$

    $\|r\| = \|r'\|$

- $r \in Entity$: then

  - $r \in S$

  - $r = s.m(p_1, p_2, \ldots p_k)$ where $\exists s' \in S : name(s') = s$ and $s'$ is an object of licd $L_S = (S_S, M_S, F_S, I_S)$

$\exists m' \in M_S : name(m') = m$ and $|arglist(m')| = k$

$\|r\| = EvalMethodSelf(s, m, \|p_1\|, \ldots \|p_k\|)$ where $EvalMethodSelf$ evaluates the method $m$ in object $s$ with the parameter-list and returns itself

- $r = s.m(p_1, p_2, \ldots p_k).t$ where $\exists s' \in S : name(s') = s$ and $s'$ is an object of licd $L_S = (S_S, M_S, F_S, I_S)$

  $\exists m' \in M_S : name(m') = m$ and $|arglist(m')| = k$ and $\exists t' \in S_S : name(t') = t$

  $\|r\| = EvalMethod(s, m, \|p_1\|, \ldots \|p_k\|, t)$ where $EvalMethod$ evaluates the method $m$ in object $s$ with the parameter-list and returns the state $t$

- $r = s.f(p_1, p_2, \ldots p_k)$ where $\exists s' \in S : name(s') = s$ and $s'$ is an object of licd $L_S = (S_S, M_S, F_S, I_S)$

  $\exists f' \in F_S : name(f') = f$ and $|arglist(f')| = k$

  $\|r\| = EvalFunction(s, f, \|p_1\|, \ldots \|p_k\|)$ where $EvalFunction$ evaluates the function $f$ in object $s$ with the parameter-list and returns the computed value

- $r = f(p_1, p_2, \ldots p_k) : f$ is a mathematical function with $k$ parameters, $p_i \in ArithExpr$ for $1 \le i \le k$.

  $\|r\| = f(\|p_1\|, \ldots \|p_k\|)$

- $r$ is a constant

  $\|r\| = r$

- $r \in SetExpr$: then

  - $r = (r' \, a \, r'')$ where $e \in SetExpr$ and $a \in SetOp$ and $r' \in SetExpr$

    $\|r\| = \begin{cases} \|r'\| \cup \|r''\| & \text{if } a = " \cup " \\ \|r'\| \cap \|r''\| & \text{if } a = " \cap " \\ \|r'\| \setminus \|r''\| & \text{if } a = " - " \end{cases}$

  - $r = \{r'\}$ where $r' \in ArithExpr$

    $\|r\| = \{\|r'\|\}$

  - $r = \{(r', r'')\}$ where $r', r'' \in ArithExpr$

    $\|r\| = \{(\|r'\|, \|r''\|)\}$ (this is an extension for safe iterators)

  - $r = \{(r', r''|e)\}$ where $r', r'', e \in ArithExpr$

    $\|r\| = \{(\|r'\|, \|r''\| \, \|e\|)\}$

- $r \in BoolExpr$: then

  - $r = (e \, a \, r')$ where $e \in Entity$ and $a \in BoolOp$ and $r' \in BoolExpr$

    $\|r\| = \begin{cases} \|e\| \wedge \|r'\| & \text{if } a = " \wedge " \\ \|e\| \vee \|r'\| & \text{if } a = " \vee " \end{cases}$

– $r = (\neg r')$

  $\|r\| = \neg \|r'\|$

– $r = \exists x \in i : r'$

  where $\|r'\| \in ArithExpr$, $i \in SetExpr$ and $x$ appears in $r'$

  $\|r\| = \begin{cases} \texttt{true} & \text{if } \exists y \in \|i\| : r'[x/y] = \text{"true"} \\ \texttt{false} & \text{otherwise} \end{cases}$ where $r[a/b]$

  substitutes $a$ with $b$ in $r$

– $r = \forall x \in i : r'$

  where $\|r'\| \in ArithExpr$, $i \in SetExpr$ and $x$ appears in $r'$

  $\|r\| = \begin{cases} \texttt{true} & \text{if } \forall y \in \|i\| : r'[x/y] = \text{"true"} \\ \texttt{false} & \text{otherwise} \end{cases}$ where $r[a/b]$

  substitutes $a$ with $b$ in $r$

– $r = (e\ a\ r')$ where $e \in Entity$ and $a \in SetOp2$ and $r' \in SetExpr$

  $\|r\| = \begin{cases} \|e\| \in \|r'\| & \text{if } a = \text{"}\in\text{"} \\ \|e\| \notin \|r'\| & \text{if } a = \text{"}\notin\text{"} \end{cases}$   $\|r\| =$

  $\begin{cases} \|e\| \neq \|r'\| & \text{if } a = \text{"}\neq\text{"} \\ \|e\| = \|r'\| & \text{if } a = \text{"}=\text{"} \\ \|e\| < \|r'\| & \text{if } a = \text{"}<\text{"} \\ \|e\| > \|r'\| & \text{if } a = \text{"}>\text{"} \\ \|e\| \leq \|r'\| & \text{if } a = \text{"}\leq\text{"} \\ \|e\| \geq \|r'\| & \text{if } a = \text{"}\geq\text{"} \end{cases}$

– $r = r' \in BoolConst$

  $\|r\| = \begin{cases} \texttt{true} & \text{if } r' = \text{"true"} \\ \texttt{false} & \text{if } r' = \text{"false"} \end{cases}$

- $ArithOp = \{\text{"}+\text{"}, \text{"}-\text{"}\}$

- $BoolOp = \{\text{"}\wedge\text{"}, \text{"}\vee\text{"}\}$

- $CompOp = \{\text{"}\neq\text{"}, \text{"}=\text{"}, \text{"}<\text{"}, \text{"}>\text{"}, \text{"}\leq\text{"}, \text{"}\geq\text{"}\}$

- $SetOp = \{\text{"}-\text{"}, \text{"}\cap\text{"}, \text{"}\cup\text{"}\}$

- $SetOp2 = \{\text{"}\in\text{"}, \text{"}\notin\text{"}\}$

- $BoolConst = \{\text{"}true\text{"}, \text{"}false\text{"}\}$

- $SetConst = \{\text{"}\emptyset\text{"}\}$

## 7.2   Additional Classes

### STL Wrapper

There is a **wrapper** (design pattern, see [GHJV95]) class that has the interface of iterators from the STL. Basically, "`operator*()`" is redefined in the right way, i.e. an additional data accessor is required to return an appropriate value (see section 2.4).

**L.I.C.Definition 7.1 STL iterator** *for $G = (V, E)$*

1. *it is an iterator*

2. *da is a data accessor*

3. `Creation`$(it', da').it \equiv it'$

4. `Creation`$(it', da').da \equiv da'$

5. `Value`$() \equiv da(it)$

For example one may write an STL–like iterator for the node set of a graph and has given a data accessor *distance*:

> $distance :=$ data accessor that returns the color of a node
> $it :=$ node iterator for $G = (V, E)$
> $it.$`Creation`(first node of $G$)
> $stl :=$ stl iterator for $G = (V, E)$
> $stl.$`Creation`$(it, distance)$
> **while** $stl.v \neq \varepsilon$
>        **if** $stl.$`Value`$() \equiv red$
>          do something with $stl$
>        $stl.$`Succ`$()$

Now, *stl* is an iterator that is associated with one single attribute. Although single attributed iterators are not always sufficient, they may be useful in certain algorithms. An STL–like iterator can be applied to STL algorithms, as long as the STL is available for the implementation language.

## 7.3   Some LEDA Manual Pages

All manual pages are available at [LEDA].

### 7.3.1 Node Iterators (NodeIt)

**1. Definition**

a variable *it* of class *NodeIt* is a linear node iterator that iterates over the node set of a graph; the current node of an iterator object is said to be "marked" by this object.

**2. Creation**

*NodeIt it;*   introduces a variable *it* of this class associated with no graph.

*NodeIt it(graph G);*

introduces a variable *it* of this class associated with *G*. The graph is initialized by *G*. The node is initialized by *G.first_node( )*.

*NodeIt it(graph G, node n);*

introduces a variable *it* of this class marked with *n* and associated with *G*.
Precondition: *n* is a node of *G*.

**3. Operations**

*void*     *it*.init(*graph G*)

associates *it* with *G* and marks it with *G.first_node( )*.

*void*     *it*.init(*graph G, node v*)

associates *it* with *G* and marks it with *v*.

*void*     *it*.reset( )

resets *it* to *G.first_node( )*, where *G* is the associated graph.

*void*     *it*.make_invalid( )

makes *it* invalid, i.e. *it.valid( )* will be false afterwards and *it* marks no node.

*void*     *it*.reset_end( )

resets *it* to *G.last_node( )*, where *G* is the associated graph.

| | |
|---|---|
| *void* | *it*.update(*node n*) |

         *it* marks *n* afterwards.

| | |
|---|---|
| *void* | *it*.insert( ) |

         creates a new node and *it* marks it afterwards.

| | |
|---|---|
| *void* | *it*.del( ) |

         deletes the marked node, i.e. *it.valid*( ) returns false afterwards.
Precondition: *it.valid*( ) returns true.

| | |
|---|---|
| *NodeIt&* | *it = it2* |

         *it* is afterwards associated with the same graph and node as *it2*. This method returns a reference to *it*.

| | |
|---|---|
| *bool* | *it == it2* |

         returns true if and only if *it* and *it2* are equal, i.e. if the marked nodes are equal.

| | |
|---|---|
| *node* | *it*.get_node( ) |

         returns the marked node or nil if *it.valid*( ) returns false.

| | |
|---|---|
| *graph* | *it*.get_graph( ) |

         returns the associated graph.

| | |
|---|---|
| *bool* | *it*.valid( ) |

         returns true if and only if end of sequence not yet passed, i.e. if there is a node in the node set that was not yet passed.

| | |
|---|---|
| *bool* | *it*.eol( ) |

         returns !*it.valid*( ) which is true if and only if there is no successor node left, i.e. if all nodes of the node set are passed (eol: end of list).

| | |
|---|---|
| *NodeIt&* | *++it* |

         performs one step forward in the list of nodes of the associated graph. If there is no successor node, *it.eol*( ) will be true afterwards. This method returns a reference to *it*.
Precondition: *it.valid*( ) returns true.

*NodeIt&* − − *it*

performs one step backward in the list of nodes of the asso­ciated graph. If there is no predecessor node, *it.eol*( ) will be true afterwards. This method returns a reference to *it*. Precondition: *it.valid*( ) returns true.

## 4. Implementation

Creation of an iterator and all methods take constant time.

### 7.3.2  Adjacency Iterators for leaving edges (OutAdjIt)

## 1. Definition

a variable *it* of class *OutAdjIt* is an adjacency iterator that marks a node (which is fixed in contrast to linear node iterators) and iterates over the edges that leave this node.

## 2. Creation

*OutAdjIt   it*;        introduces a variable *it* of this class associated with no graph.

*OutAdjIt   it(graph& G)*;

introduces a variable *it* of this class associated with *G*. The node is initialized by *G.first_node*( ) and the edge by *G.first_adj_edge*(*n*) where *n* is the marked node.

*OutAdjIt   it(graph& G, node   n)*;

introduces a variable *it* of this class marked with *n* and associated with *G*.  The marked edge is initialized by *G.first_adj_edge*(*n*). Precondition: *n* is a node of *G*.

*OutAdjIt   it(graph& G, node   n, edge e)*;

introduces a variable *it* of this class marked with *n* and *e* and associated with *G*. Precondition:  *n* is a node and *e* an edge of *G* and *source*(*e*) = *n*.

### 3. Operations

| | | |
|---|---|---|
| *void* | *it*.init(*graph G*) | associates *it* with *G* and marks it with $n' = G.first\_node(\ )$ and $G.first\_adj\_edge(n')$. |

*void*    *it*.init(*graph G, node n*)

associates *it* with *G* and marks it with *n* and $G.first\_adj\_edge(n)$.
Precondition: *n* is a node of *G*.

*void*    *it*.init(*graph G, node n, edge e*)

associates *it* with *G* and marks it with *n* and *e*.
Precondition: *n* is a node and *e* an edge of *G* and $source(e) = n$.

| | | |
|---|---|---|
| *void* | *it*.update(*edge  e*) | *it* marks *e* afterwards. |
| *void* | *it*.reset(\ ) | resets *it* to $G.first\_adj\_edge(n)$ where *G* and *n* are the marked node and associated graph. |
| *void* | *it*.insert(*OutAdjIt other*) | |

creates a new leaving edge from the marked node of *it* to the marked node of *other*. *it* is marked with the new edge afterwards. The marked node of *it* does not change.

| | | |
|---|---|---|
| *void* | *it*.del(\ ) | deletes the marked leaving edge, i.e. *it.valid*(\ ) returns false afterwards.<br>Precondition: *it.valid*(\ ) returns true. |
| *void* | *it*.reset_end(\ ) | resets *it* to $G.last\_adj\_edge(n)$ where *G* and *n* are the marked node and associated graph. |
| *void* | *it*.make_invalid(\ ) | makes *it* invalid, i.e.  *it.valid*(\ ) will be false afterwards and *it* marks no node. |
| *void* | *it*.update(*node  n*) | |

*it* marks *n* and the first leaving edge of *n* afterwards.

*void*    *it*.update(*node  n, edge  e*)

*it* marks *n* and *e* afterwards.

| | | |
|---|---|---|
| *OutAdjIt& it = it2* | | assigns *it2* to *it*. This method returns a reference to *it*. |
| *bool* | *it == it2* | returns true if and only if *it* and *it2* are equal, i.e. if the marked nodes and edges are equal. |
| *bool* | *it*.has_node( ) | returns true if and only if *it* marks a node. |
| *bool* | *it*.eol( ) | returns !*it.valid*( ) which is true if and only if there is no successor edge left, i.e. if all edges of the edge set are passed (eol: end of list). |
| *bool* | *it*.valid( ) | returns true if and only if end of sequence not yet passed, i.e. if there is an edge in the edge set that was not yet passed. |
| *edge* | *it*.get_edge( ) | returns the marked edge or nil if *it.valid*( ) returns false. |
| *node* | *it*.get_node( ) | returns the marked node or nil if *it.has_node*( ) returns false. |
| *graph* | *it*.get_graph( ) | returns the associated graph. |
| *OutAdjIt* | *it*.curr_adj( ) | returns a new adjacency iterator that is associated with $n' = target(e)$ and $G.first\_adj\_edge(n')$ where $G$ is the associated graph.<br>Precondition: *it.valid*( ) returns true. |
| *OutAdjIt& ++it* | | performs one step forward in the list of outgoing edges of the marked node. If there is no successor edge, *it.eol*( ) will be true afterwards. This method returns a reference to *it*.<br>Precondition: *it.valid*( ) returns true. |
| *OutAdjIt& − − it* | | performs one step backward in the list of outgoing edges of the marked node. If there is no predecesssor edge, *it.eol*( ) will be true afterwards. This method returns a reference to *it*.<br>Precondition: *it.valid*( ) returns true. |

## 4. Implementation

Creation of an iterator and all methods take constant time.

### 7.3.3   Node Array Data Accessor (node_array_da)

## 1. Definition

An instance *da* of class *node_array_da<T>* is instantiated with a LEDA `node_array<T>`.

The data in the node array can be accessed by the functions *get(da, it)* and *set(da, it, value)* that take as parameters an instance of *node_array_da<T>* and an iterator, see below.

For *node_map<T>* there is the variant *node_map_da<T>* which is defined completely analogous to *node_array_da<T>*.

## 2. Creation

*node_array_da<T>   da*;

                introduces a variable *da* of this class that is not bound.

*node_array_da<T>   da(node_array<T>& na)*;

                introduces a variable *da* of this class bound to *na*.

## 3. Operations

| *T* | get(*node_array_da<T> da, Iter it*) |
| --- | --- |
| | returns the associated value of *it* for this accessor. |
| *void* | set(*node_array_da<T>& da, Iter it, T val*) |
| | sets the associated value of *it* for this accessor to the given value. |

## 4. Implementation

Constant Overhead.

**5. Example**

We count the number of 'red nodes' in a parameterized graph G.

```
int count_red(graph G, node_array<color> COL) {
  node_array_da<color> Color(COL);
  int counter=0;
  NodeIt it(G);
  while (it.valid()) {
    if (get(Color,it)==red) counter++;
    it++; }
  return counter;
}
```

Suppose we want to make this 'algorithm' flexible in the representation of colors. Then we could write this version:

```
template<class DA>
int count_red_t(graph G, DA Color) {
  int counter=0;
  NodeIt it(G);
  while (it.valid()) {
    if (get(Color,it)==red) counter++;
    it++; }
  return counter;
}
```

With the templatized version it is easily to customize it to match the interface of the version:

```
int count_red(graph G, node_array<color> COL) {
  node_array_da<color> Color(COL);
  return count_red_t(G,Color); }
```

## 7.4  Additional Figures

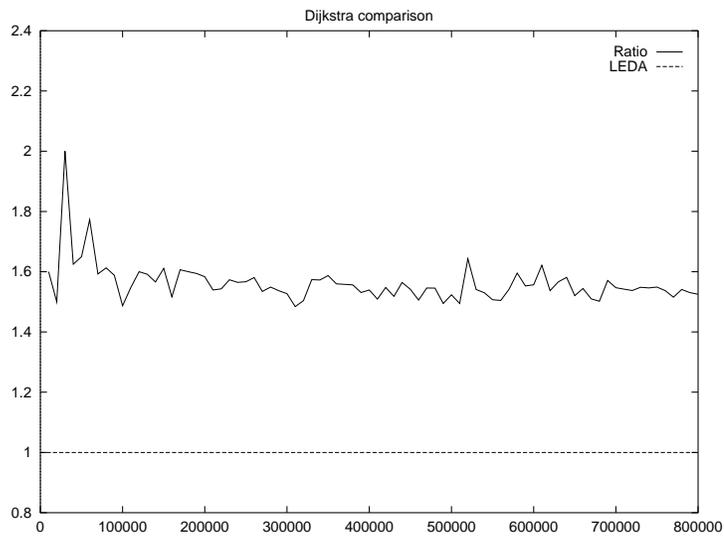

*Fig. 7.1:* Runtime comparison of different versions of Dijkstra (LEDA and iterator version) on randomly generated graphs with 1.000 nodes and edges from 10.000 to 800.000). The graphs are generated by choosing arbitrarily 3 nodes and connecting them with edges until the number of edges is reached; no parallel edges are generated.

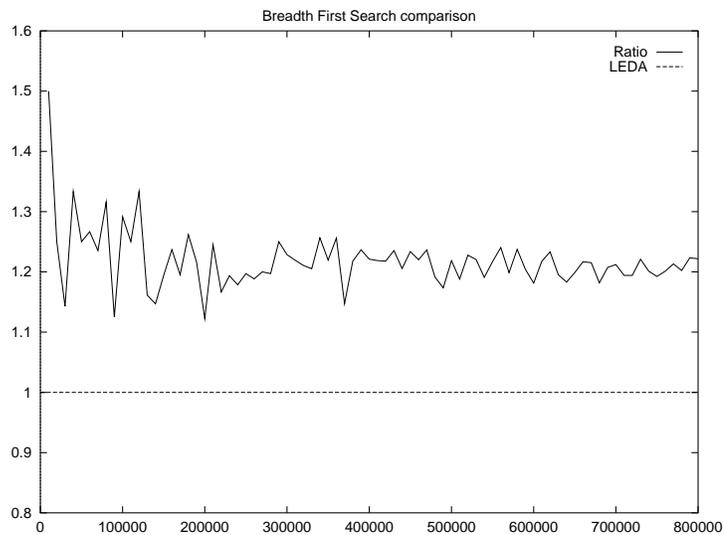

*Fig. 7.2:* Runtime comparison of different versions of Breadth First Search (LEDA and iterator version) on randomly generated graphs with 1.000 nodes and edges from 10.000 to 800.000). (see figure 7.1)

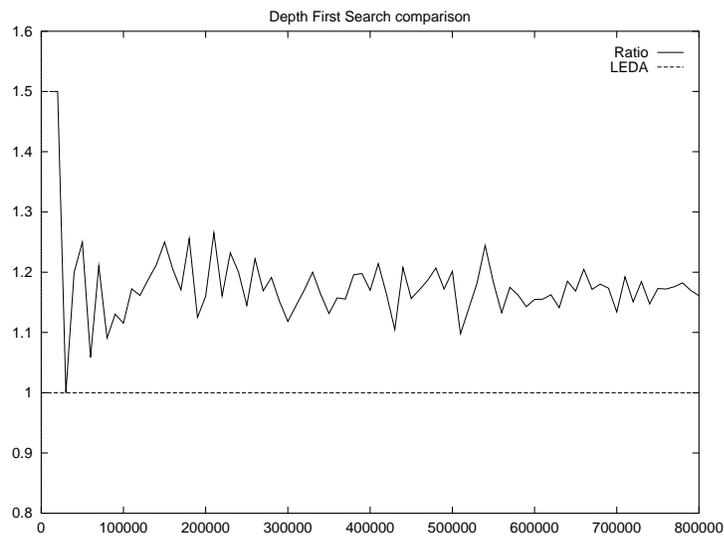

*Fig. 7.3:* Runtime comparison of different versions of Depth First Search (LEDA and iterator version) on randomly generated graphs with 1.000 nodes and edges from 10.000 to 800.000). (see figure 7.1)

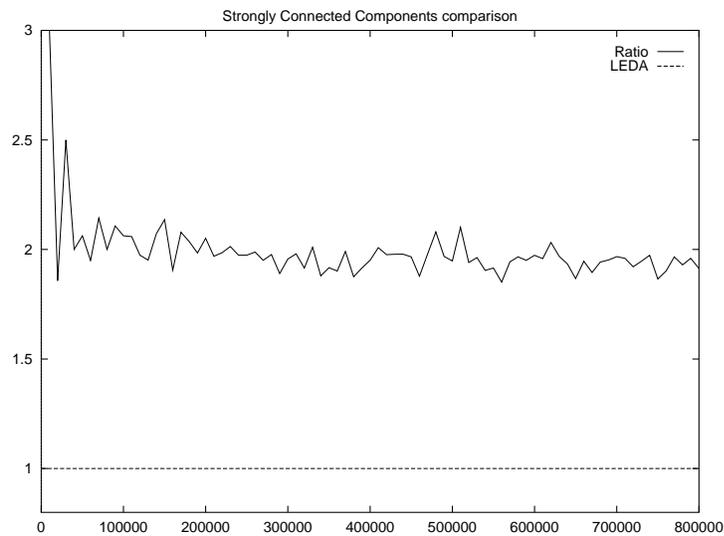

*Fig. 7.4:* Runtime comparison of different versions of Strongly Connected Components (LEDA and iterator version) on randomly generated graphs with 1.000 nodes and edges from 10.000 to 800.000). (see figure 7.1)

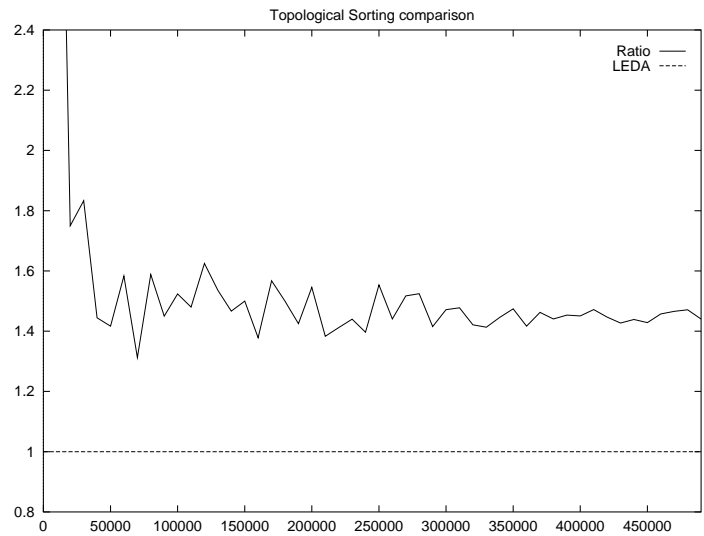

*Fig. 7.5:* Runtime comparison of different versions of Topological Sorting (LEDA
and iterator version) on randomly generated graphs with 1.000 nodes and
edges from 10.000 to 480.000). (see figure 7.1)

Konstanzer Schriften in Mathematik und Informatik are available at `http://www.informatik.uni-konstanz.de/Preprints`.